\documentclass{aa}  

\usepackage{graphicx}
\usepackage{txfonts}
\bibpunct[; ]{(}{)}{;}{a}{}{;}

\begin{document}

    \title{Simulations of imaging the event horizon of Sagittarius A* from space}

   \author{Freek Roelofs\inst{1} \and Heino Falcke\inst{1}\inst{2}\inst{3} \and Christiaan Brinkerink\inst{1} \and Monika Mo\'scibrodzka\inst{1} \and Leonid I. Gurvits\inst{4}\inst{5} \and Manuel Martin-Neira\inst{6} \and Volodymyr Kudriashov\inst{1} \and Marc Klein-Wolt \inst{1} \and Remo Tilanus \inst{1}\inst{7} \and Michael Kramer\inst{2} \and Luciano Rezzolla\inst{8}}

   \institute{Department of Astrophysics, Institute for Mathematics, Astrophysics and Particle Physics, Radboud University, PO Box 9010, 6500 GL Nijmegen, The Netherlands\\
              \email{f.roelofs@astro.ru.nl}
         \and Max-Planck-Institut f\"{u}r Radioastronomie, Auf dem H\"{u}gel 69, D-53121 Bonn, Germany
         \and ASTRON, The Netherlands Institute for Radio Astronomy, Postbus 2, NL-7990 AA Dwingeloo, The Netherlands 
        \and Joint Institute for VLBI ERIC, P.O. Box 2, 7990 AA Dwingeloo, The Netherlands  
        \and Department of Astrodynamics and Space Missions, Delft University of Technology, 2629 HS Delft, The Netherlands
        \and European Space Research and Technology Centre (ESTEC), The European Space Agency, Keplerlaan 1, 2201 AZ Noordwijk, The Netherlands   
        \and Leiden Observatory, Leiden University, P.O. Box 9513, 2300 RA Leiden, The Netherlands 
        \and Institut f\"ur Theoretische Physik, Goethe-Universit\"at, Max-von-Laue-Str. 1, 60438 Frankfurt, Germany}

   \date{\today}

 
  \abstract
   {It has been proposed that Very Long Baseline Interferometry (VLBI) at sub-millimeter waves will allow us to image the shadow of the black hole in the center of our Milky Way, Sagittarius\,A* (Sgr\,A*), and thereby test basic predictions of the theory of general relativity.}
   {This paper presents imaging simulations of a new Space VLBI (SVLBI) mission concept. An initial design study of the concept has been presented in the form of the Event Horizon Imager (EHI). The EHI may be suitable for imaging Sgr\,A* at high frequencies (up to $\sim$\,690\,GHz), which has significant advantages over performing ground-based VLBI at 230\,GHz. The concept EHI design consists of two or three satellites in polar or equatorial circular Medium-Earth Orbits (MEOs) with slightly different radii. Due to the relative drift of the satellites along the individual orbits over the course of several weeks, this setup will result in a dense spiral-shaped $uv$-coverage with long baselines (up to $\sim$\,60\,G$\lambda$), allowing for extremely high-resolution and high-fidelity imaging of radio sources.}
   {We simulate observations of general relativistic magnetohydrodynamics (GRMHD) models of Sgr\,A* for the proposed configuration and calculate the expected noise based on preliminary system parameters. On long baselines, where the signal-to-noise ratio may be low, fringes could be detected if the system is sufficiently phase stable and the satellite orbits can be reconstructed with sufficient accuracy. Averaging visibilities accumulated over multiple epochs of observations could then help improving the image quality. With three satellites instead of two, closure phases could be used for imaging.}
   {Our simulations show that the EHI could be capable of imaging the black hole shadow of Sgr\,A* with a resolution of 4\,$\mu$as (about 8\,\% of the shadow diameter) within several months of observing time.}
   {Our preliminary study of the EHI concept shows that it is potentially of high scientific value, as it could be used to measure black hole shadows much more precisely than with ground-based VLBI, allowing for stronger tests of General Relativity and accretion models.}


   \maketitle
%

\section{Introduction}

\subsection{Sagittarius A*}

Sagittarius A* (Sgr\,A*) is a strong radio source at the center of the Milky Way. Proper motion measurements of Sgr\,A* with respect to two extragalactic radio sources close in angular separation have confirmed that it is the dynamical center of the Galaxy \citep{Reid1999, Reid2004}. By monitoring stellar orbits at the Galactic Center, it was found that at the position of the radio source there is an object with a mass of $4.3 \pm 0.4 \times 10^6 M_{\odot}$ at $8.3 \pm 0.4$\,kpc from Earth \citep{Ghez2008, Gillessen2009}. From the monitoring data, it has been estimated that only up to $\sim 4\times 10^5M_{\odot}$ within a region of 1 pc could be attributed to an extended mass distribution, leaving a supermassive black hole as the only physically feasible explanation. The radio emission is therefore believed to be produced by accretion onto and outflow from the black hole.  

Since its discovery by \citet{Balick1974}, Sgr\,A* has been monitored frequently at various wavelengths. The broad-band radio spectrum is flat-to-inverted up to the ``sub-mm bump'' at $\sim$\,10$^{12}$\,Hz, which is interpreted as a transition from optically thick to optically thin synchrotron emission \citep[]{Falcke1998, Bower2015}. The sub-mm bump was explained by a compact synchrotron-emitting region with a radius comparable to that of the event horizon of Sgr\,A*, which led to the prediction of the appearance of a roughly circular ``shadow'' of the event horizon surrounded by gravitationally lensed emission from an accretion flow at sub-mm wavelengths \citep{Falcke2000}.

Very Long Baseline Interferometry (VLBI) observations of Sgr\,A* at centimeter wavelengths have confirmed that the apparent size of the radio source becomes smaller towards higher frequencies \citep{Bower2004, Shen2005} as is expected for a scatter-dominated source. Due to this blurring by interstellar scattering, the source looks like a Gaussian that increases in size with the observing wavelength squared. The intrinsic source structure starts to contribute to the measured size at $\lambda \lesssim 6$\,cm, and dominates over the scattering effects in the millimeter regime \citep{Bower2006, Doeleman2008}. Hence, only at mm (and shorter) wavelengths, an image of the intrinsic structure of the central object showing the black hole shadow can be obtained.

Sgr\,A* has also been detected in the near-infrared and X-ray regime \citep[e.g.][]{Yusef-Zadeh2006, Porquet2008, Do2009, Meyer2009}, where it shows simultaneous flares on $\sim$ hour time scales. In the submillimeter regime, variability seems to occur on slightly longer time scales \citep{Marrone2008, Dexter2014}. With light crossing time arguments, the size of the flare emission region can be constrained to $\lesssim 10$\,Schwarzschild radii.

\subsection{(Sub-)millimeter ground-based VLBI of Sgr\,A*}
\label{sec:eht}

The apparent angular size of the black hole shadow of Sgr\,A* assuming zero spin is $2\sqrt{27} GM/c^2D \approx 53$\,$\mu$as, where $G$ is Newton's gravitational constant, $M\approx 4.3\times10^6 M_{\odot}$ is the black hole mass, $c$ is the speed of light, and $D\approx 8.3$\,kpc is the distance to the observer \citep{Falcke2000, Johannsen2012}. This makes Sgr\,A* the largest black hole in the sky in angular size, and therefore the most promising candidate to image a black hole shadow. Another prime candidate is the black hole in the center of the giant elliptical galaxy M\,87. This black hole is about 2000 times further away \citep{Bird2010}, but also about 1000-1500 times more massive than Sgr\,A* \citep{Gebhardt2011, Walsh2013}, so that its angular size is comparable to that of Sgr\,A*.

At 1.3\,mm, the black hole shadow of Sgr\,A* can be resolved with Earth-size baselines of $\sim$\,9\,G$\lambda$, yielding an angular resolution of $\sim$\,23\,$\mu$as. Resolving the black hole shadow is the main aim of the Event Horizon Telescope (EHT), a VLBI array consisting of (sub-)mm stations across the globe \citep{Fish2013}. Observations with the Combined Array for Research in Millimeter-wave Astronomy (CARMA) in California, the Sub-Millimeter Telescope (SMT) in Arizona, and the Sub-Millimeter Array (SMA) in Hawaii have resolved structure of Sgr\,A* and M\,87 on event horizon scales \citep{Doeleman2008,Fish2011, Doeleman2012,Akiyama2015,Johnson2015pol,Lu2018}. With this number of stations the $uv$-coverage is insufficient to image the source, but the size of Sgr\,A* was determined to be $\sim$\,40\,$\mu$as, indicating structure on scales smaller than the event horizon. The intrinsic source size was measured to be $(120\pm34)\times (100 \pm 18)$\,$\mu$as at 3.5\,mm \citep{Shen2005, Ortiz2016, Issaoun2019}, and $(354\pm4)\times (126^{+55}_{-41})$\,$\mu$as at 7\,mm \citep{Bower2014}.
 
From the measurement of non-zero closure phases, the sum of interferometric phases on a triangle of baselines, \citet{Fish2016} concluded that the source is asymmetric at 1.3\,mm, which can be attributed to either the intrinsic source structure or scattering effects. Asymmetric structure due to scattering or instrinsic source morphology was also found in closure phase and amplitude measurements at 3.5\,mm \citep{Ortiz2016, Brinkerink2016, Brinkerink2019}. In April 2017, 1.3\,mm observations of Sgr\,A* have been performed with 8 stations as part of the EHT: the IRAM 30-meter telescope on Pico Veleta in Spain, the Large Millimeter Telescope (LMT) in Mexico, the Atacama Large Millemeter Array (ALMA), the Atacama Pathfinder Experiment (APEX) telescope in Chile, the SMT in Arizona, the SMA and James Clerk Maxwell Telescope (JCMT) in Hawaii, and the South Pole Telescope (SPT). With the increased $uv$-coverage and sensitivity of e.g. ALMA and the LMT, these observations may be suitable for image reconstruction. 

Imaging the black hole shadow could provide a strong-field test of general relativity as it predicts its size and shape \citep[e.g.][]{Bambi2009, Johannsen2010, Goddi2016, Psaltis2018}. \citet{Psaltis2015} show that Sgr\,A* is the optimal target for a general relativistic null hypothesis test because of strong constraints on the opening angle of one gravitational radius $m=GM/Dc^2$, which is known to within $\sim$\,$4$\,\% from stellar monitoring observations \citep{Ghez2008, Gillessen2009}. Uncertainties in our prior knowledge of the effect of interstellar scattering, which is still severe at 230\,GHz, pose limitations to the accuracy of shadow opening angle measurements (see also Section \ref{sec:scat}). \citet{Psaltis2015} infer that the general relativistic null hypothesis can in principle be tested down to the $\sim$\,10\,\% level at 230\,GHz. \citet{Mizuno2018} compare synthetic EHT observations of a general relativistic magnetohydrodynamics (GRMHD, see also Section \ref{sec:simul}) model with accretion onto a Kerr black hole to synthetic observations of a GRMHD model with accretion onto a dilaton black hole, the latter of which is taken as a representative solution of an alternative theory of gravity. They show that, with the observational setup of the 2017 EHT observations, it could be extremely difficult to distinguish between these cases. Similarly, \citet{Olivares2018} show that it could also be difficult for the EHT to distinguish between a black hole and a boson star, although the differences in source size are slightly larger here. 

EHT observations could also shed light on the nature of the accretion flow, which may be dominated by emission from an accretion disk or relativistic jet \citep[e.g.][]{Falcke2000b, Yuan2003, Fish2009, Dexter2010, Gold2017}. \citet{Moscibrodzka2014} generate ray-traced GRMHD images for different electron temperature prescriptions leading to disk- or jet-dominated models, the latter providing a better fit to the radio spectrum of Sgr\,A*. \citet{Chan2015} argue that with EHT images, one may be able to distinguish disk-dominated from jet-dominated models, but that additional information would be needed to measure plasma and black hole properties within these models. \citet{Broderick2016} use EHT closure phases to constrain the dimensionless black hole spin parameter $a_*$, disfavoring values larger than $\sim$\,0.5. A value of 1 would correspond to a maximally spinning black hole. They also determine the black hole inclination and position angle to within $\sim$\,a few tens of degrees, all within the context of semi-analytic radiatively inefficient disk models. The model fits show consistency over several observation epochs spanning seven years.

EHT observations could thus produce the first image of a black hole shadow and put constraints on different accretion flow models. However, due to interstellar scattering effects for Sgr\,A* and limited $uv$-coverage, it will likely be difficult to perform high-precision tests of general relativity and measure black hole and plasma parameters with high accuracy. Observations performed at substantially higher frequencies would be less affected by interstellar scattering and increase the image resolution, allowing for more precise tests of general relativity and accretion models. For example, a resolution of $\lesssim$\,10\,$\mu$as would start to make it possible to visually distinguish between the Kerr and dilaton black hole shadows in \citet{Mizuno2018}, and between a black hole and boson star in \citet{Olivares2018}.

\subsection{Space VLBI}
VLBI is not only carried out from the ground, but also from space. This allows one to observe with longer baselines and thus obtain a higher angular resolution. Also, there are no atmospheric corruptions for space-based antennas. The first Space VLBI (SVLBI) observations were done by \citet{Levy1986}, who detected fringes for three Active Galactic Nuclei (AGN) at 2.3\,GHz on baselines of up to 1.4 Earth diameters between the Tracking and Data Relay Satellite System (TDRSS) and ground-based telescopes in Australia and Japan. The first dedicated SVLBI mission was VSOP \citep{Hirabayashi1998, Hirabayashi2000}, with an 8-meter antenna carried by the satellite HALCA, orbiting between 560 (perigee) and 21,000\,km (apogee) above the Earth's surface. It was operational between 1997 and 2003, imaging bright AGN and masers at 1.6 and 5.0\,GHz with a network of ground-based telescopes.

The second and currently only operational SVLBI mission is RadioAstron, which has a 10-meter antenna carried by the Spektr-R spacecraft, operating at wavelengths between 1.3 and 92\,cm \citep{Kardashev2013}. It has an orbital perigee altitude of about 10,000\,km and apogee of about 350,000 km, making it the largest interferometer to date. At the maximum frequency of 22\,GHz, the resolution achievable with RadioAstron (7\,$\mu$as) is in principle high enough to resolve event horizon-scale structures of Sgr\,A*. However, at this frequency the intrinsic structure of Sgr\,A* is blurred by interstellar scattering too severely to image the black hole shadow (see Sec. \ref{sec:scat}). Also, the high resolution is only achievable in one direction because of the highly elliptical orbit of Spektr-R.

Studies of two-element SVLBI setups were performed in the iARISE project \citep{Ulvestad1999, Murphy2005} and studies for the Chinese space Millimeter-wavelength VLBI array \citep[][Ji Wu, private communication]{Hong2014}. 

There are some examples of space-based submillimeter telescopes operating at the high frequencies (up to $\sim$\,690\,GHz) considered in this work. The ESA Herschel satellite had a 3.5-meter dish, and its on-board spectrometer HIFI covered wavelengths between $\sim$\,0.16 and 0.6\,mm \citep[480-1250 and 1410-1910\,GHz;][]{Graauw2010}. The Swedish-led Odin satellite had a 1.1-meter dish and operated between 486 and 580\,GHz and at 119 GHz \citep{Frisk2003}. The ESA Planck satellite had a 1.6 by 1.9-meter primary mirror and instruments sensitive to frequencies between 30 and 857\,GHz \citep{Tauber2010}.

\subsection{Outline}
This paper investigates the possible imaging capabilities of a new SVLBI system concept consisting of two or three satellites in polar or equatorial circular Medium Earth Orbits (MEOs). With the individual satellites in slightly different orbits, this configuration has the capability to image Sgr\,A* and other black holes (e.g. M\,87) with a resolution that is an order of magnitude higher than the resolution that can be obtained from Earth. We perform simulated observations of Sgr\,A* with realistic source and system parameters in order to assess the expected image quality that could be obtained with this setup. The system concept will be introduced in Section \ref{sec:mission}. Our source models are described in Section \ref{sec:simul}. Section \ref{sec:sim} describes our simulated observations. The simulation results are presented in Section \ref{sec:results}, and our conclusions and ideas for future directions are summarized in Section \ref{sec:sum}.
 
\section{System setup}
\label{sec:mission}
\subsection{Antennas, orbits and $uv$-coverage}
The SVLBI setup considered in this paper consists of two or three satellites orbiting Earth in circular MEOs at slightly different radii. We first consider the setup of the initial design study conducted by \citet{Martin2017} and \citet{Kudriashov2017} for the purpose of the Event Horizon Imager (EHI). They propose launching satellites equipped with $\sim$\,3-meter reflectors into MEOs with radii of around $\sim$\,14,000\,km. The MEOs should be circular and either polar or equatorial for stability purposes. In contrast to VSOP and RadioAstron, the EHI concept is a pure space-space interferometer rather than a ground-space interferometer, observing at frequencies up to $\sim$\,690\,GHz. 

Observing at high frequencies is possible in space because there is no phase corruption or signal attenuation by the atmosphere. There are several reasons for increasing the observing frequency. Firstly, the angular resolution of the array increases with frequency as the baseline length measured in wavelengths increases. Secondly, the effect of interstellar scattering on the observed image (discussed in Sec. \ref{sec:scat}) will be considerably smaller. Also, the emitted radiation will originate from closer to the black hole, tracing the lensed photon ring more closely \citep[e.g.][see also Sec. \ref{sec:simul}]{Falcke1993, Moscibrodzka2009}. The latter effect also causes the image variability to be generally less profound at high frequencies, since it is confined to a smaller region that is more dominated by general relativistic effects.

\begin{figure*}
\centering
\includegraphics[width=.49\hsize]{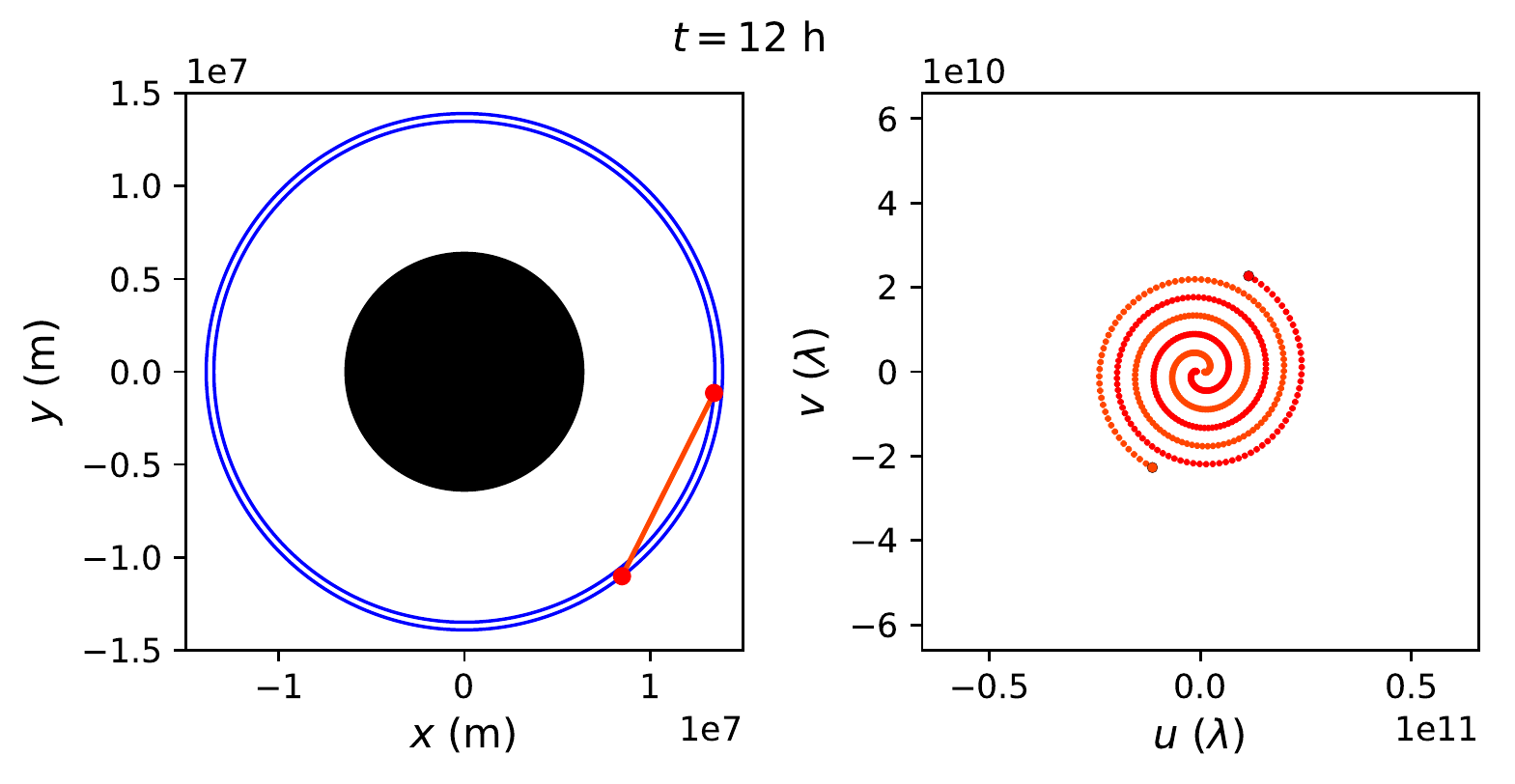}
\includegraphics[width=.49\hsize]{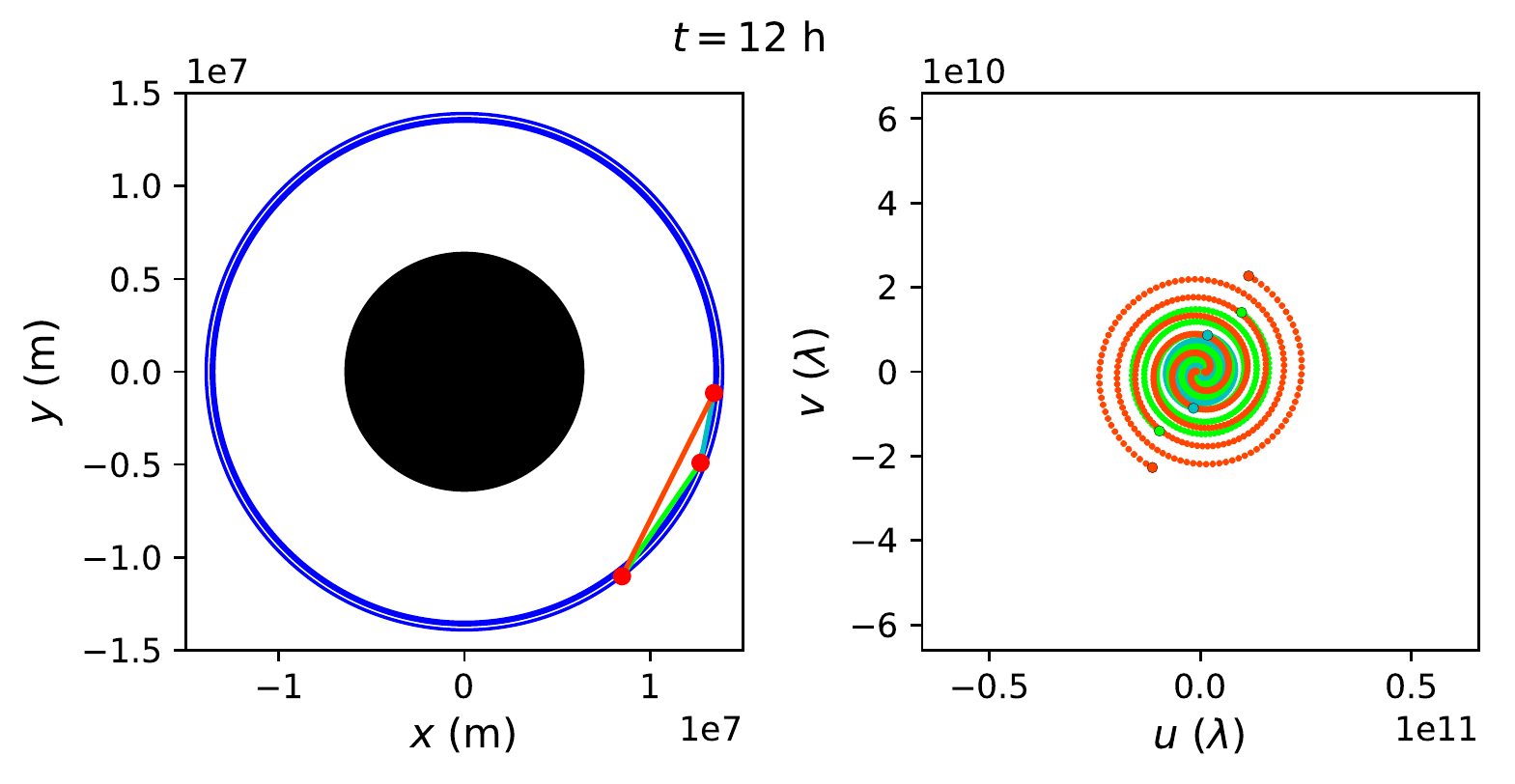}
\includegraphics[width=.49\hsize]{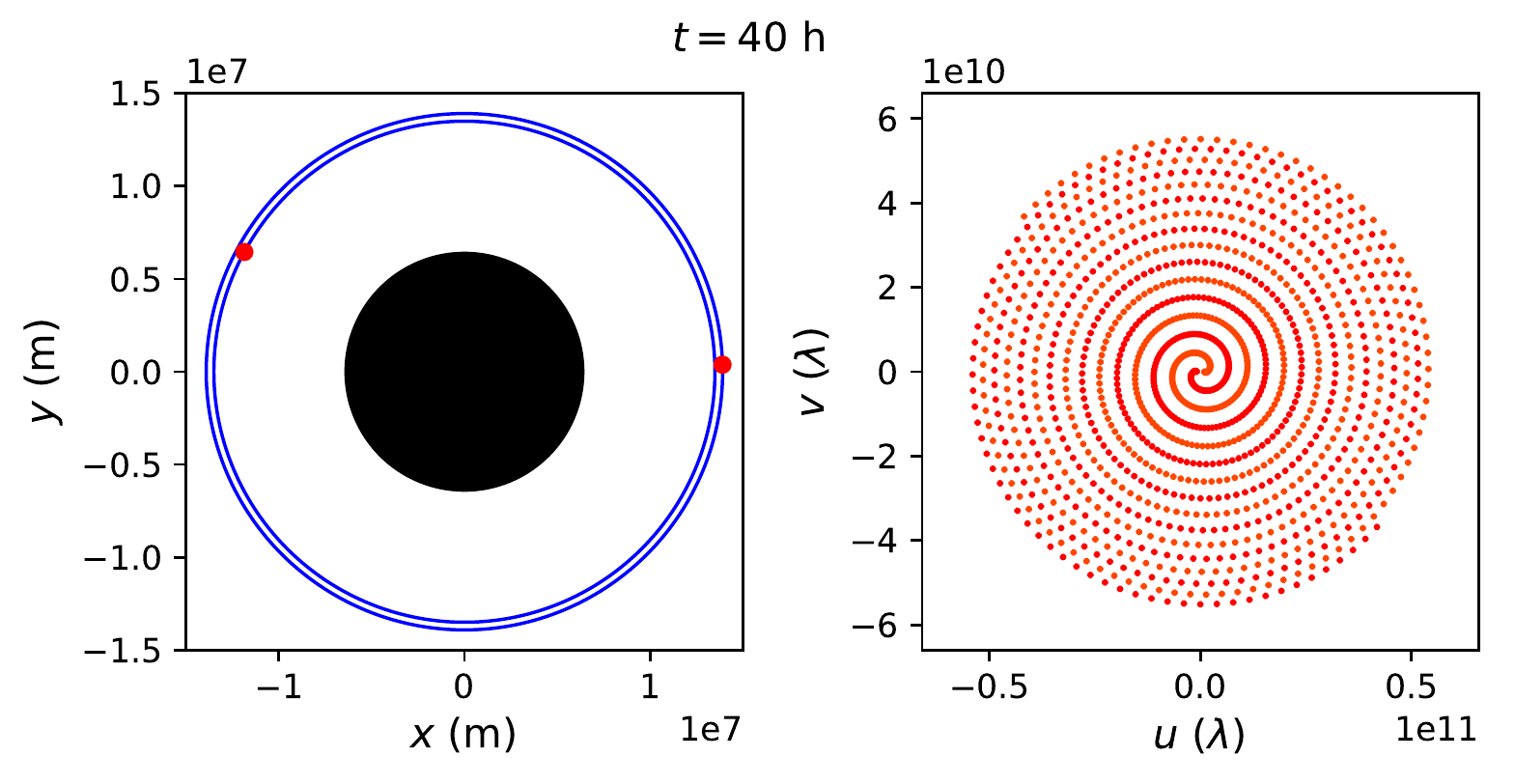}
\includegraphics[width=.49\hsize]{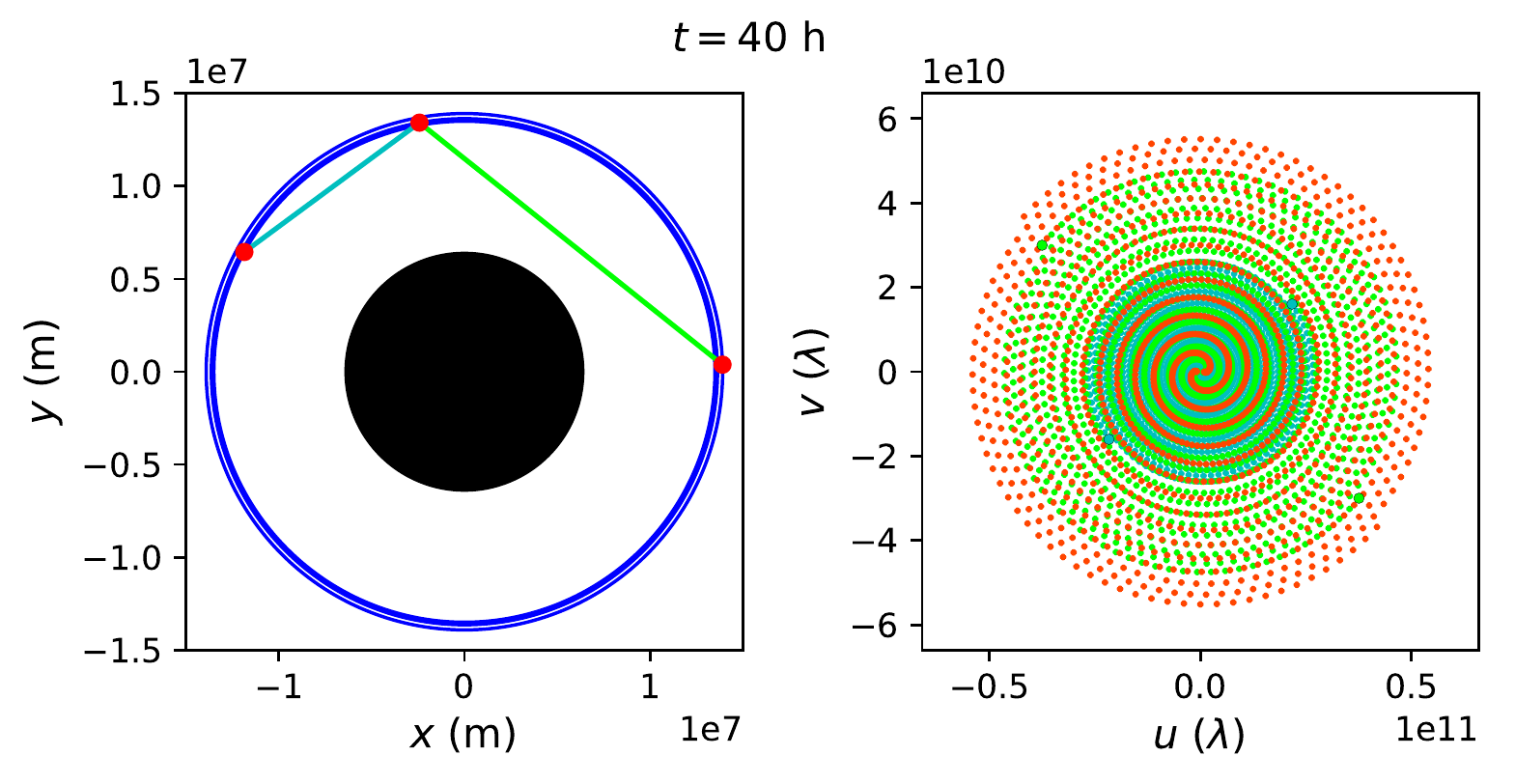}
\includegraphics[width=.49\hsize]{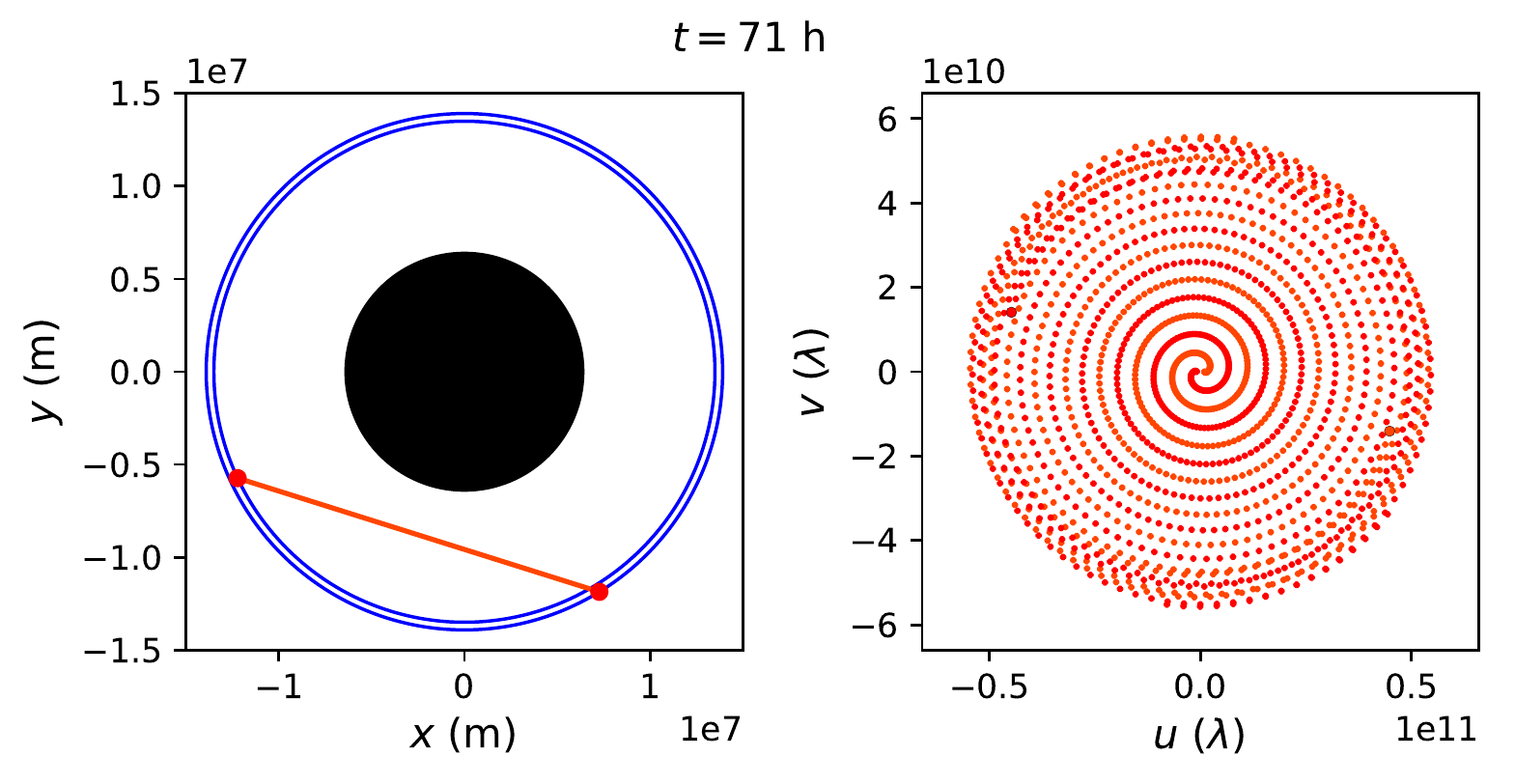}
\includegraphics[width=.49\hsize]{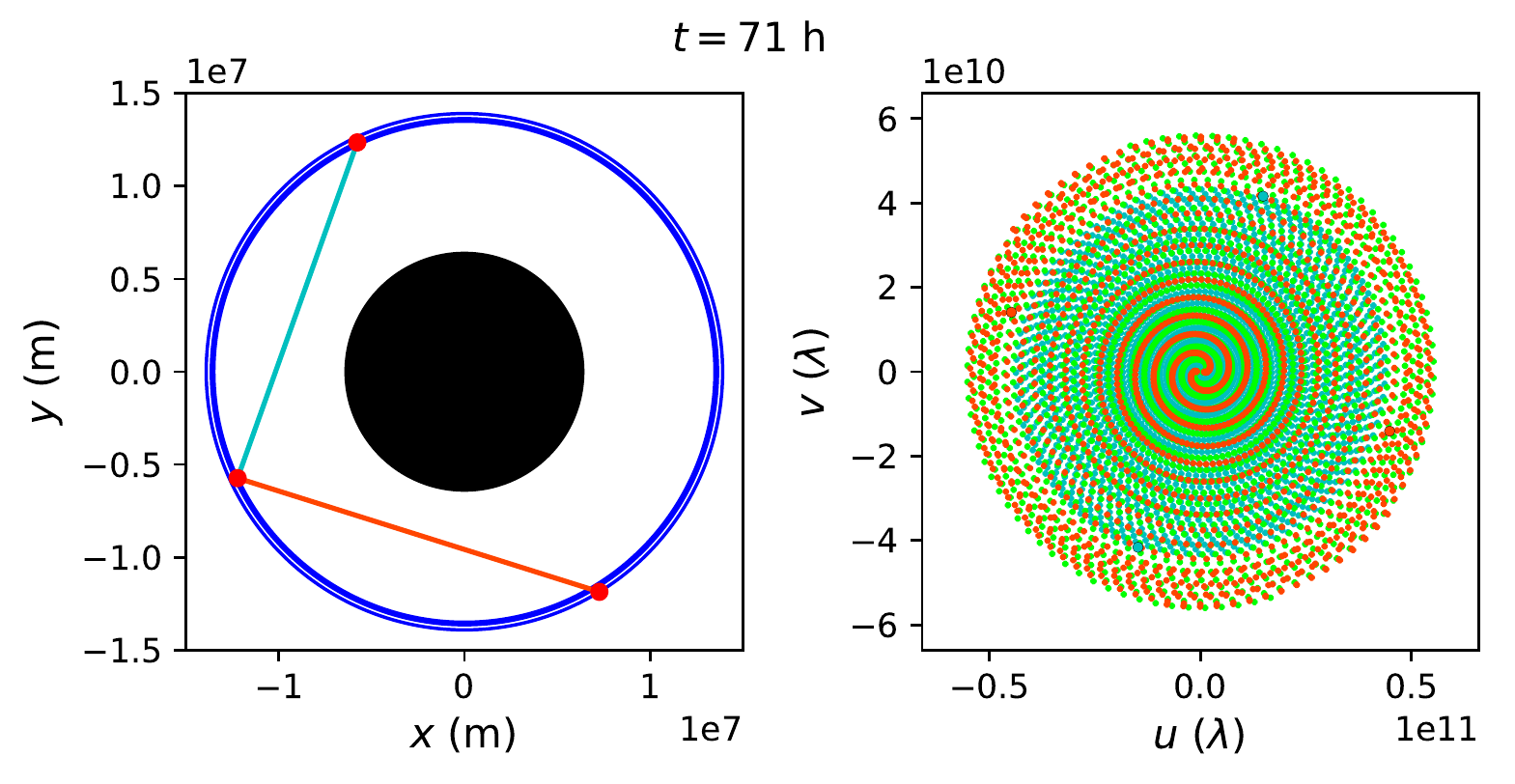}
\caption{Example of satellite positions and $uv$-coverage for circular MEOs at different time stamps (vertical) for a system consisting of two (columns 1 and 2) and three (columns 3 and 4) satellites. The orbital radii are 13500 and 13913\,km for the two-satellite system, and for the three-satellite system a satellite is added at an orbital radius of 13638 km, at one third of the distance between the two. In order to make the spiral structure visually clearer, the inner radii were set smaller than the ones used in our imaging simulations. In column 1, the satellites are shown in red and the corresponding baseline in orange. The orbits are shown in blue, and the Earth is represented by the black disk. In column 2, the red and orange points show the past $uv$-track for the two directions along the baselines. The current $uv$-coordinates are shown as larger dots. In columns 3 and 4, three baselines and $uv$-tracks are shown in corresponding colors. From the initial satellite positions on the line between them and the Earth's center, the distance between them increases and the $uv$-spiral spreads outwards (upper panels) until the Earth atmosphere occults the line of sight between two satellites (middle panels). As the inner satellite catches up with an outer satellite, the spiral is traversed inwards (lower panels).}
\label{fig:uv}
\end{figure*}

Since there is a small difference between the orbital radii of the satellites, the inner satellite orbits slightly faster than the outer one, slowly increasing the distance between the two as they move from their initial positions on the line intersecting them and the Earth's center. As the baseline between them constantly changes orientation as seen from a fixed source, the resulting $uv$-coverage will have the shape of a spiral, with a dense and isotropic sampling of the $uv$-plane, allowing for high-fidelity image reconstructions. The angular resolution of the reconstructed image is determined by the maximum baseline length, which is limited not only by the orbital radius, but also by the occultation of the required intersatellite link (ISL, Sec. \ref{sec:technical}) by the Earth. Figure \ref{fig:uv} illustrates the concept with an example. It also shows a three-satellite configuration. Such a system would allow the use of closure phase, which could relax technical system requirements (Sec. \ref{sec:technical}). Also, a three-satellite system has a faster $uv$-plane filling rate, and measurements could continue with two baselines when one baseline is occulted by the Earth.

With full understanding that many SVLBI system parameters mentioned above are extremely challenging, we take them as given and as input into the analysis presented in this paper. Some engineering aspects are discussed in Section \ref{sec:technical}. We emphasize that the aim of this work is not to provide a technical justification for the concept, but merely to investigate the imaging capabilities, which could serve as input for future engineering design studies.

\subsection{Directions for future engineering system analysis}
\label{sec:technical}
\citet{Martin2017} and \citet{Kudriashov2017} have conducted a first design study addressing the engineering domain of the two-satellite EHI. Although this paper is not an engineering study, we summarize some of the technical aspects and challenges here. They should be addressed in more detail in future engineering studies.

The concept assumes that each satellite generates the local oscillator signal by combining a sufficiently stable reference produced on-board with the one received over the ISL from the other satellite. It further assumes that the cross correlation of the data streams from both satellites is conducted on-board in real time within a delay window compatible with the on-the-fly relative positioning of a sufficient accuracy. It should be investigated whether the on-board correlation and processing would not have to be prohibitively complicated in order to reach a sufficiently low rate for the data transfer to the ground, where the actual fringe fitting would take place. The interferometer will need to be phase stable and phase calibrated in such a way that long integrations can be performed without coherence loss. It may be necessary to perform an acceleration search in fringe fitting as is done for RadioAstron.

For real-time cross-correlation, the baseline vector and its time derivatives should be known with an accuracy that scales down with the observing wavelength. This defines the requirements for the state vector knowledge of the EHI satellites. Traditionally, in all Space VLBI systems implemented to date (TDRSS, VSOP and RadioAstron), the correlator delay model is based on estimates of the spacecraft vector provided by ground-based orbit determination assets. In the EHI concept we consider a different approach in which the baseline vector and its time derivatives are obtained (measured) directly between the EHI spacecraft in real time. As demonstrated by RadioAstron, orbit determination measurements (not to confuse with orbit reconstruction) are provided by means of radio measurements at the level that correspond to $\sim$\,20 to 1500 observing wavelengths at 92 to 1.3\,cm, respectively, and radial velocity (Doppler) measurements at the mm/s accuracy level \citep{Zakhvatkin2014, Duev2015}. Escalating similar requirements for the EHI sub-mm wavelength range leads to the baseline measurement requirements at a precision level of $\sim$\,20\,cm for the baseline vector and 0.01\,mm/s for its time derivative. 

While these values are challenging for an onboard real-time system, they are achieved by modern ground-based systems supporting interplanetary missions \citep[e.g.][]{Iess2009, Duev2016}. One should expect further improvement of the real-time space-borne baseline measurement accuracy based on the relative position determination from the Global Navigation Satellite Systems (GNSS) satellites, which orbit further out at $\sim$\,20,000\,km altitude. These may be able to determine position at centimeter precision \citep[e.g.][]{Allende2016, Park2016, Jaggi2016}. Also, ranging measurements with an accuracy down to $\sim$\,30\,$\mu$m could possibly be performed with the ISL \citep[][J. Perdigues, private communication]{Zech2014}. Lessons could also be learned from the technological preparation of the LISA mission \citep[][and references therein]{Johann2008}. The European Space Agency is currently undertaking a study into the precise relative positioning of MEO satellites.

The idea currently explored in the EHI engineering study is to carry out the fringe-search among the measured correlations on the ground, using for this a finer orbit reconstruction to get coherent phases for visibilities with a low signal-to-noise ratio. While the satellites are close together, Sgr\,A* could be detected as a strong point source, which could be used as a starting point for fringe finding and orbit determination. This is a new way of doing space VLBI, and it still needs to be demonstrated that this is feasible.

Furthermore, a stable frequency reference will be needed to perform VLBI at 690\,GHz. There are two active hydrogen masers on-board RadioAstron \citep{Kardashev2013}, but these have served as frequency references up to observing frequencies of 22\,GHz. There is room for improvement of such instrumentation \citep{Rodrigo2018}. 

The 3-meter reflectors considered by \citet{Martin2017} and \citet{Kudriashov2017} would fit in a medium-sized space launcher such as a Soyuz fairing. ESA's larger space launcher Ariane 6 will have a useable payload diameter of 4.6\,m \citep{Ariane}, so monolithic reflectors up to this diameter can in principle be launched. Larger reflectors would have to be deployable, although a large (up to 25\,m in diameter) monolithic reflector has been considered with side-mounting on a super-heavy launcher for the ESA International VLBI Satellite (IVS) study \citep[][B. Ye. Chertok 1989, private communication]{Pilbratt1991}.

The choice of orbits is limited by several factors. The maximum orbital radius, and hence the array resolution and filling speed of the $uv$-plane, is limited in particular by the visibility of GNSS satellites. Three GNSS satellites should be visible at the same time for a reliable real time position determination. Assuming two navigation antennas with a field of view of $\pm$\,30$^{\circ}$, the maximum orbital radius is then 13913\,km \citep{Kudriashov2017}. The minimum orbital radius is determined by the inner Van Allen radiation belt, which is confined to a radius of $\sim$\,12400\,km \citep[e.g.][]{Bakhtiyarov2014}. The maximum baseline length then ranges from 21252 to 24726 km \citep{Kudriashov2017}, corresponding to a resolution of to 4.2 to 3.6\,$\mu$as at 690\,GHz. All orbits in this range would thus allow one to image the black hole shadow with a resolution that is about an order of magnitude higher than the resolution that can be obtained from the Earth. The time it takes for the spiral to be completed depends on the radial separation of the satellite orbits. Placing one of the satellites at the maximum radius, the spiral is completed in 1 month for a radial separation of 20\,km and in 6 months for a radial separation of 3\,km.

Note that in practice measurements can most likely not be performed at all points in the $uv$-spiral due to functional constraints. For example, the Sun or Moon may be in the line of sight to the observed object or perturbing the measurements \citep[a detailed discussion on similar constraints for the RadioAstron mission is presented by][]{Gurvits1991}. Also, the observations may have to be carried out during multiple months, with possible interruptions due to attitude control, command and communication, orbit determination and correction, and other operational activities. These effects should be given close attention in the project design. There would be no further geometric seasonal effects because polar (and equatorial) circular MEOs have no nodal precession: the line of nodes (i.e. the intersection of the orbital plane and the Earth's equatorial plane) will always be perpendicular to the line of sight towards the observed object. The orbits are thus fixed against e.g. the Galactic Center.

\subsection{Other SVLBI setups for imaging Sgr\,A*} 
Adding a third satellite to the MEO system would triple the number of baselines at each time, so that the $uv$-plane can be filled much faster. Such a system would also enable the use of closure phases, which are immune to station-based phase corruptions and useful for non-imaging analysis. However, adding a third satellite would increase the complexity and cost of the already challenging mission concept. The data would need to be exchanged and correlated for three baselines instead of one. Since there are no atmospheric corruptions in space and the local oscillator signal is exchanged between the satellites, the advantage of having closure phases and an increase in $uv$ sampling speed may not weigh up against the increase in mission complexity and cost. The necessity for closure quantities will need to be assessed as the concept develops further and the expected satellite-based phase corruptions due to e.g. the orbit determination are better quantified.

The EHI MEO concept is not the only setup one could consider for an SVLBI mission. Instead of two or three satellites forming a space-space interferometer, one could launch one or multiple satellites and observe together with ground-based stations, similarly to RadioAstron. Using a high-sensitivity station like ALMA, one could significantly reduce the required integration time and track the time evolution of the source using dynamical imaging \citep{Johnson2017, Bouman2017} from Low Earth Orbits \citep[LEOs,][]{Palumbo2018, Palumbo2019}. The angular resolution could also be increased by launching satellites into MEOs or Geosynchronous Equatorial Orbits (GEOs) \citep{Fish2019}. However, with this setup observing at the high frequencies we consider here will be difficult, as the raw data from the satellites would have to be sent to the ground, and the ground data would still be affected by atmospheric attenuation and phase corruption, which is severe at high frequencies. Also, the $uv$-coverage would not be as dense and uniform as with two satellites in MEOs. An advantage of this method would be that it has been done successfully at low frequencies for RadioAstron, so that less advanced technology would have to be developed than for the two- or three-satellite space-space interferometer. 

Although we consider a space-ground system focusing on resolving source dynamics valuable for understanding black hole accretion, we focus on space-space systems in this work. These are more suitable for high-resolution static imaging.   

\section{Synthetic image generation}
\label{sec:simul}
In this section, we describe the generation of the general relativistic hydrodynamics (GRMHD) simulations, and present ray-traced images. These images are used as input for our simulated observations in the follow-up sections, where we present reconstructions of these images under the assumption of different variants and parameters of the space-space interferometer concept outlined in Section \ref{sec:mission}.

\subsection{Theoretical emission maps of Sgr\,A* at 230 and 690\,GHz}
\label{sec:grmhd}
Sgr\,A* is a quiescent galactic nucleus. Its bolometric luminosity is low in units of the Eddington luminosity: $L_{\mathrm{Bol}}/L_{\mathrm{Edd}}$\,$\sim$\,$10^{-9}$. The mass accretion rate onto Sgr\,A* is estimated to be $10^{-9}$\,$\lesssim$\,$\dot{M}$\,$\lesssim$\,10$^{-7}M_{\odot}$\,$\mathrm{yr}^{-1}$ \citep{Bower2005, Marrone2007, Bower2018}. The spectral energy distribution of Sgr\,A* can be fit with models of radiatively inefficient accretion flows \citep[RIAFs,][]{Yuan2014} coupled to a jet model \citep{Falcke1993, Falcke2000b, Yuan2002, Moscibrodzka2014}.

Because in RIAFs there are few particle interactions and electron cooling is inefficient, the accretion flow becomes a two-temperature  advection dominated accretion flow \citep{Narayan1998}, where ions and electrons are described by different temperatures, $T_p$ and $T_e$. The behavior of the infalling magnetized plasma can be modeled numerically with GRMHD simulations. Starting from an initial RIAF-type plasma density and magnetic field configuration, these simulations solve the equations of magnetohydrodynamics within a specified spacetime metric \citep[e.g.][]{Gammie2003, Moscibrodzka2009, Porth2017}.

From the physical GRMHD quantities such as the plasma density, magnetic field, and temperatures, the synchrotron emission and absorption are calculated. The gas pressure in the simulations is dominated by protons, so $T_p$ is computed from the simulations. For the electron temperature $T_e$ additional assumptions need to be made. Radiative transfer equations with source and sink terms are then integrated along the geodesics running from the pixels of a virtual ``camera'' located far away from the source to each point in the simulation domain. This results in an image of the source as seen by a distant observer. 

In this work, we use models of Sgr\,A* presented in \citet{Moscibrodzka2014}. They generated radiative transfer models based on the 3D GRMHD simulation $\texttt{b0-high}$ from \citet{Shiokawa2013}. This GRMHD simulation starts with a Fishbone-Moncrief torus \citep{Fishbone1976} with inner radius $12 \,{\rm GM/c^2}$ and pressure maximum at $24 \,{\rm GM/c^2}$ in Keplerian orbit at the equator of a rotating supermassive black hole. The black hole spin parameter $a_*$ is set to 0.94 and its mass to $4.5 \times 10^6 {\rm M_{\odot}}$. \citet{Moscibrodzka2014} consider various electron temperature models. As examples, we use their models 16, 24, 31, and 39 (see Table 1 in \citealt{Moscibrodzka2014}) as these models represent different electron heating scenarios within the same physical model of the accretion flow. In models 16 and 31, the electrons are heated mainly in the turbulent accretion disk and hence only the disk around the black hole is visible. In models 24 and 39, the electrons are hot in the the magnetized jet outflow while the electrons in the disk are cooler, so that the jet is visible in the images. These different heating scenarios are motivated by more detailed collisionless plasma models \citep{Ressler2015, Kawazura2018}.

We present total intensity images time-averaged over $810 \,{\rm GM/c^3}$ (about 5 hours) in Figure \ref{fig:models} (left column). All images were generated with the relativistic ray tracing radiative transfer code {\tt ibothros} \citep{Noble2007, Moscibrodzka2018}. We assume that the source is at a distance of 8.5 kpc from Earth. The inclination angle between the black hole spin axis and line of sight is $60^{\circ}$ for models 16 and 24, and $30^{\circ}$ for models 31 and 39. Images were generated with a field of view of 210.44 $\mu$as (corresponding to $40\times 40 \,{\rm GM/c^2}$ at the distance of the black hole) and a resolution of $256\times 256$ pixels at the two frequencies of 230\,GHz (EHT frequency) and 690\,GHz that EHI aims to use. In this work, we demonstrate the space VLBI array performance in reconstructing images of the black hole shadow in case of these four, quite distinct models of plasma around the black hole.

The images show emission in a region close to the event horizon. Gravitational lensing causes emission originating close to the black hole to bend around it, leading to the appearance of a lensed photon ring and the ``shadow'' \citep{Falcke2000} in the center. Doppler boosting of emission from plasma moving towards the observer causes the apparent asymmetry. Near the horizon, the emission pattern of the disk (models 16 and 31) and jet (models 24 and 39) models is similar: especially at 690\,GHz the reconstructed image will be dominated by general relativistic effects that are not strongly dependent on the exact nature of the accretion flow. At 230\,GHz, the difference between the models is more pronounced. Due to strong lensing of the emission originating in a small region close to the black hole, the observed variability in the image plane is generally less profound at higher frequencies. At lower frequencies, larger moving structures that are less easily averaged out can be seen further out in the accretion flow, especially for the jet models.

The total flux density of Sgr\,A* at 690\,GHz is variable on intra-day time scales \citep{Dexter2014}. In the simulations, the accretion rate was set such that the total flux density at 230\,GHz is within 30\,\% of the 2.4\,Jy measured by \citet{Doeleman2008}. As the 1.3\,mm flux density varies as well \citep[e.g.][]{Fish2011, Bower2018}, all models shown are considered to be reasonable representations of the expected brightness of Sgr\,A*. 

\subsection{Interstellar scattering}
\label{sec:scat}
Scattering of radio waves due to electron density fluctuations in the interstellar medium between the Earth and the Galactic Center causes phase fluctuations of the incoming plane wave. The effect of scattering can usually be treated as a random phase-changing screen described by a spatial structure function $D_{\phi}(\vec{x})\equiv\langle[\phi(\vec{x}_0+\vec{x})-\phi(\vec{x}_0)]^2\rangle$, where $\phi$ is the change in phase and $\vec{x}$ is a transverse screen coordinate \citep{NG89, GN89, Johnson2015}. Scattering occurs in two main regimes. Diffractive scattering is dominated by fluctuations on the phase coherence length, which is the length scale $r_0$ on the scattering screen corresponding to a change in phase of 1 radian, i.e. $D_{\phi}(r_0)\equiv 1$. It is quenched for sources larger than $r_0$ and has therefore only been relevant in observations of extremely compact sources such as pulsars. Refractive scattering is dominated by fluctuations on the refractive scale $r_\mathrm{R}$, which corresponds to the apparent size of a scattered point source. For Sgr\,A*, refractive scattering is expected to affect observations at frequencies up to about 2\,THz \citep{Johnson2015}.

\begin{figure*}
\centering
\includegraphics[width=\hsize]{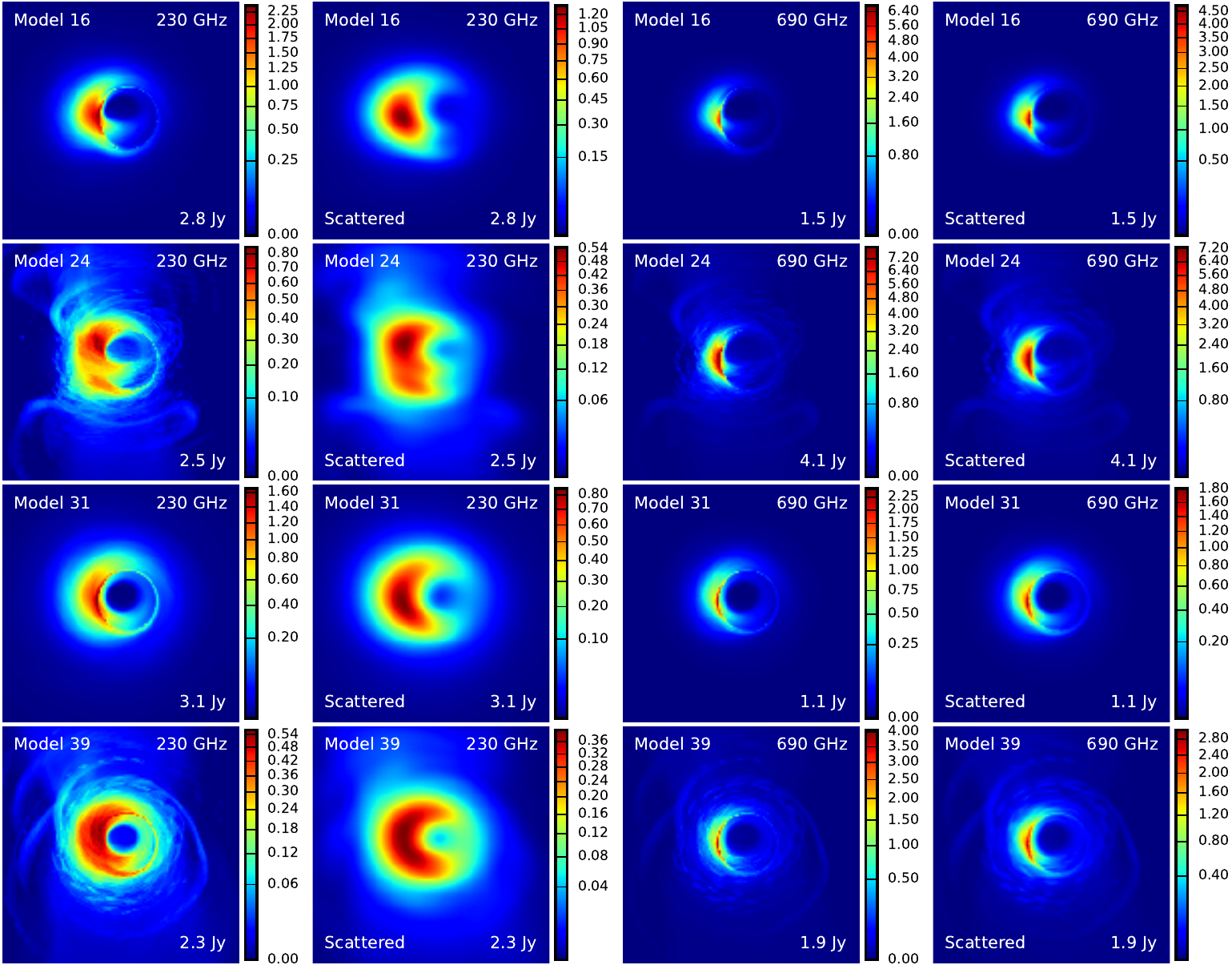}
\caption{Time-averaged GRMHD source models used for simulated observations of Sgr\,A*. Images with note `scattered' were convolved with the scattering kernel from \citet{Bower2006}. The total flux of the model is given in the bottom right corner of each image. The field of view is 210\,$\mu$as for all images. Colors indicate brightness/pixel in mJy (square root scale).}
\label{fig:models}
\end{figure*}

The scattering screen is generally assumed to be frozen, with a transverse velocity $v_{\bot}$ with respect to Sgr\,A* as the only source of variations in time. Three averaging regimes introduced by \citet{NG89} and \citet{GN89} are important for interferometric imaging of a scattered source. In the snapshot regime, source and background noise are averaged over for a single scattering realization. Observing a source extending over scales larger than $r_0$ or time scales longer than $t_{\mathrm{dif}}=r_0/v_{\bot}$ brings one into the average image regime, where only the refractive noise is relevant. In the average image regime, the source image contains spurious refractive substructure. This substructure is quenched, but not smoothed for a source size exceeding the refractive scale \citep{Johnson2015}. The ensemble average is the average over many realizations of the (refractive) scattering screen. Averaging scattering screen realizations over a time $t\gg t_{\mathrm{R}}=r_\mathrm{R}/v_{\bot}$ brings one into the ensemble average regime. For Sgr\,A*, $t_{\mathrm{R}}$ is expected to be about a day at 230\,GHz and about 3 hours at 690\,GHz. In the ensemble average regime, the source appears as the unscattered source convolved with a scattering kernel that effectively blurs the image. The size of the kernel, which is Gaussian at least down to centimeter wavelengths, increases with the square of the observing wavelength $\lambda$. \citet{Bower2006} measured the full-width-at-half-maximum (FWHM) of the scattering kernel major axis to be $1.309 \pm 0.015$\,mas\,cm$^{-2}$, the FWHM of the minor axis to be $0.64^{+0.04}_{-0.05}$\,mas\,cm$^{-2}$, and the position angle to be $78^{+0.8}_{-1.0}$ degrees East of North. The right column of Figure \ref{fig:models} shows our model images blurred with this scattering kernel. The blurring effect is significant at 230\,GHz, but hardly visible at 690\,GHz due to the $\lambda^2$ size law. Using a physically motivated scattering model including refractive effects, \citet{Johnson2018} infer that the scattering kernel at millimeter wavelenghts may be smaller than predicted by extrapolating the kernel from \citet{Bower2006}.

\section{Simulated observations}
\label{sec:sim}
In this section, we outline the process of simulating observations of the Sgr\,A* source models described in Section \ref{sec:simul}.

\subsection{$uv$-sampling}
\label{sec:uvsampling}
For calculating model visibilities, we used the $\texttt{eht-imaging}$\footnote{https://github.com/achael/eht-imaging} software \citep{Chael2016}. This software package calculates the observed complex visibilities corresponding to a given source model and $uv$-coverage. As an example, we adopt the orbital parameters consistent with the two-satellite MEO setup as discussed by \citet{Martin2017} and \citet{Kudriashov2017}. The two satellites are in circular orbits with radii of 13,892 and 13,913 km, the latter of which is the maximum based on the requirement of having simultaneous visibility of at least three GNSS satellites (Sec. \ref{sec:technical}). With this setup, the maximum baseline length is $1.9 \times 10^{10}\lambda$ for 230\,GHz and $5.7\times 10^{10}\lambda$ for 690\,GHz, corresponding to an angular resolution of 11 and 3.6\,$\mu$as, respectively. We also performed simulations with a three-satellite system. In this case, the third satellite was placed at a radius of 13899\,km, which is at one third of the distance between the inner- and outermost satellite. 

The orbital period of the satellites is $\sim$\,4.5\,h. The completion time of the full spiral, starting with the satellites at their minimum distance and ending when the line of sight between them is occulted by the Earth, is set by the radial difference between the satellite orbits, which is 21\,km in this case. The full spiral is then traversed in 29 days. The orbital plane is initially set perpendicular to the line of sight to the observed source in order to keep the simulations free from any preferential directions initially (we study different geometries in Sec. \ref{sec:dec}). 

The integration time per measurement $t_{\mathrm{int}}$ should be set short enough to avoid image corruption by $uv$-smearing. If within an integration time a displacement is made in the $uv$-plane that corresponds to an angle on the sky that is smaller than the source size $\theta_{\mathrm{source}}$, the reconstructed image will be affected \citep{TMS2017, Palumbo2019}. In the case of our SVLBI system, motions in the $uv$-plane are dominated by the azimuthal component since the spiral contains many (29\,days\,/\,4.5\,h\,=\,155) loops. The $uv$-vector rotates fastest at the edge of the spiral, where the satellites are at their maximum separation. Here, the $uv$-separation per integration time is given by
\begin{equation}
ds=\frac{2\pi B_{\mathrm{max}}t_{\mathrm{int}}}{P},
\end{equation}
where $B_{\textrm{max}}$ is the maximum baseline length and $P$ is the orbital period. The $uv$-smearing limit on the integration time is then
\begin{equation}
\label{eq:tint}
t_{\mathrm{int}} < \frac{P}{2\pi B_{\mathrm{max}}\theta_{\mathrm{source}}}.
\end{equation}
With $B_{\mathrm{max}}=5.7\times 10^{10}$\,G$\lambda$, $P=4.5$\,h, and $\theta_{\mathrm{source}}=150$\,$\mu$as (the important emission features in our model images are within this field of view), we get $t_{\mathrm{int}} < 62$\,s. Since this limit only holds when the satellites are at their maximum separation, the integration time can be made longer when they are closer together. In calculating our $uv$-spirals, we set a $uv$-distance-dependent integration time that is equal to half the limit from Equation \ref{eq:tint}. The integration time is then well within the $uv$-smearing limit everywhere while sufficient signal-to-noise can be accumulated. At the shortest baselines, we set the maximum integration time to 454 seconds in order to limit the $uv$-arcs to 10 degrees.

The total integration time required for imaging is at least one iteration of the spiral (29 days), which is much longer than both the expected source variability time scale and the expected refractive time scale. We comment on mitigating source and scattering variability in Sections \ref{sec:tv} and \ref{sec:deblur}, respectively. Simulated observations involving a third satellite are presented in Section \ref{sec:alt}.

\subsection{System noise calculation}
\label{sec:noise}
Thermal noise can be characterized by a circular Gaussian in the visibility plane with zero mean and standard deviation $\sigma$. The value of $\sigma$ can be calculated with the standard noise equation used in radio interferometry
\begin{equation}
\label{eq:sigma}
\sigma=\frac{1}{0.88}\sqrt{\frac{\mathrm{SEFD}_1\mathrm{SEFD}_2}{2\Delta\nu t_{\mathrm{int}}}},
\end{equation}
where $\Delta\nu$ is the observing bandwidth and $t_{\mathrm{int}}$ is the integration time \citep{TMS2017}. The factor 1/0.88 results from 2-bit quantization losses. $\mathrm{SEFD}_{1,2}$ are the System Equivalent Flux Densities of the antennas, and can be calculated as
\begin{equation}
\label{eq:sefd}
\mathrm{SEFD}=\frac{2k_{\mathrm{B}}T_{\mathrm{sys}}}{\eta A},
\end{equation}
where $k_{\mathrm{B}}$ is Boltzmann's constant, $T_{\mathrm{sys}}$ is the system temperature, and $A=\pi(D/2)^2$ is the area of an antenna with diameter $D$. In our simulations, $\eta = \eta_{\mathrm{ap}} \eta_{\mathrm{cor}} \eta_{\mathrm{clock}}$ includes the efficiencies of the aperture, correlator, and clock, respectively. The system parameters adopted in this work are consistent with the EHI setup considered by \citet{Martin2017} and \citet{Kudriashov2019}. The parameters and resulting noise for the most important system setups considered in this paper are shown in Table \ref{tab:params}. In addition to these, we also perform simulations for a system consisting of three 4.0-meter satellites (Sec. \ref{sec:alt}).

The system temperature is consistent with single side band (SSB) SIS receivers as installed in the Herschel HIFI instrument \citep{Graauw2008}, with a 10\,K antenna temperature. The 4.4-meter antennas would fit in the Ariane 6 spacecraft \citep{Ariane}. In the case of a three-satellite system, three 4.0-meter dishes would fit. The aperture, correlator, and clock efficiencies are the current baseline figures for the EHI design study. The estimate for $\eta_{\mathrm{clock}}$ is based on the coherence of the Space Hydrogen Maser of the Atomic Clock Ensemble in Space \citep[ACES,][]{Goujon2010}, at a time scale of 1 second. Since the clock connection time over the ISL would be of order $10^2$\,ms for the longest EHI baselines, this estimate may be pessimistic. The bandwidth of the inter-satellite link is 10\,GHz, which is the sum of the bandwidths in two polarizations, assumed to be 5\,GHz each.

\begin{table}
\centering
\caption{System parameters and resulting noise}
\label{tab:params}
\begin{tabular}{l|lllll}
\hline
$\nu$\,(GHz)                   & 230           & 230   & 690 & 690                \\
$D$\,(m)                       & 4.4           & 25    & 4.4 & 25             \\ \hline
$T_{\mathrm{sys}}$\,(K)        & 150            & 150   & 150 & 150              \\
$\eta_{\mathrm{ap}}$           & 0.58          & 0.58  & 0.58 & 0.58              \\
$\eta_{\mathrm{cor}}$          & 0.97          & 0.97  & 0.97 & 0.97             \\
$\eta_{\mathrm{clock}}$        & 0.87          & 0.87  & 0.87 & 0.87             \\
$\Delta\nu$\,(GHz)             & 5             & 5     & 5    & 5          \\
$t_{\mathrm{int, center}}$\,(s) & 453           & 453   & 453  & 453  \\
$t_{\mathrm{int, edge}}$\,(s) & 94            & 94    & 32   & 32              \\ \hline
SEFD\,(Jy)   & $5.6\times 10^4$ & $1.7\times 10^3$  & $5.6\times 10^4$ & $1.7\times 10^3$  \\
$\sigma_\mathrm{center}$\,(Jy) & 0.030  & 0.00092    & 0.030  & 0.00092            \\
$\sigma_\mathrm{edge}$\,(Jy) & 0.065 & 0.0020     & 0.11 & 0.0035            \\
\hline
\end{tabular}
\tablefoot{$\sigma$-values were calculated with equations \ref{eq:sigma} and \ref{eq:sefd}, using different frequencies and dish sizes at the center (long integration time) and edge (short integration time) of the $uv$-spiral.}
\end{table}

For all simulations in this paper, we have assumed that the baseline vector is known exactly, and hence no phase corruptions due to uncertainties in the orbital model are introduced. In practice, the baseline vector and its time derivatives will only be known up to a certain precision. Future engineering studies should determine the accuracy that can be reached, so that the phase corruptions and their effect on the images can be modeled properly.

\subsection{Deblurring}
\label{sec:deblur}
Interstellar scattering introduces variable refractive substructure as the scattering screen traverses in front of the source \citep[][see also Sec. \ref{sec:scat}]{Johnson2015}. The average of different realizations of the scattering screen is the scattering kernel described in Section \ref{sec:scat}, which is convolved with the background source image and has a blurring effect. The refractive time scale of $\sim$ a day at 230\,GHz and 3 hours for 690\,GHz is much shorter than the planned observation time ($\sim$months) in which visibilities are averaged, so that the scattering effect in the reconstructed average image can be approximated by the ensemble average kernel. If the size and orientation of the scattering kernel are known \citep{Bower2006}, its effect may be mitigated \citep{Fish2014}. Calculating the Fourier transform of the convolution of two functions $I(\xi, \eta)$ and $G(\xi, \eta)$ is equivalent to pointwise multiplication of the Fourier transforms $\tilde{I}(u, v)$ and $\tilde{G}(u, v)$ of those functions:
\begin{equation}
I(\xi, \eta)*G(\xi, \eta) \rightleftharpoons \tilde{I}(u, v)\tilde{G}(u, v),
\end{equation}
where $*$ denotes convolution and $\rightleftharpoons$ denotes a Fourier transform. The visibilities of a scattered source may thus be corrected by dividing them by the Fourier transform of the scattering kernel. Assuming that the scattering kernel is known, the remaining effect on the deblurred visibilities is that their signal-to-noise ratio is smaller than for an unscattered source. As the scattering kernel from \citet{Bower2006} is Gaussian, so is its Fourier transform, and the scattering causes the measured S/N to drop steeply as a function of baseline length. This effect is relevant for the setup discussed here as the array performance is sensitivity-limited and the baselines are long (although recent modeling by \citet{Johnson2018} suggests that the kernel may in fact be non-Gaussian and smaller at millimeter wavelengths, mitigating this effect). At 230\,GHz, where the scattering is much stronger than at 690\,GHz (Fig. \ref{fig:models}), deblurring will become a problem at long baselines as the low-S/N visibilities are divided by small numbers, blowing up the errors. In order for the deblurring to work, we impose an S/N cutoff of 4.5. In order to remove some remaining outliers, we impose the requirement that the error on the visibility in a grid cell should be smaller than the flux of the source on the shortest baseline. The S/N cutoff of 4.5 was determined empirically by selecting the value that gave the best image reconstruction quality (Sec. \ref{sec:nrmse}).

\subsection{Image quality metric}
\label{sec:nrmse}
In order to quantify the quality of our image reconstructions, we calculated the Normalized Root-Mean-Squared Error (NRMSE). The NRMSE is defined as \citep{Chael2016}
\begin{equation}
\mathrm{NRMSE}=\sqrt{\frac{\sum_{i=1}^{n^2}\left|I'_i-I_i\right|^2}{\sum_{i=1}^{n^2}\left|I_i\right|^2}},
\end{equation}
where $I_i$ is the $i$th pixel of the $n\times n$ pixels model image (Fig. \ref{fig:models}), and $I'_i$ is the same for the reconstructed image. The NRMSE is thus a pixel-by-pixel comparison of two images: the lower the NRMSE, the more similar the two images are. Since the model and reconstructed images do not have the same number of pixels in our case, the reconstructed image was regridded to the pixel size of the model image and the two images were aligned before calculating the NRMSE. The NRMSE is only a coarse image comparison metric: it does not compare e.g. image smoothness or contrast, or the reconstruction of specific features.

\section{Simulation results}
\label{sec:results}
In this section, we describe the outcome of the simulations for which the setup is described in the previous sections. We start with a two-satellite system, first reconstructing images using conventional VLBI techniques and then using the assumption that the system can be made to behave like a connected interferometer, which requires sharing the LO signal and obtaining an excellent orbit reconstrucion in post-processing. We then describe imaging techniques and results for a three-satellite system. 

\subsection{Two-satellites: conventional VLBI techniques}
\begin{figure*}
\centering
\includegraphics[width=\textwidth]{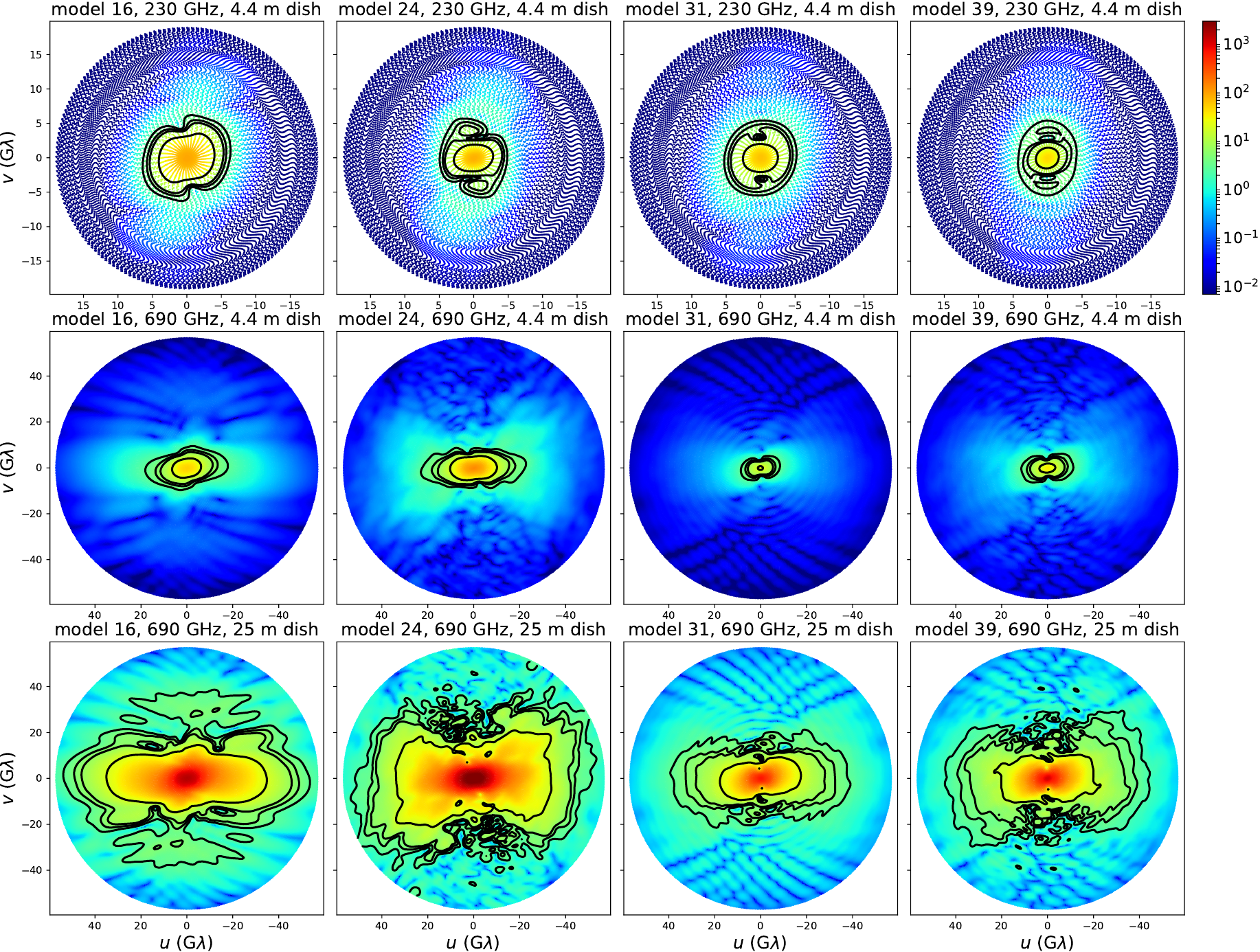}
\caption{Signal-to-noise ratio (S/N, colored) for the spiral $uv$-points calculated with the system parameters in Table \ref{tab:params} (but with a 25-meter reflector diameter in the bottom row) and scattered source models in Figure \ref{fig:models}. Contours indicate S/N values of 3, 5, 7, and 20.}
\label{fig:snr_spiral}
\end{figure*}

Figure \ref{fig:snr_spiral} shows the expected visibility signal-to-noise ratio (S/N) at the spiral points for the scattered source models (Fig. \ref{fig:models}) and system noise (Table \ref{tab:params}) described above, assuming two 4.4 or 25-meter reflectors. The S/N is highest at short baselines sampling the integrated large-scale source structure. The S/N drops more steeply towards longer baselines at 230\,GHz than at 690\,GHz due to blurring by interstellar scattering. The contours show that the region in $uv$-space where an S/N\,$>$\,7 can be reached is confined to $\sim$\,10-20\,G$\lambda$ for a 4.4-meter reflector, whereas the proposed orbital and frequency setup allows for baselines up to $\sim$\,60\,G$\lambda$ at 690\,GHz. With a 25-meter reflector, S/N\,$>$\,7 can be reached near the maximum baseline length for most models.

Note that in practice the $uv$-coverage will not be circular, but elliptical due to the declination of the source ($-29^{\circ}$ for Sgr\,A*). The decrease of the structural source information that can be obtained as the baselines get shortened in one direction will depend on the shape and orientation of the source on the sky (see Sec. \ref{sec:dec} for additional discussion).

\begin{figure*}[!h]
\centering
\includegraphics[scale=0.31]{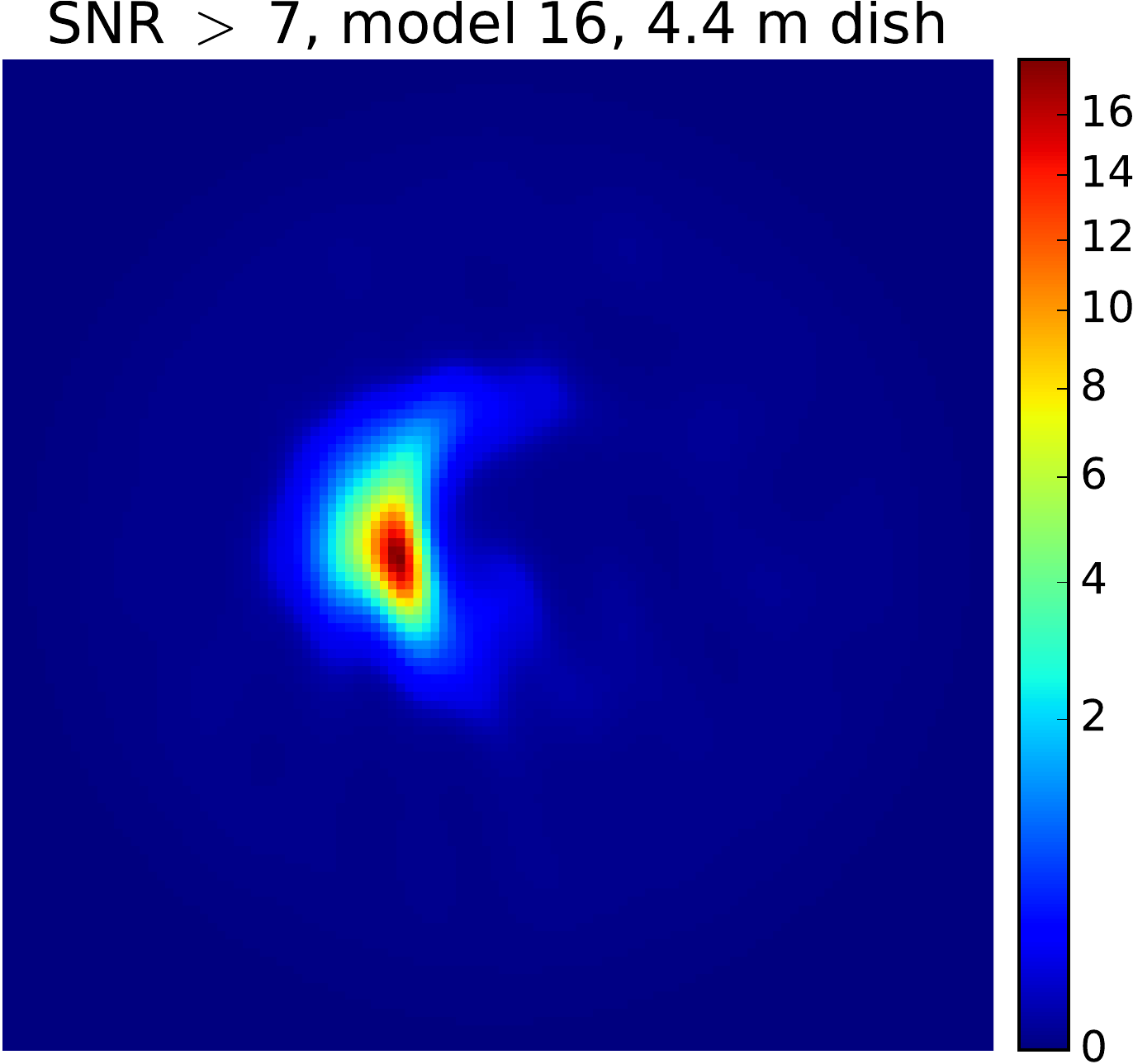}
\includegraphics[scale=0.31]{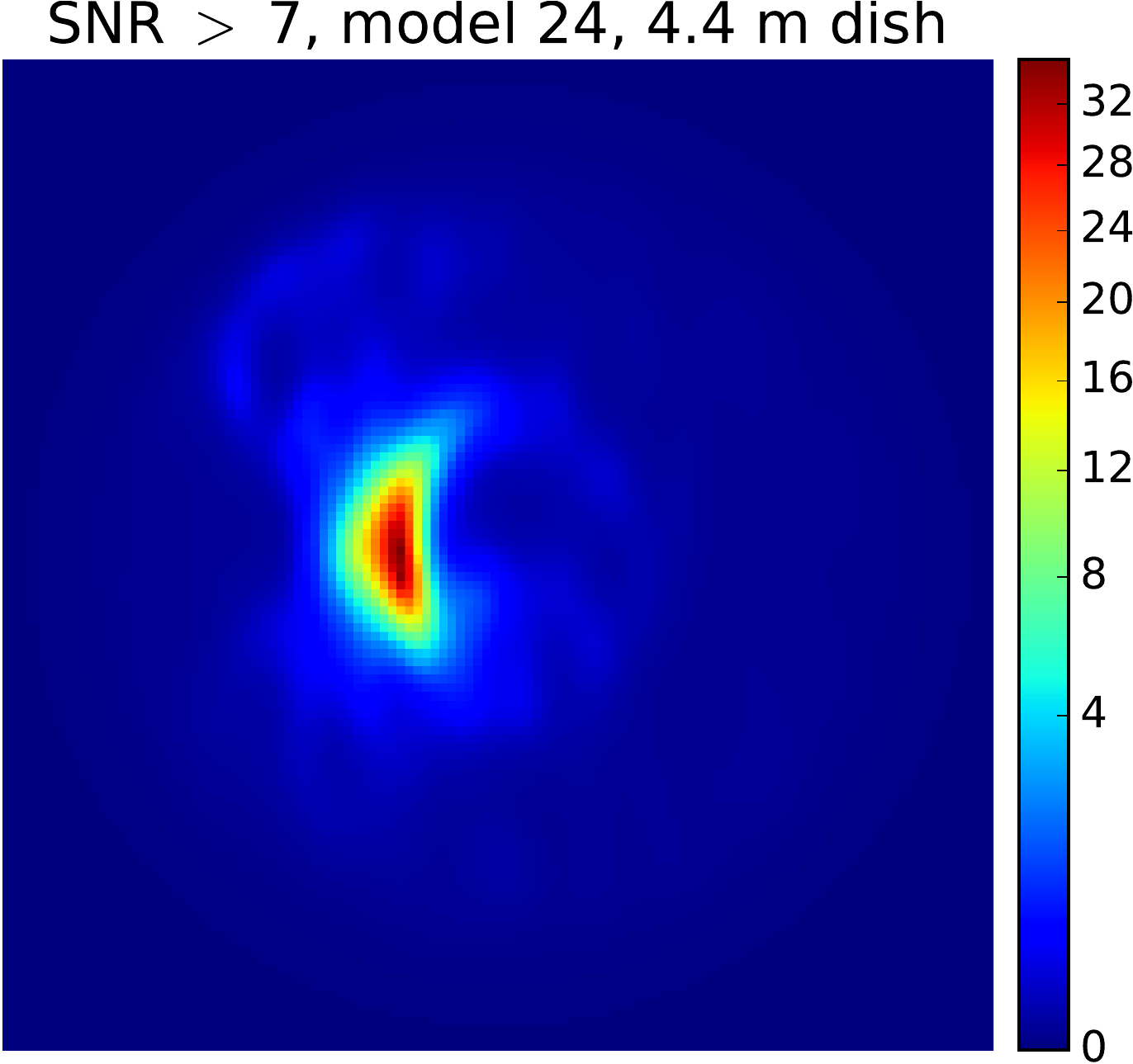}
\includegraphics[scale=0.31]{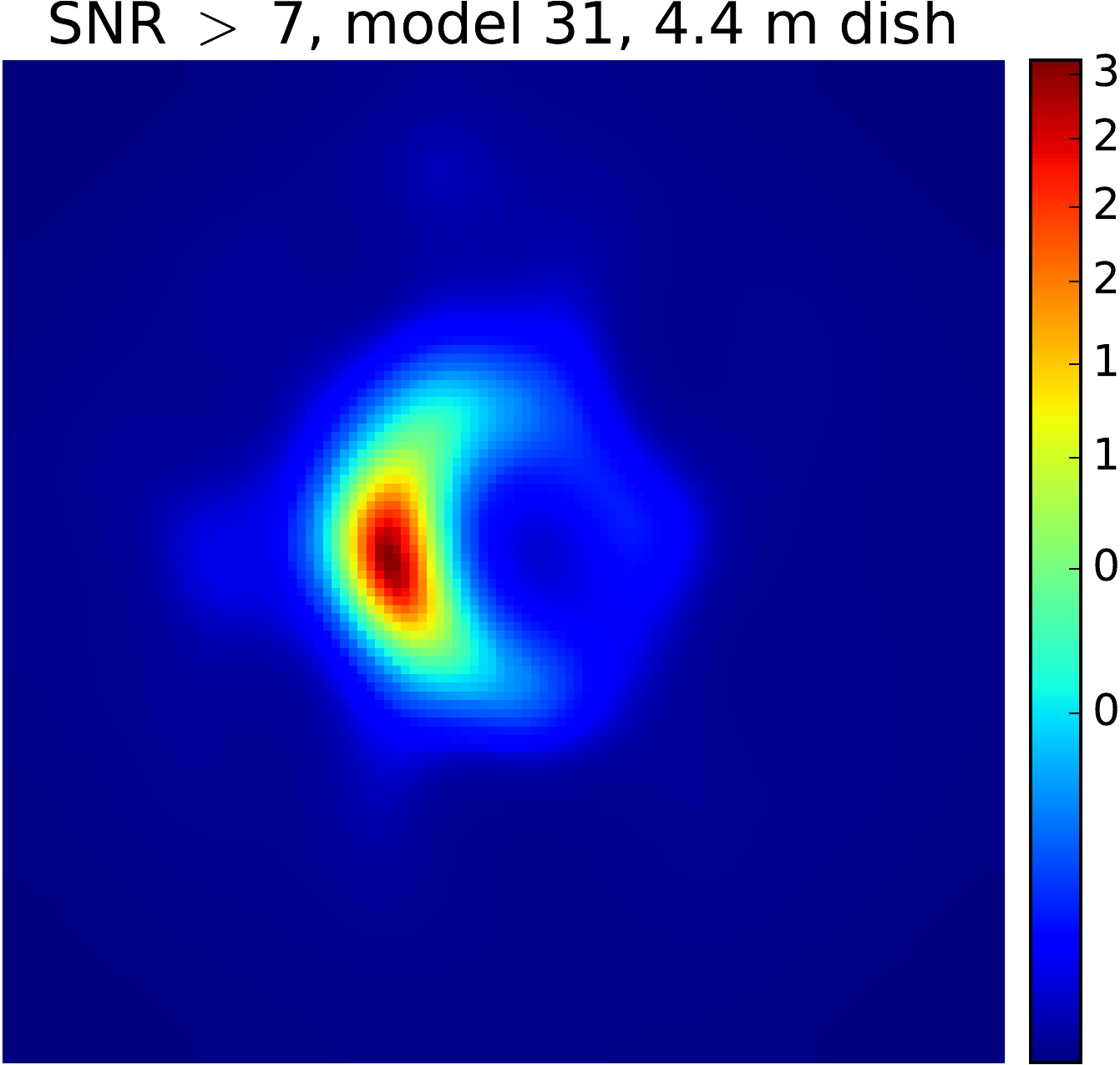}
\includegraphics[scale=0.31]{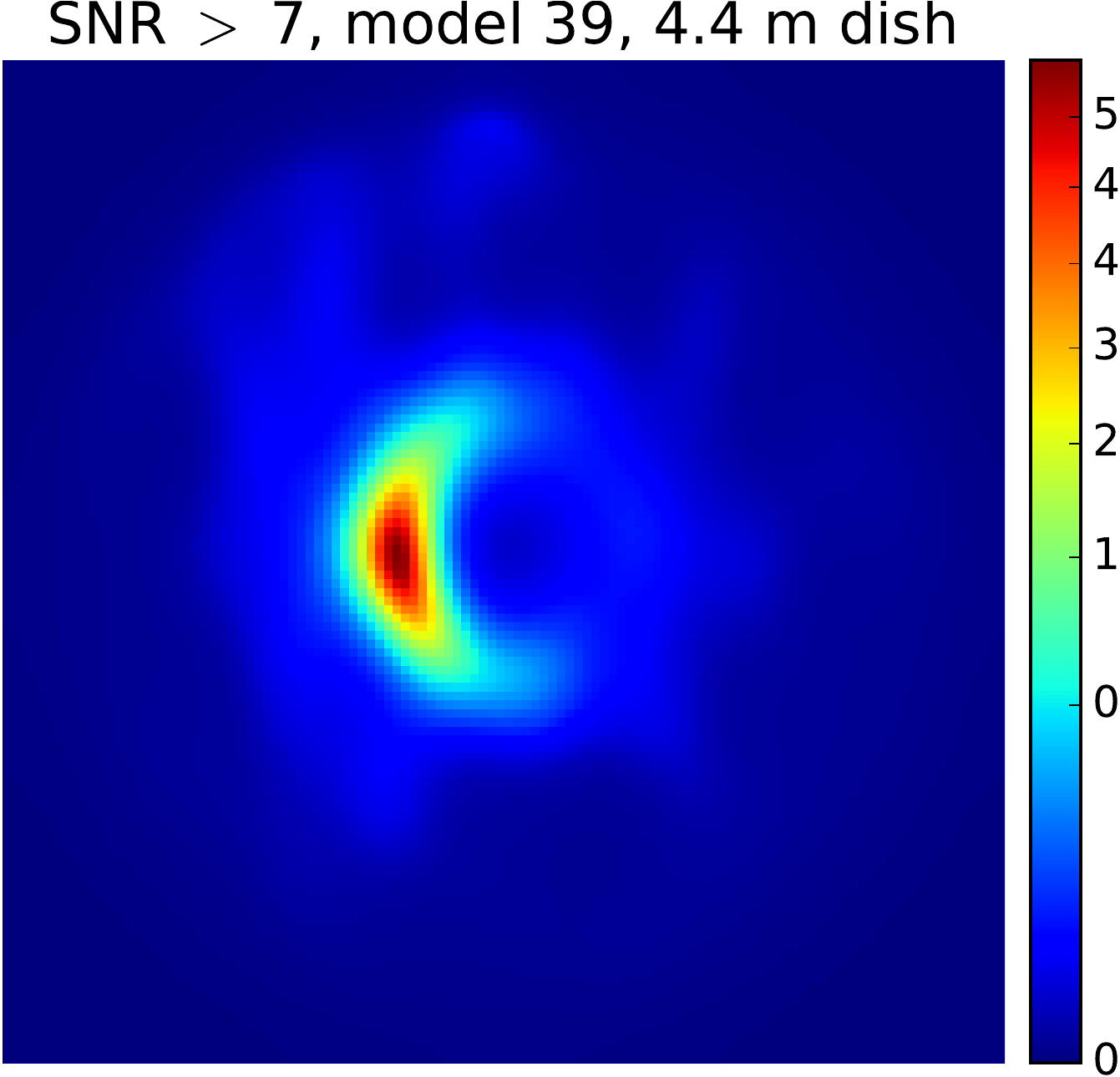} \\
\includegraphics[scale=0.31]{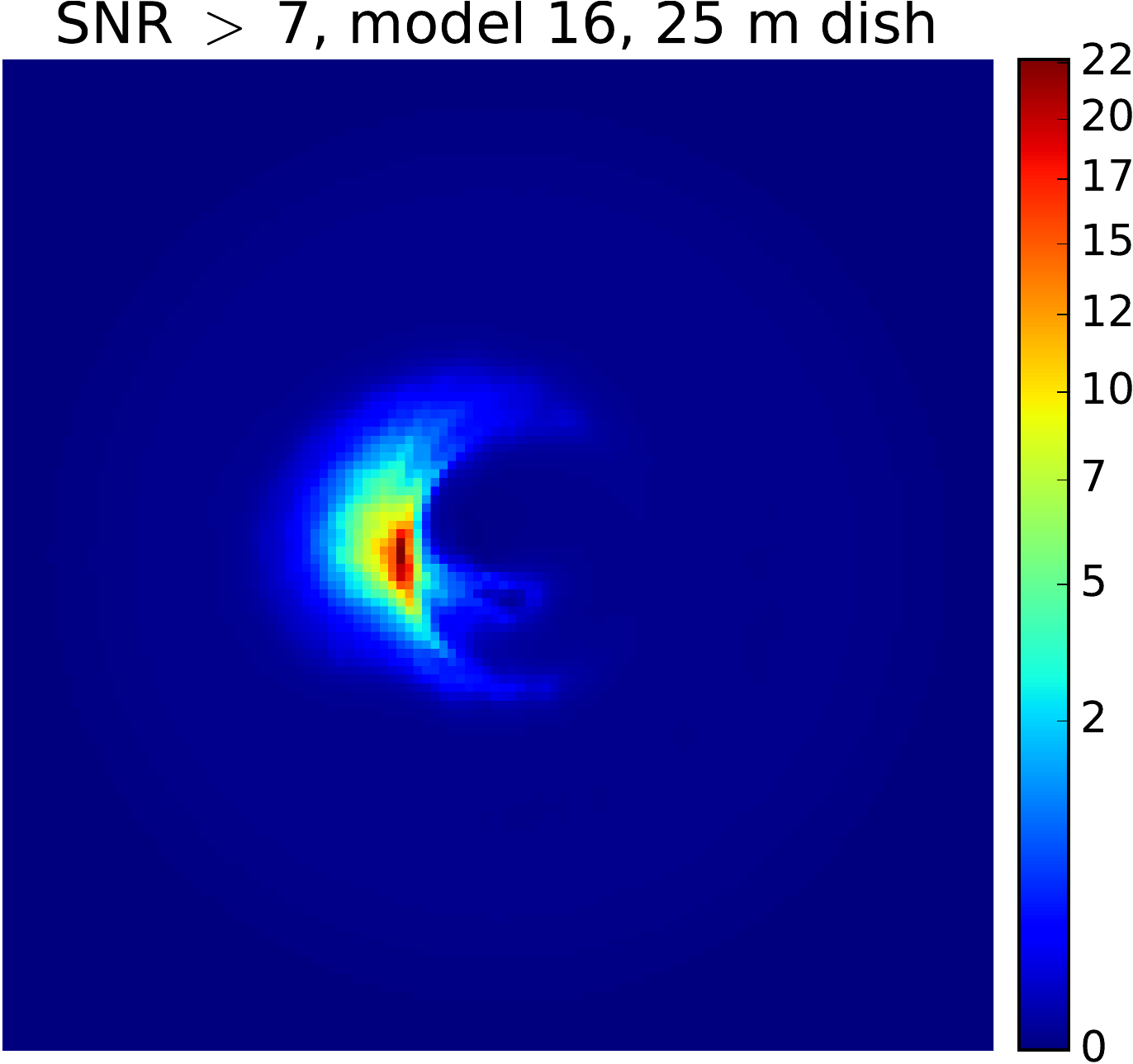}
\includegraphics[scale=0.31]{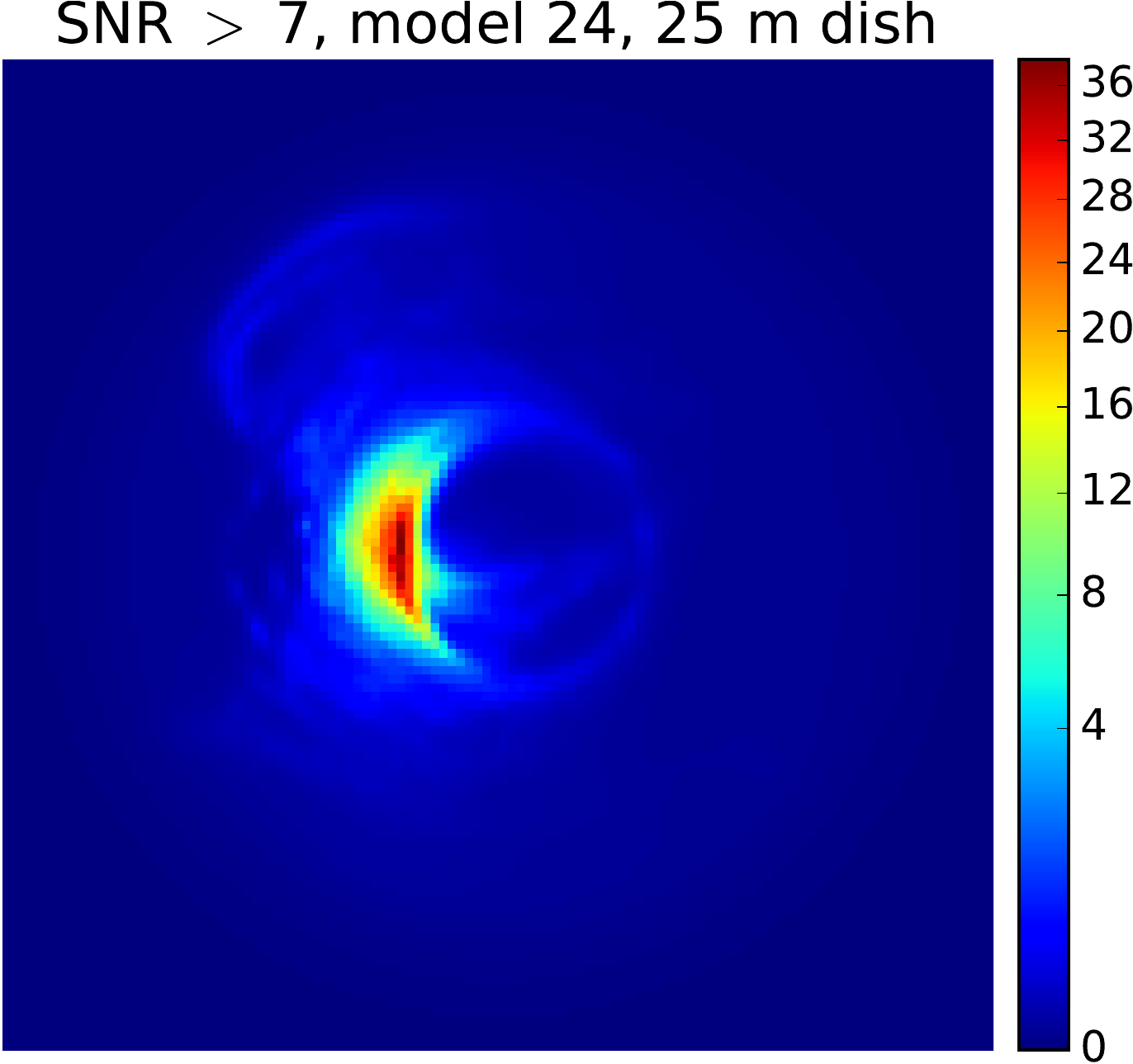}
\includegraphics[scale=0.31]{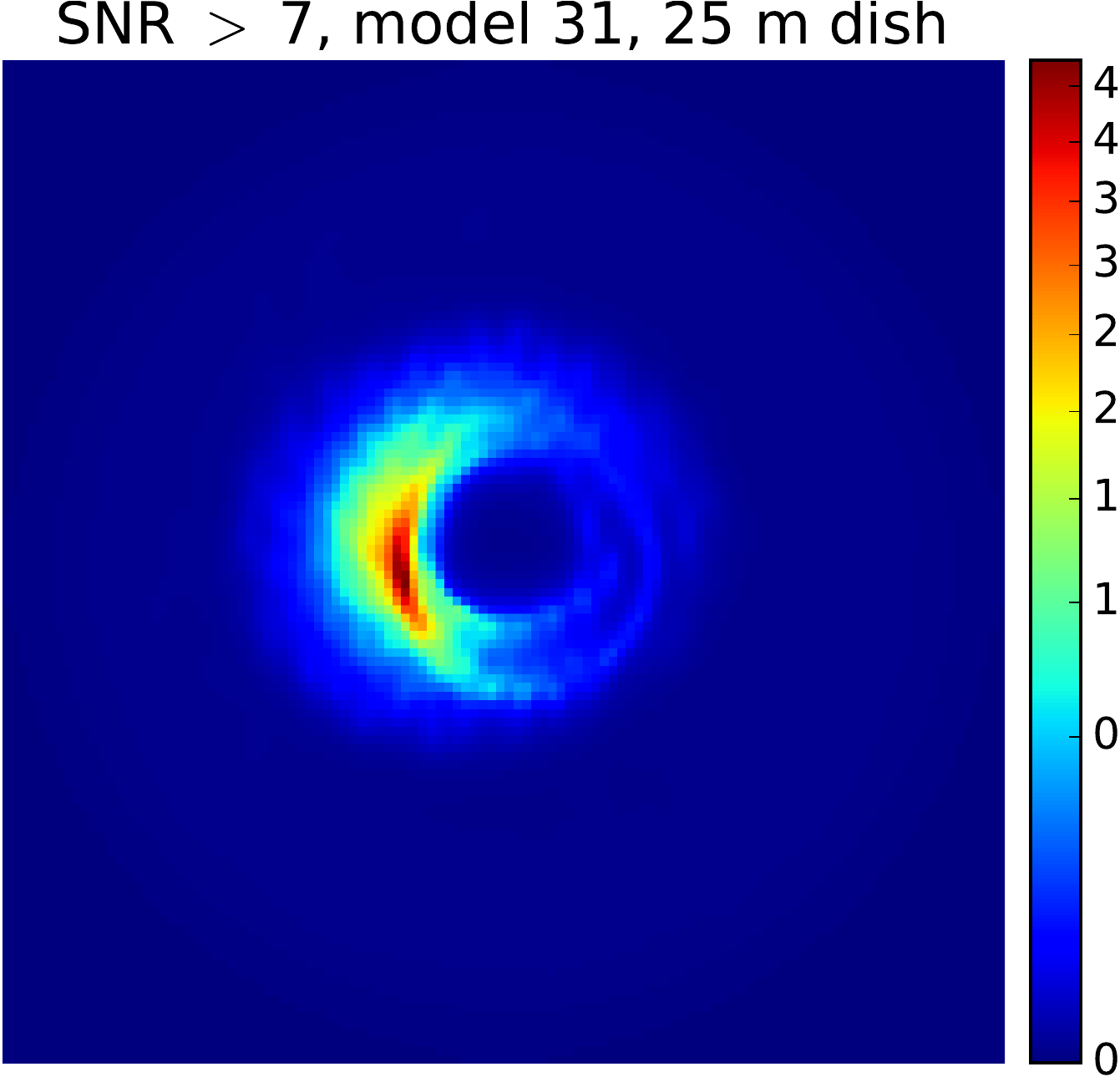}
\includegraphics[scale=0.31]{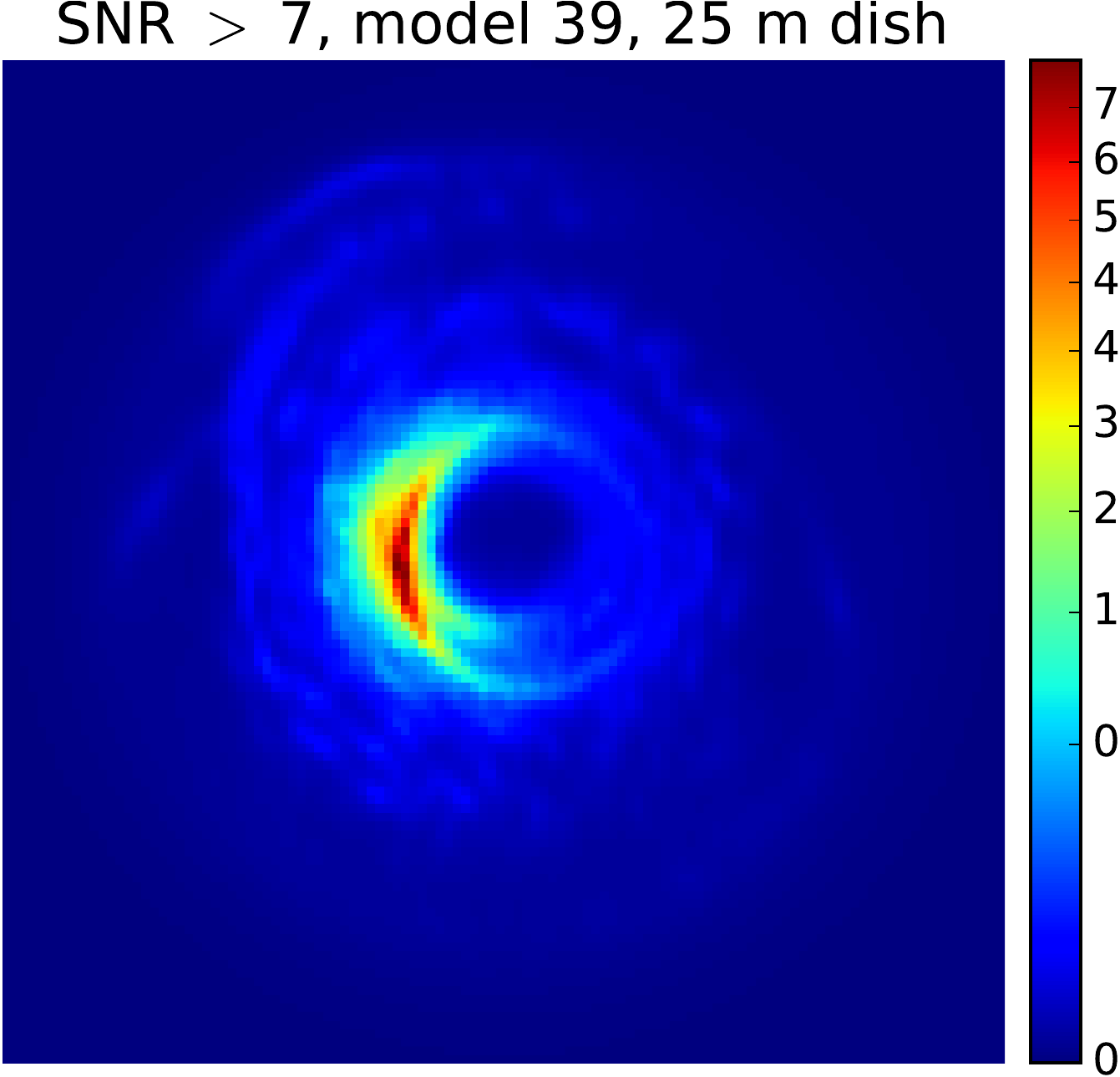} 
\caption{Image reconstructions of simulated observations of all 690\,GHz models (left to right) with a MEO system with two 4.4 (top row) or 25 (bottom row) meter dishes, using only the data points that have S/N\,$>$\,7 over a total observation time of one month. Images were reconstructed with MEM using \texttt{eht-imaging} \citep{Chael2016}. The NRMSE-values when comparing to the input models (Fig. \ref{fig:models}) are shown in Table \ref{tab:nrmsehighsnr}. The field of view is 210\,$\mu$as for all images. Colors indicate brightness/pixel in mJy (square root scale).}
\label{fig:highsnr}
\end{figure*}

In ground-based VLBI, one can use only the visibilities with sufficient S/N over a single integration time for fringe detection. The S/N-limit for fringe detection is typically set to 7, which is sufficient for a false detection rate of less than 0.01\,\% in a search of $10^6$ values of delay and delay rate \citep{TMS2017}. Figure \ref{fig:highsnr} shows images for the four source models at 690 GHz that were reconstructed using only visibilities that fulfil this requirement within the set integration time of half the $uv$-smearing limit over a total observation time of one month, as indicated in Figure \ref{fig:snr_spiral}. With 4.4-meter dishes, the image resolution is considerably lower than with 25-meter dishes. With 25-meter dishes, conventional VLBI techniques could be used to reconstruct images with a nominal resolution of 4\,$\mu$as within one month of observing time. However, launching 25-meter dishes that have sufficient surface accuracy to observe at 690\,GHz is extremely challenging, if not impossible within the next few decades. 

\begin{table}[]
\centering
\caption{NRMSE values for image reconstructions in Figures \ref{fig:highsnr}}
\label{tab:nrmsehighsnr}
\begin{tabular}{l|ll}
        & $2 \times 4.4$ m, & $2 \times 25$ m,  \\ 
Model   & S/N\,$>$\,7       & S/N\,$>$\,7       \\ \hline
16   &  0.34 & 0.22              \\
24   &  0.21 & 0.13          \\
31   &  0.40 & 0.23              \\
39   &  0.86 & 0.22     

\end{tabular}
\tablefoot{Images were compared to the unscattered images in Figure \ref{fig:models}. Lower values indicate a stronger pixel-by-pixel resemblance between the input model and reconstruction.}
\end{table}

\subsection{Two satellites: two-stage correlation}
\label{sec:conint}
The S/N detection threshold of 7 may be lowered considerably depending on the system setup. Because there is no atmosphere in space, the system could be made to behave like a connected interferometer using a two-stage correlation scheme. In this setup, the two local oscillator signals should be shared (Sec. \ref{sec:technical}) and an (a posteriori) orbital reconstruction down to the sub-wavelength level should be obtained, using e.g. the intersatellite link for high-accuracy ranging measurements. After the on-board correlation has been performed and the data has been sent to the ground, the refined orbit reconstruction could then be used to expand the fringe search to longer solution intervals. Visibilities having S/N\,<\,7 on short time intervals (set by the $uv$-smearing limit) may then still be detected because the fringe can be tracked over longer timescales.

If this behavior can be achieved, visibilities with low S/N will be coherent. Once the fringe has been detected and provided that the system is phase-stable over long timescales, multiple low-S/N visibilities can be averaged to obtain high-S/N visibilities that can be used for image reconstruction. The $uv$-spiral may be traversed for multiple iterations to build up S/N over time. If, for example, the uncertainties on the reconstructed baselines are too large, the low-S/N visibilities will be incoherent and averaging the visibilities will not yield robust higher-S/N data points.

\citet{Kudriashov2019} show that for the orbits considered in this paper, a baseline vector knowledge uncertainty (1-sigma 3D) of 0.1\,mm leads to a directivity loss 3\,dB. It is not yet clear whether it will be possible to achieve such accuracy within a reasonable budget. Another challenging task is to obtain a sufficiently accurate estimate of the baseline velocity vector and possibly its acceleration. Specific requirements for these parameters depend on the characteristics of the processing system (the size of the delay/delay-rate window) and will be addressed in other studies.

The dense $uv$-coverage of the two-satellite system allows to divide the $uv$-plane into a square grid, and then to average all the spiral points that lie within the same grid cell. The resulting uniform $uv$-coverage allows for image reconstruction by simply taking the Fourier transform of the gridded visibilities. The size of the grid cells should be kept limited in order to keep a field of view that is large enough to image the entire source (the finer the grid cell spacing, the larger the field of view of the resulting image) and avoid $uv$-smearing. In our simulations, we set the grid size to be equal to the $uv$-distance that corresponds to the field of view of the model image. With our source model and observational parameters, this results in a grid of $39\times 39$ pixels at 230\,GHz, and $116\times 116$ pixels at 690\,GHz, with a grid cell size of 0.49\,G$\lambda$. With our baseline-dependent integration time (Sec. \ref{sec:uvsampling}), each grid cell typically contains 1-2 measurements per observing epoch.

In the following subsections, we first present the simulation results for model 39 observed with the system described above while varying the total integration time (Sec. \ref{sec:tint}). We then compare results for different source models (Sec. \ref{sec:simul}). In these simulations, we observe the scattered time-averaged source models in Figure \ref{fig:models}, setting the orbital plane perpendicular to the line of sight to the source. We study the effects of source declination and time variability in Sections \ref{sec:dec} and \ref{sec:tv}, respectively. 

\begin{figure*}[!h]
\centering
\includegraphics[scale=0.441]{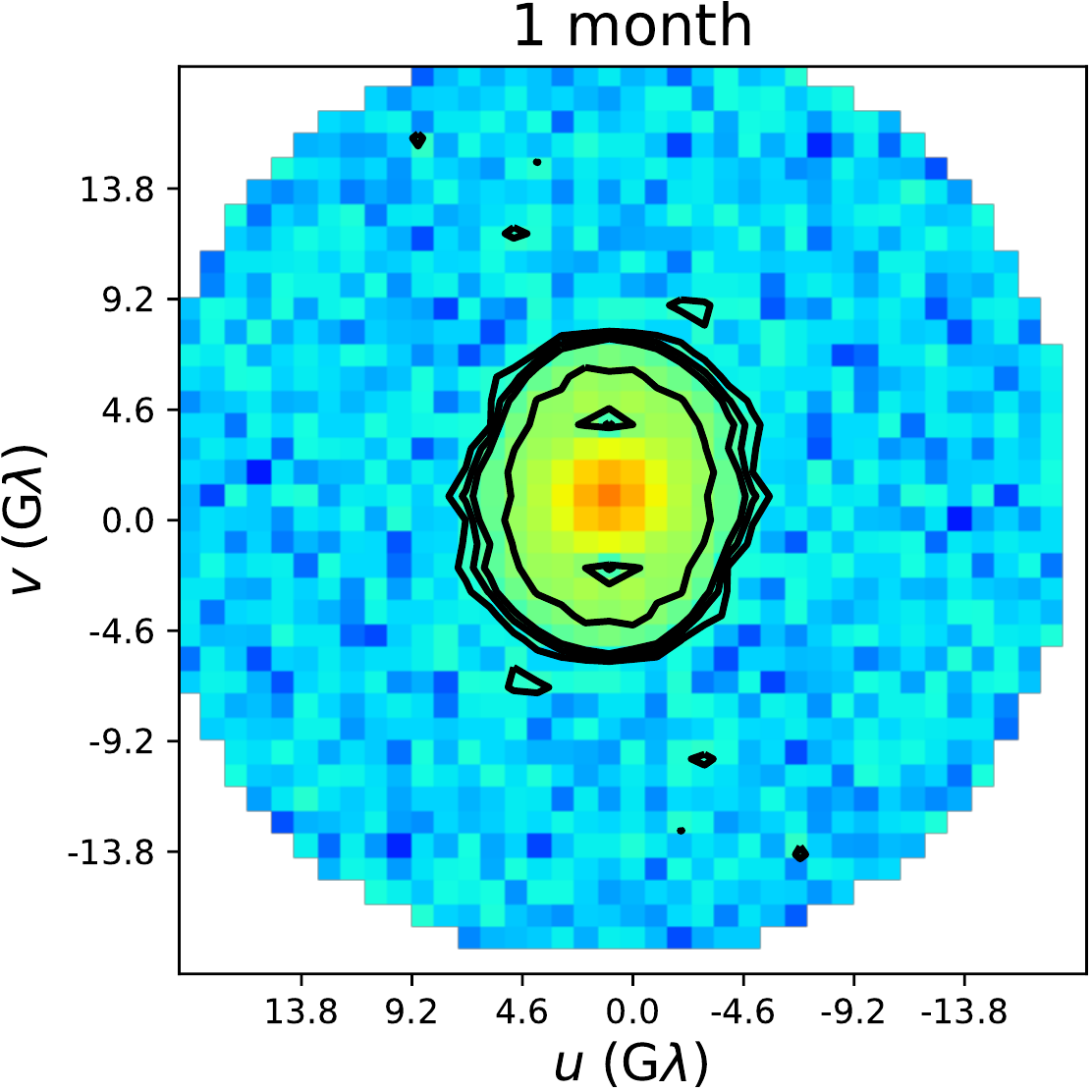}
\includegraphics[scale=0.441]{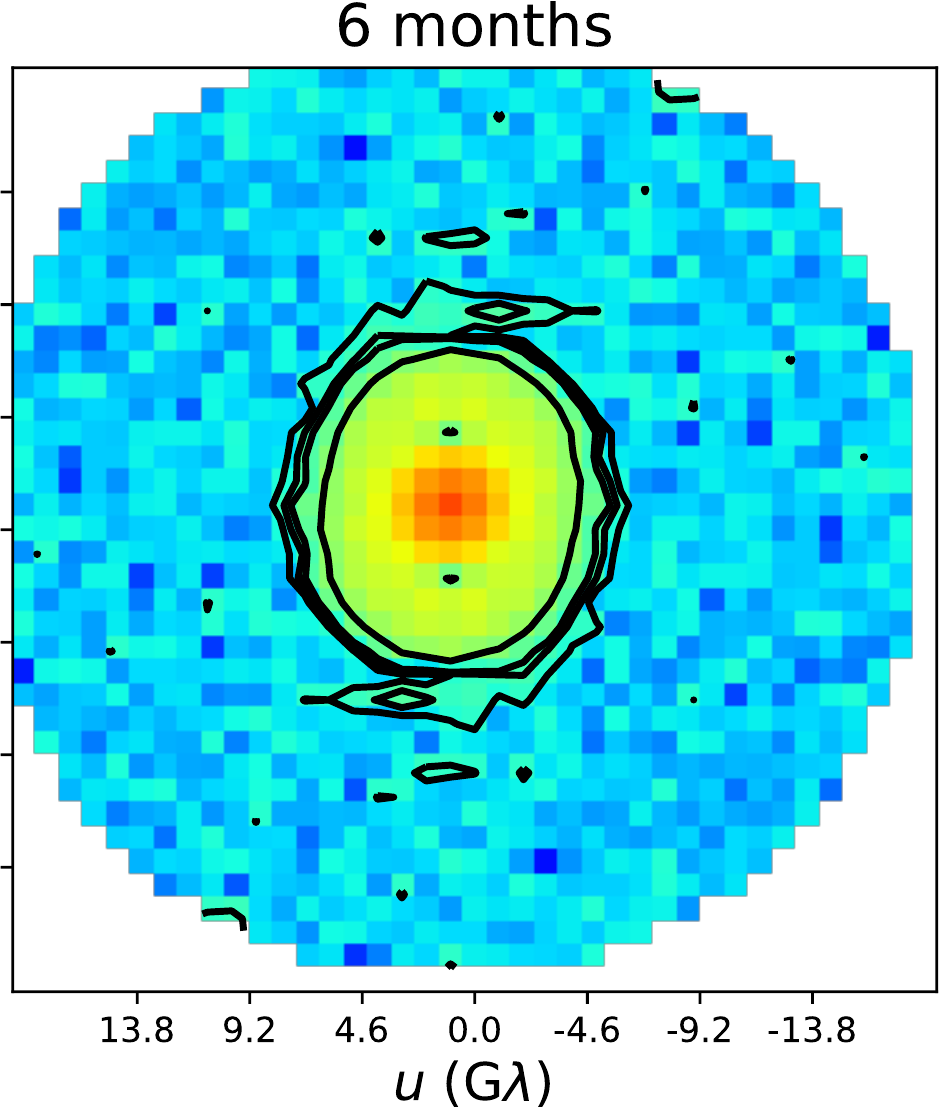}
\includegraphics[scale=0.441]{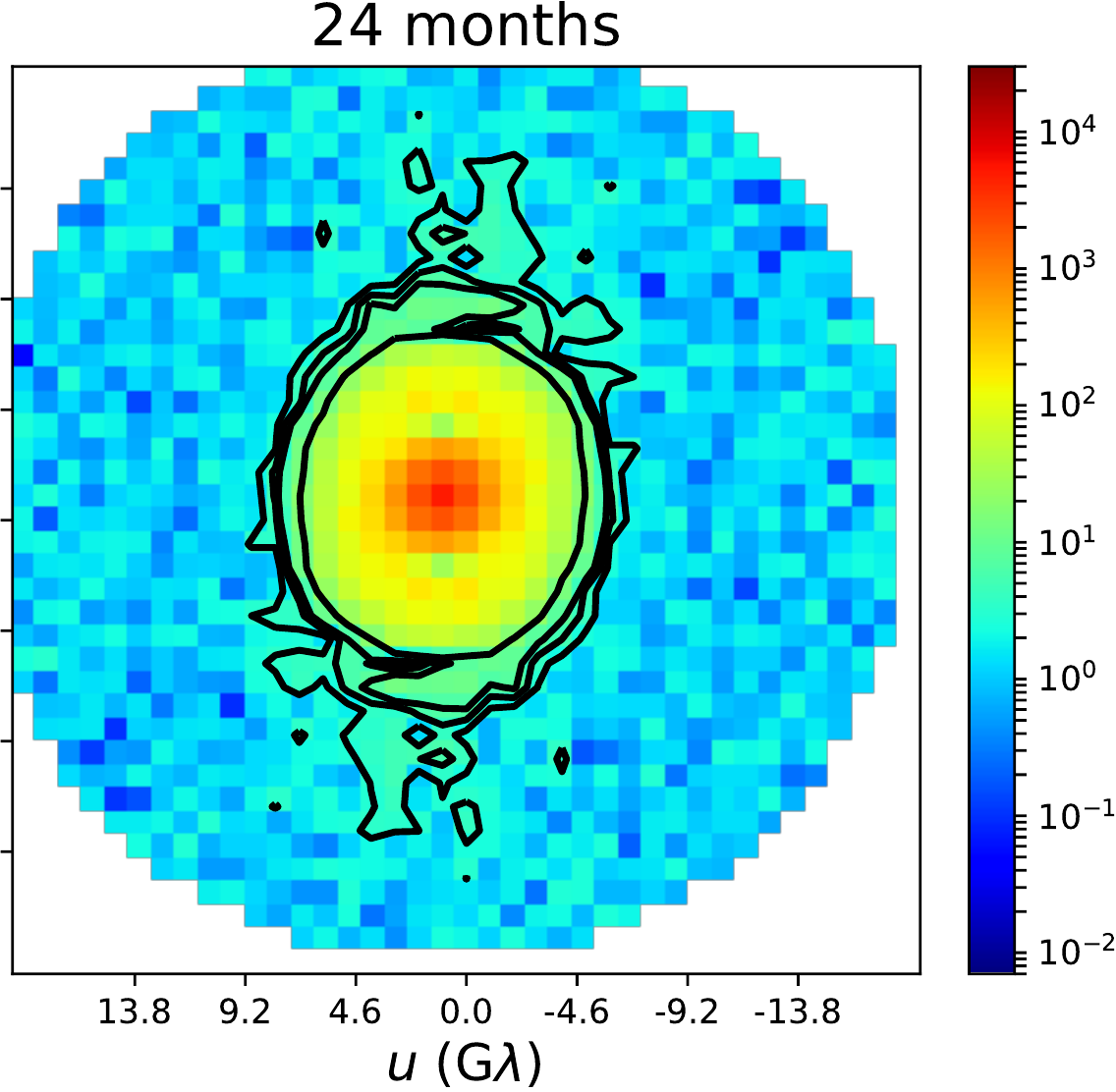}
\caption{S/N map of the gridded visibilities of model 39 (scattered) at 230\,GHz after integrating for 1, 6, and 24 months (left to right), with a reflector diameter of 4.4 meters. Contours indicate the points with an S/N of 3, 5, 7, and 20.}
\label{fig:snrgrid230}
\end{figure*}

\begin{figure*}[!h]
\centering
\includegraphics[scale=0.340]{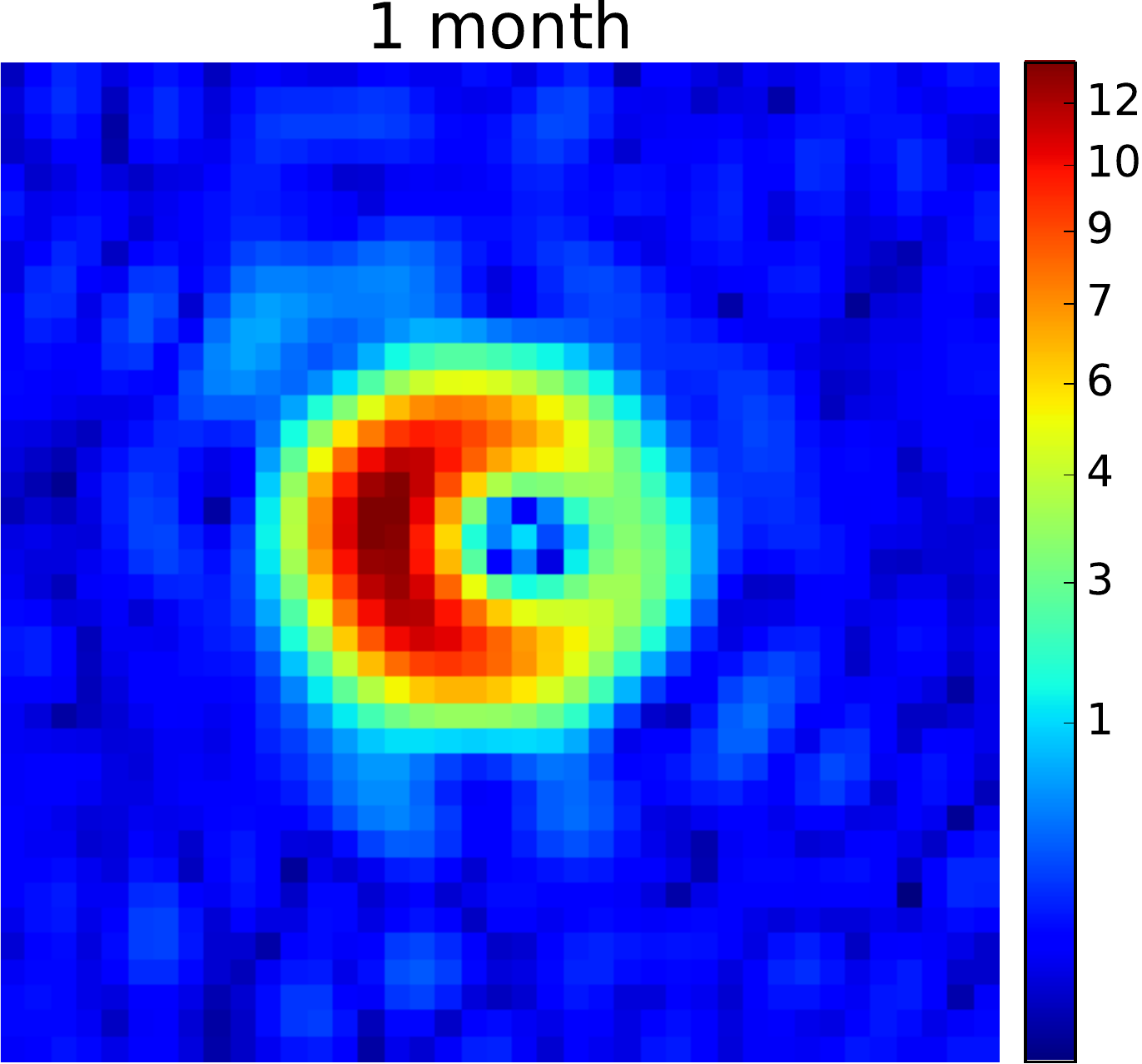}
\includegraphics[scale=0.340]{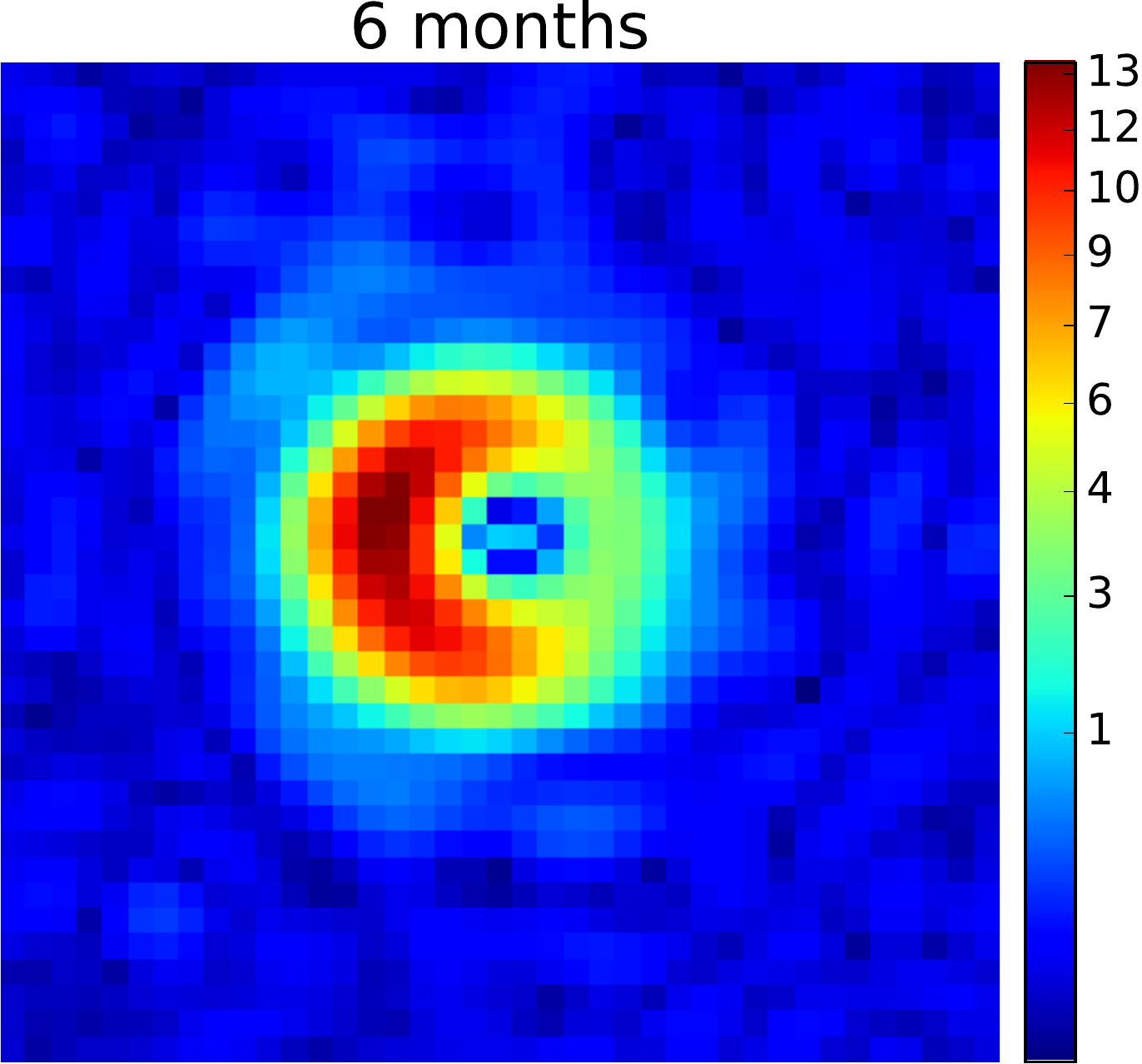}
\includegraphics[scale=0.340]{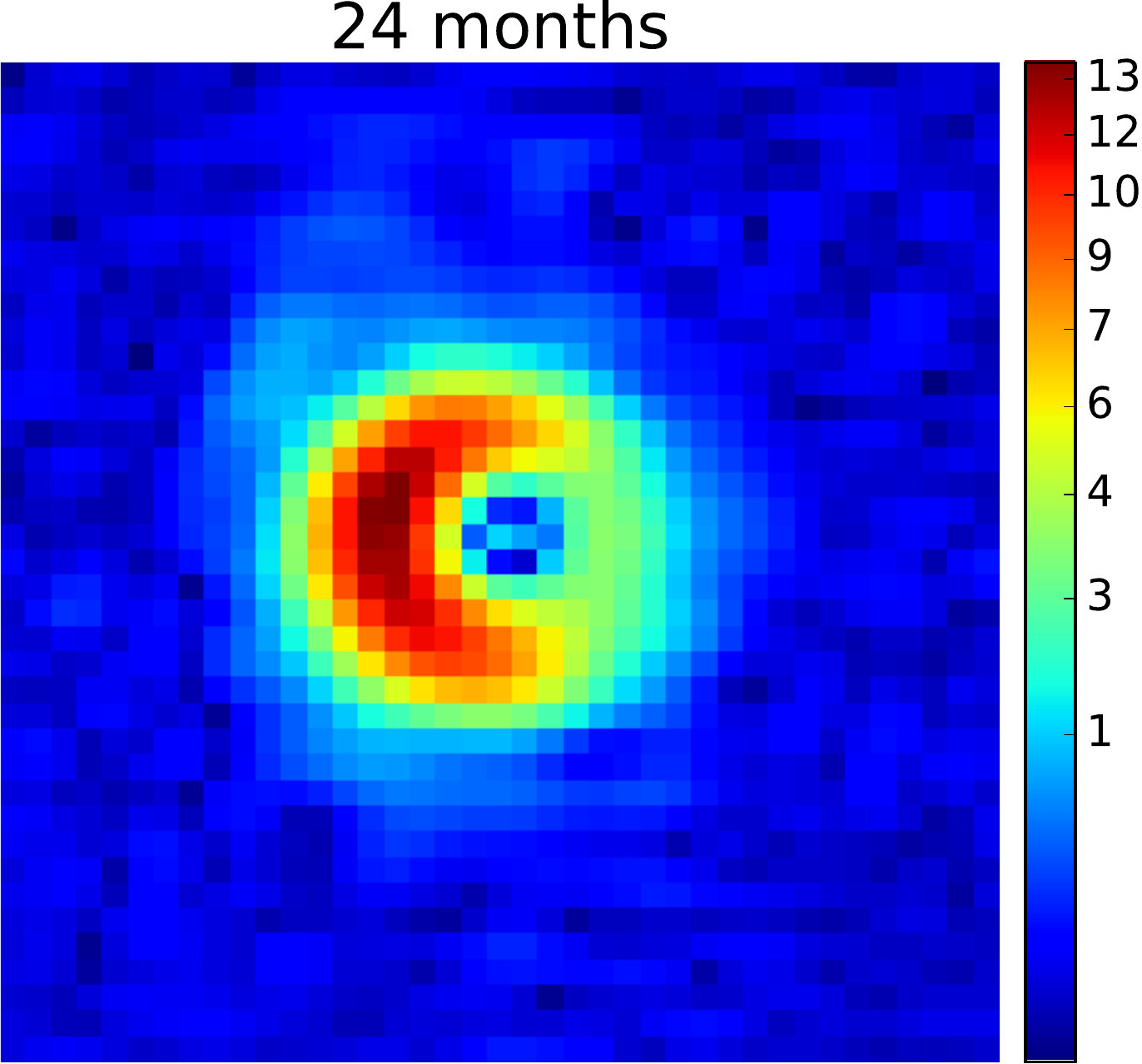}
\caption{FFT of the gridded and deblurred visibilities of model 39 (scattered) at 230\,GHz after integrating for 1, 6, and 24 months (left to right), with a reflector diameter of 4.4 meters. The field of view is 210\,$\mu$as for all images. Colors indicate brightness/pixel in mJy (square root scale). The NRMSE-values when comparing to the input model (Fig. \ref{fig:models}) are shown in Table \ref{tab:nrmse}.}
\label{fig:fft230}
\end{figure*}

\begin{figure*}[!h]
\centering
\includegraphics[scale=0.441]{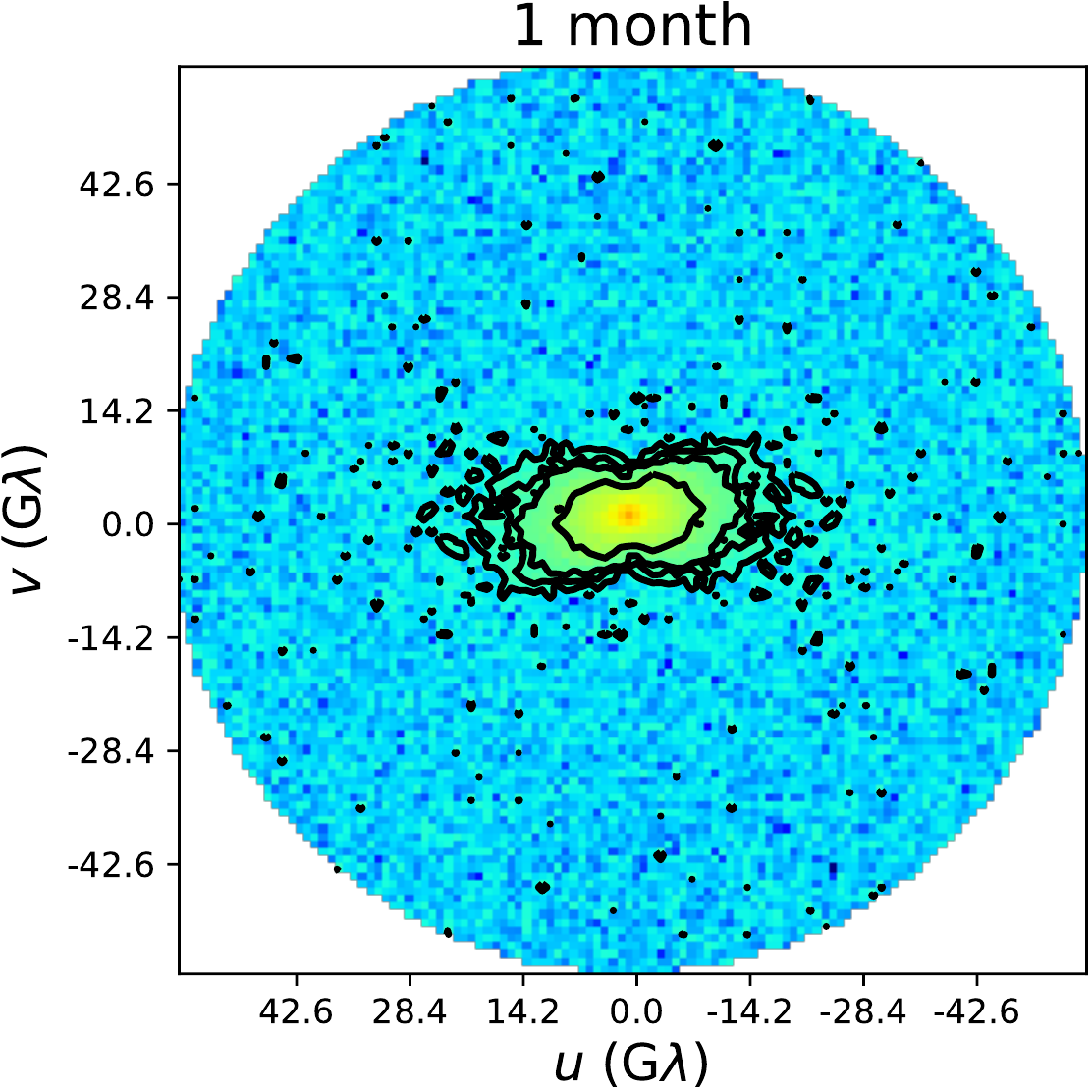}
\includegraphics[scale=0.441]{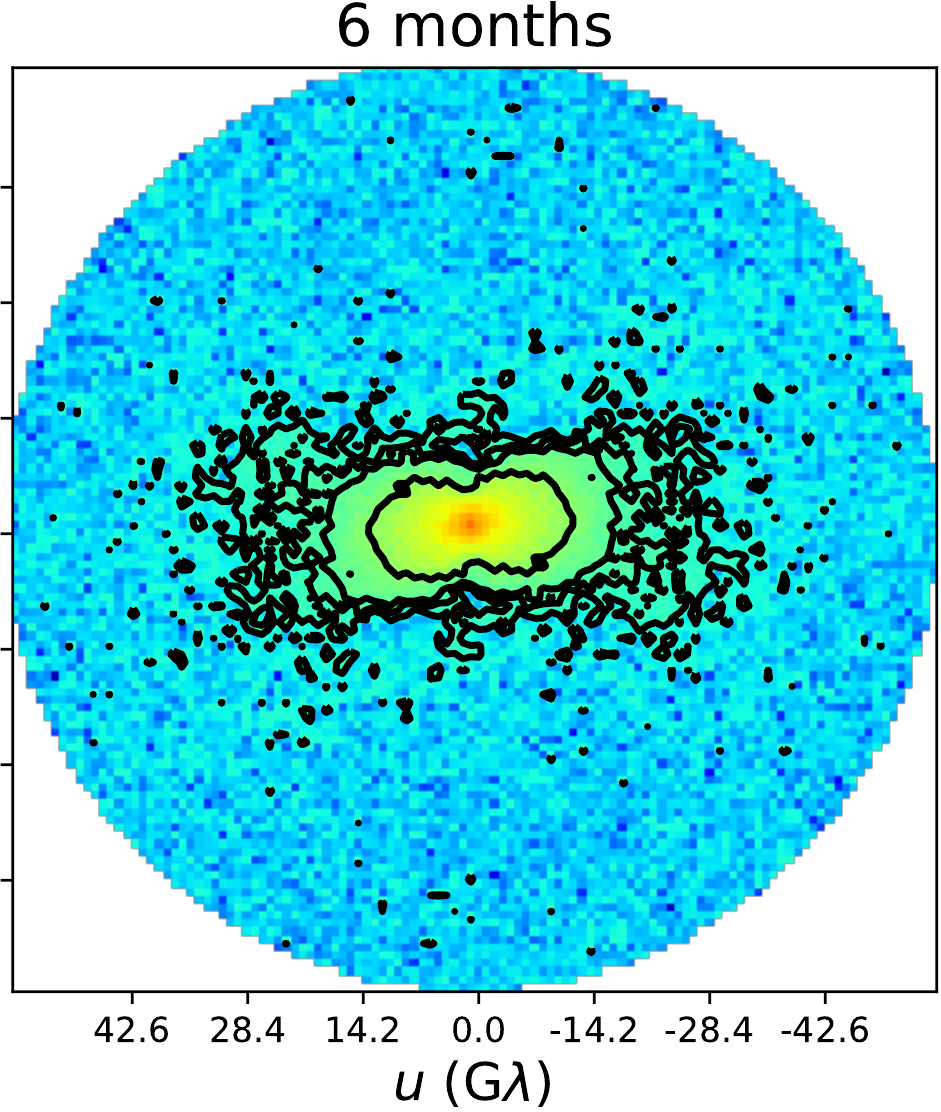}
\includegraphics[scale=0.441]{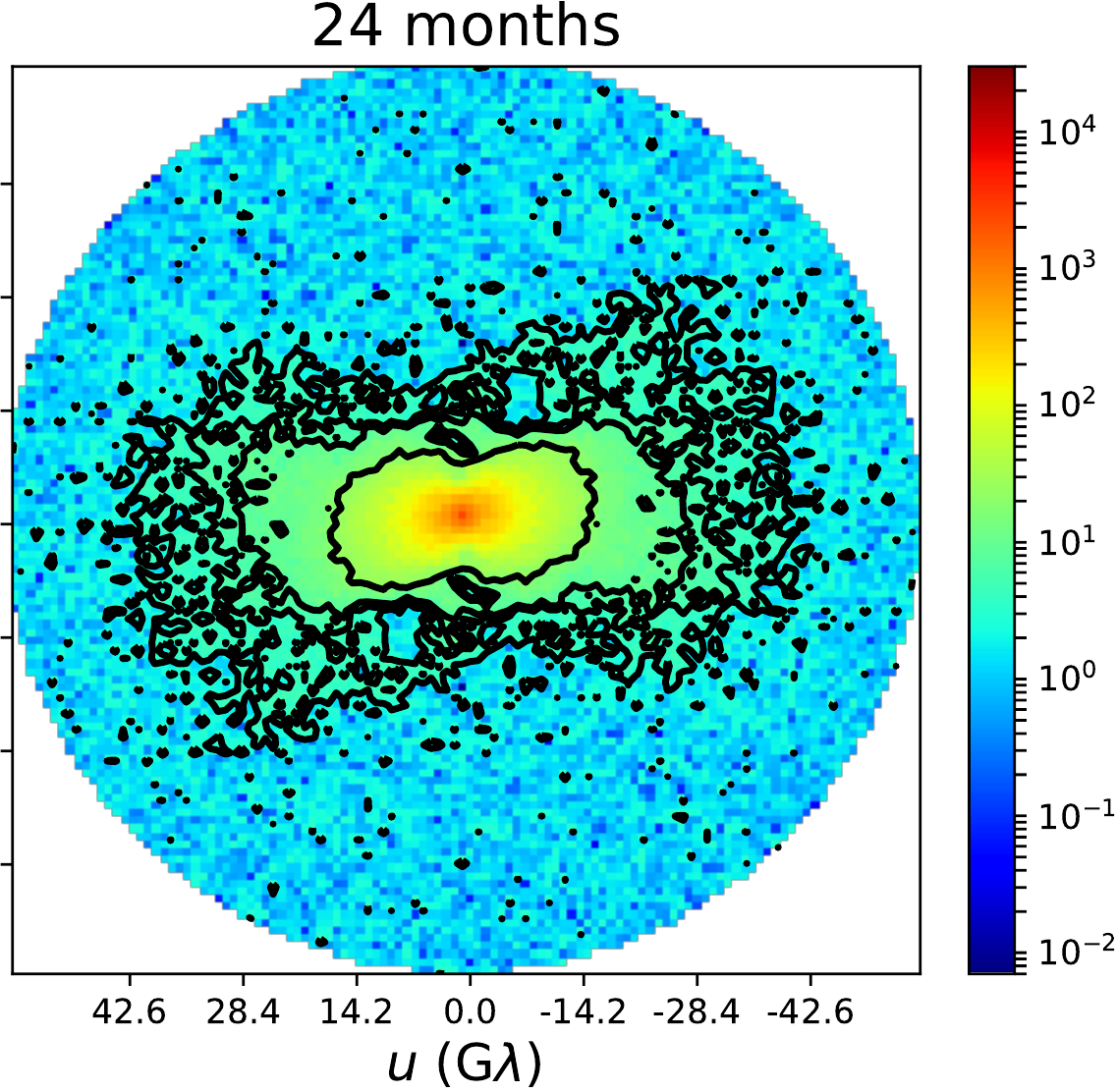}
\caption{Same as Fig. \ref{fig:snrgrid230}, but for an observation frequency of 690 GHz.}
\label{fig:snrgrid690}
\end{figure*}

\begin{figure*}[!h]
\centering
\includegraphics[scale=0.340]{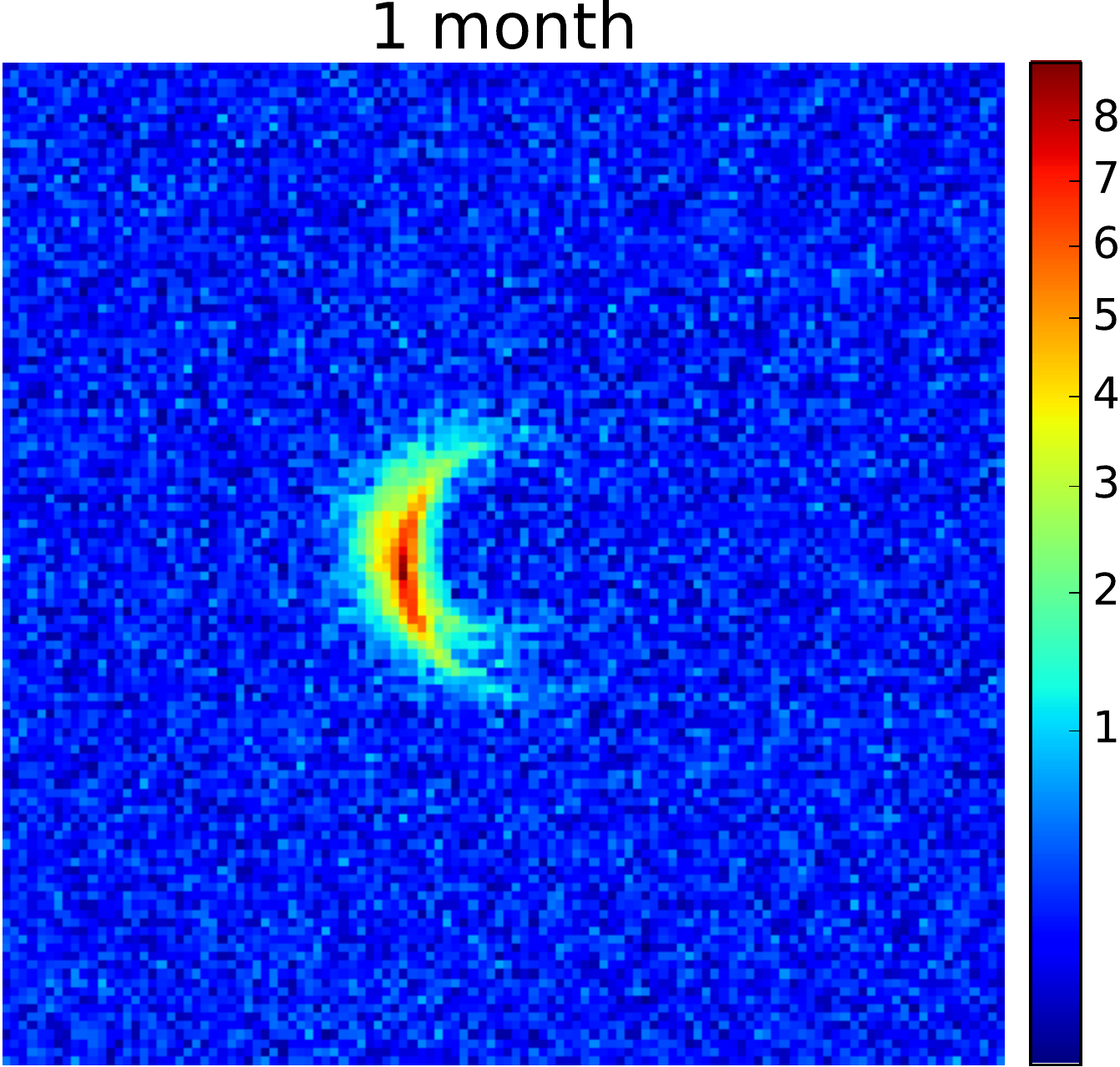}
\includegraphics[scale=0.340]{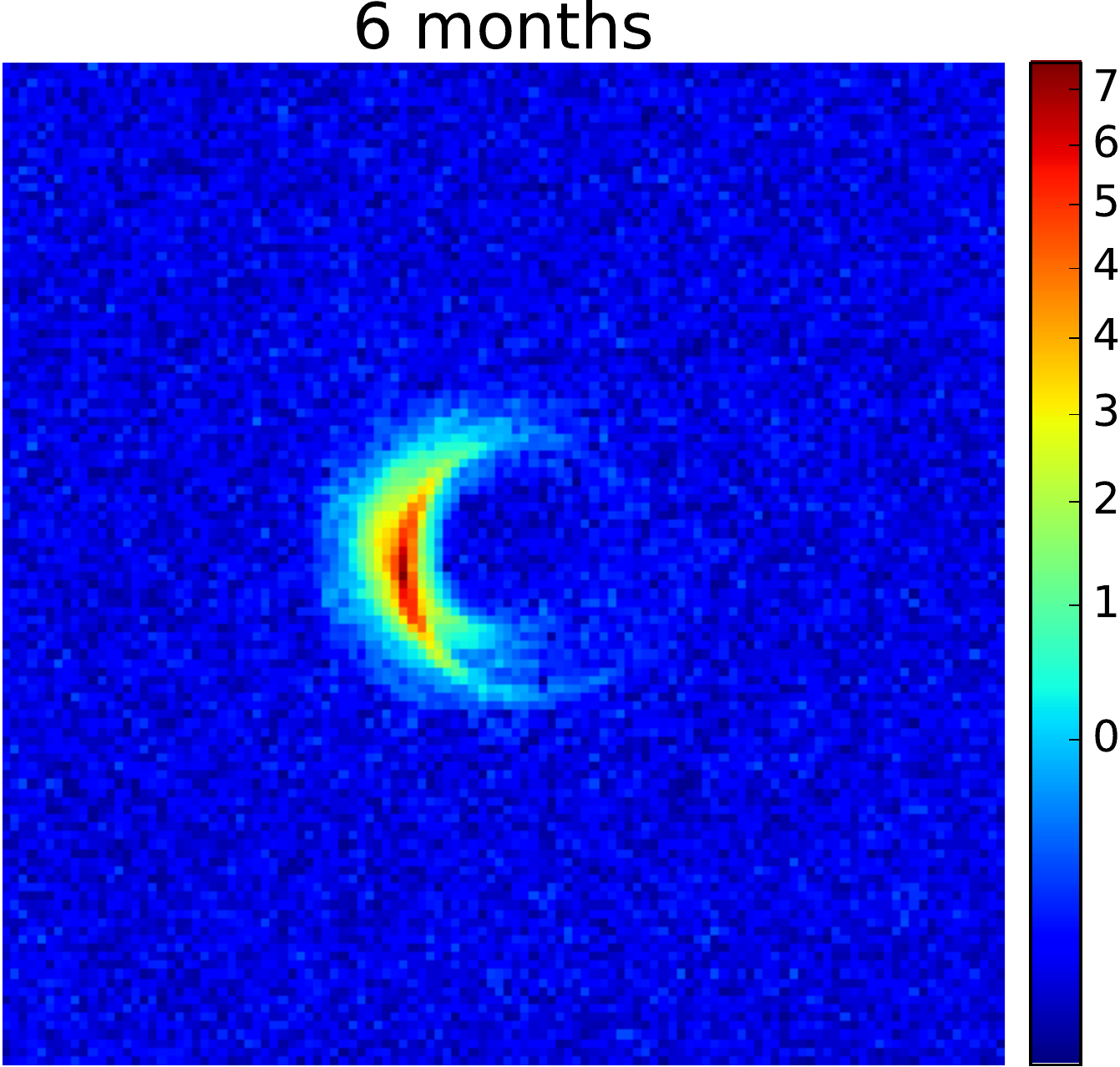}
\includegraphics[scale=0.340]{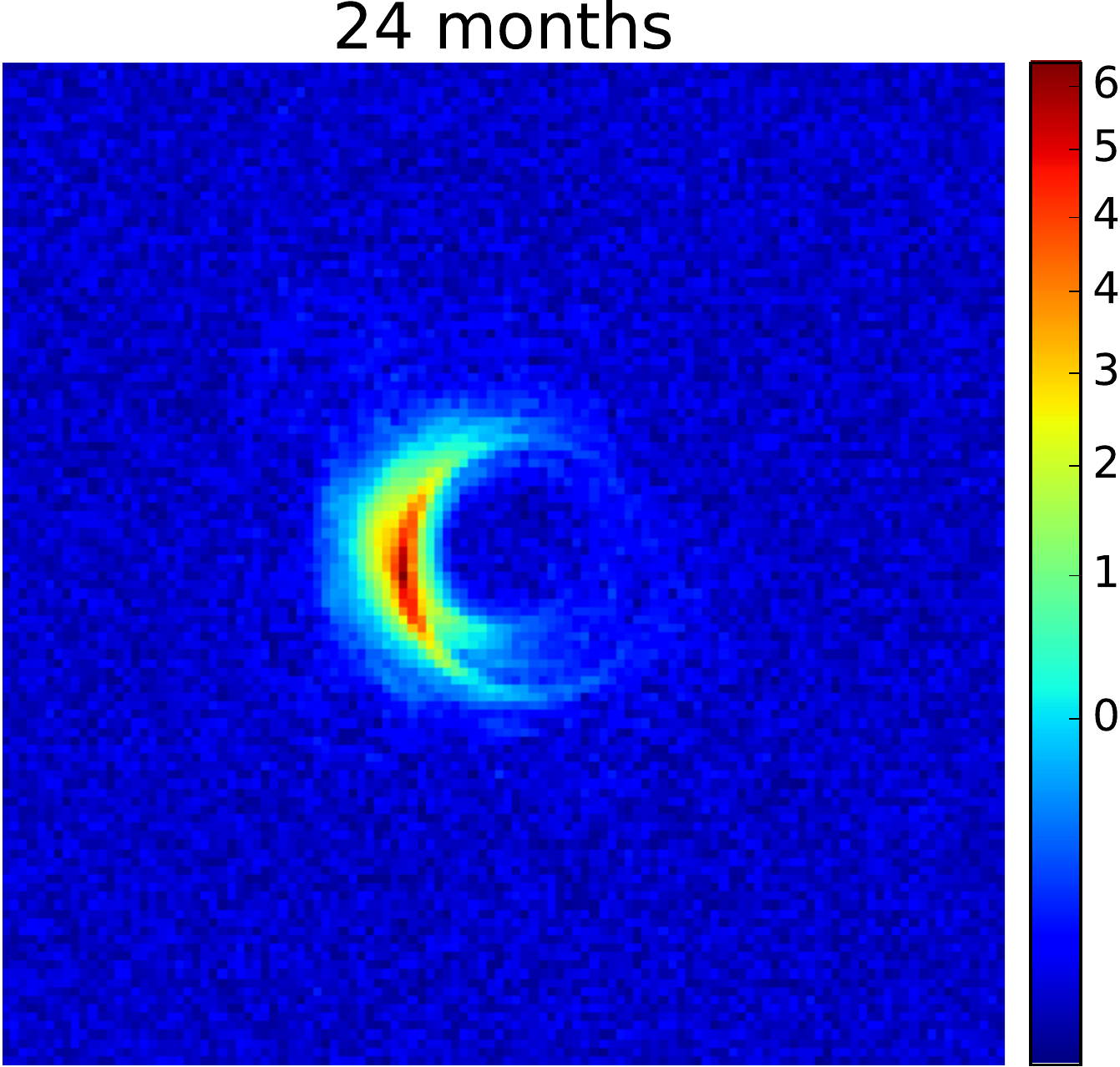}
\caption{Same as Fig. \ref{fig:fft230}, but for an observation frequency of 690 GHz.}
\label{fig:fft690}
\end{figure*}

\subsubsection{Total integration time} 
\label{sec:tint}
Figure \ref{fig:snrgrid230} shows the improvement in signal-to-noise ratio for the gridded visibilities of model 39 observed at 230\,GHz with increasing total integration time with a 4.4-meter dish. The assumed bandwidth is 10\,GHz, which one would reach by combining all polarizations (Sec. \ref{sec:noise}). Figure \ref{fig:fft230} shows the Fourier transform (dirty image) of these visibilities. The visibilities were deblurred (Sec. \ref{sec:deblur}) before performing the FFT. The same plots are shown in Figures \ref{fig:snrgrid690} and \ref{fig:fft690} for an observing frequency of 690\,GHz. Deblurring was not applied here as the extrapolated major axis of the scattering kernel at 690\,GHz is only 2\,$\mu$as, which is a factor 2 smaller than the maximum angular resolution that can be obtained with the investigated setup. NRMSE values (Sec. \ref{sec:nrmse}) comparing the reconstructed images to the original model are shown in Table \ref{tab:nrmse}. For all images, the pixel values in the reconstructed image were scaled such that the total flux matched the total flux of the input model before the NRMSE was calculated.

Comparing the S/N plots after 1 month to Figure \ref{fig:snr_spiral}, the S/N has increased due to the combining of polarizations and averaging visibilities that are in the same grid cell. The latter effect is marginal as the number of measurements in a single grid cell is of order unity in most grid cells. The S/N improvement as a function of total integration time is smaller for 230\,GHz than for 690\,GHz because blurring by interstellar scattering decreases the S/N on long baselines at this frequency. This is also reflected in the NRMSE values (Tab. \ref{tab:nrmse}), which show a plateau in image quality as the noise decreases at 230\,GHz, but keep decreasing at 690\,GHz. 

At 230\,GHz, the resolution of the reconstructed images is comparable to the resolution of the EHT (with S/N\,$>$\,7 visibilities on baselines up to $\sim$\,8\,G$\lambda$ after 1 month), despite the longer baselines available in the SVLBI setup. Due to the dense $uv$-coverage and long integration, robust image reconstructions can be obtained. Observations at 230\,GHz could also be useful for (initial) fringe detection at 690\,GHz, orbit reconstruction, and cross-comparison with EHT results. 

\begin{figure*}[!h]
\centering
\includegraphics[width=0.985\textwidth]{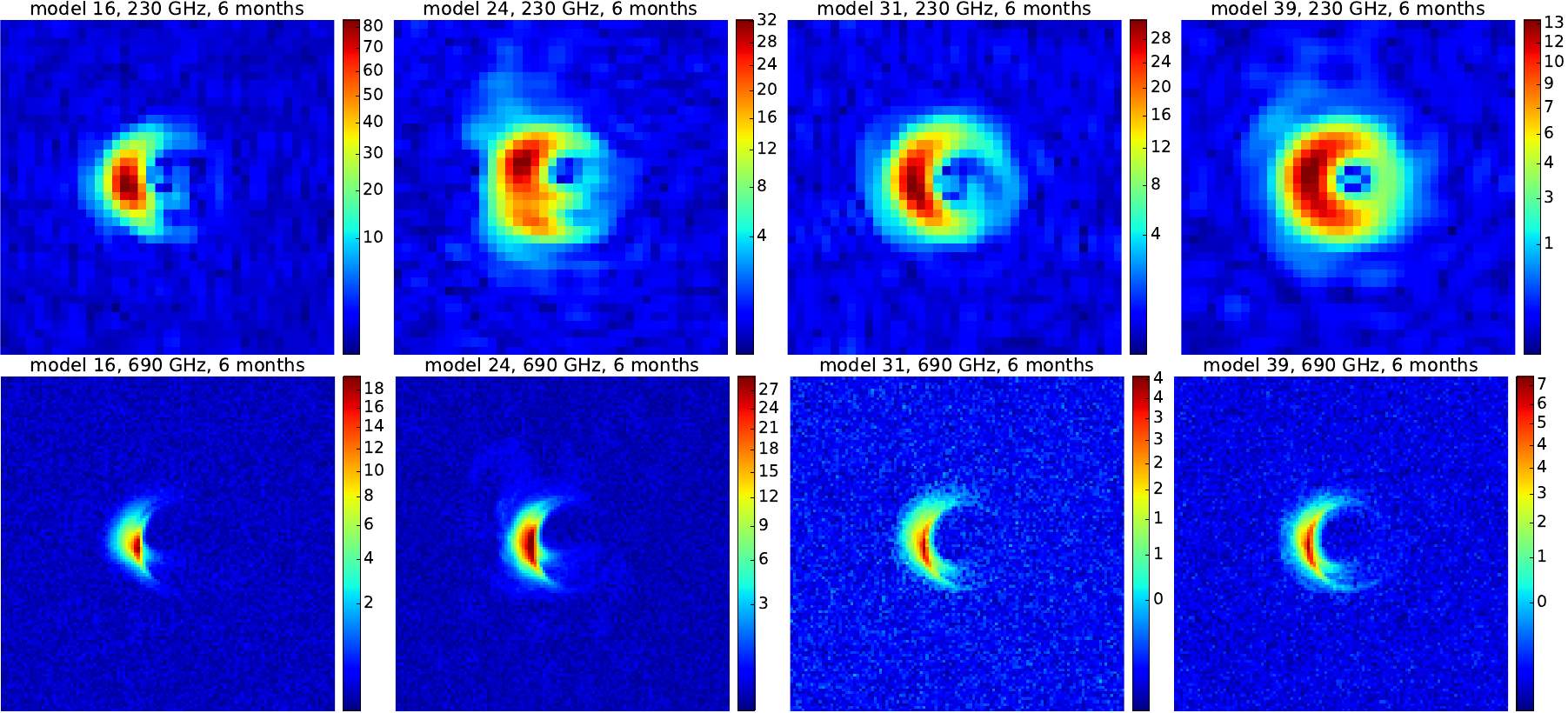}
\caption{FFT of the gridded visibilities of all models at all frequencies, integrated for 6 months with a 4.4-meter reflector. The NRMSE-values when comparing to the input models (Fig. \ref{fig:models}) are shown in Table \ref{tab:nrmsemodels}. The field of view is 210\,$\mu$as for all images. Colors indicate brightness/pixel in mJy (square root scale).}
\label{fig:fftmodels}
\end{figure*}

\begin{figure*}[!h]
\centering
\includegraphics[scale=0.32]{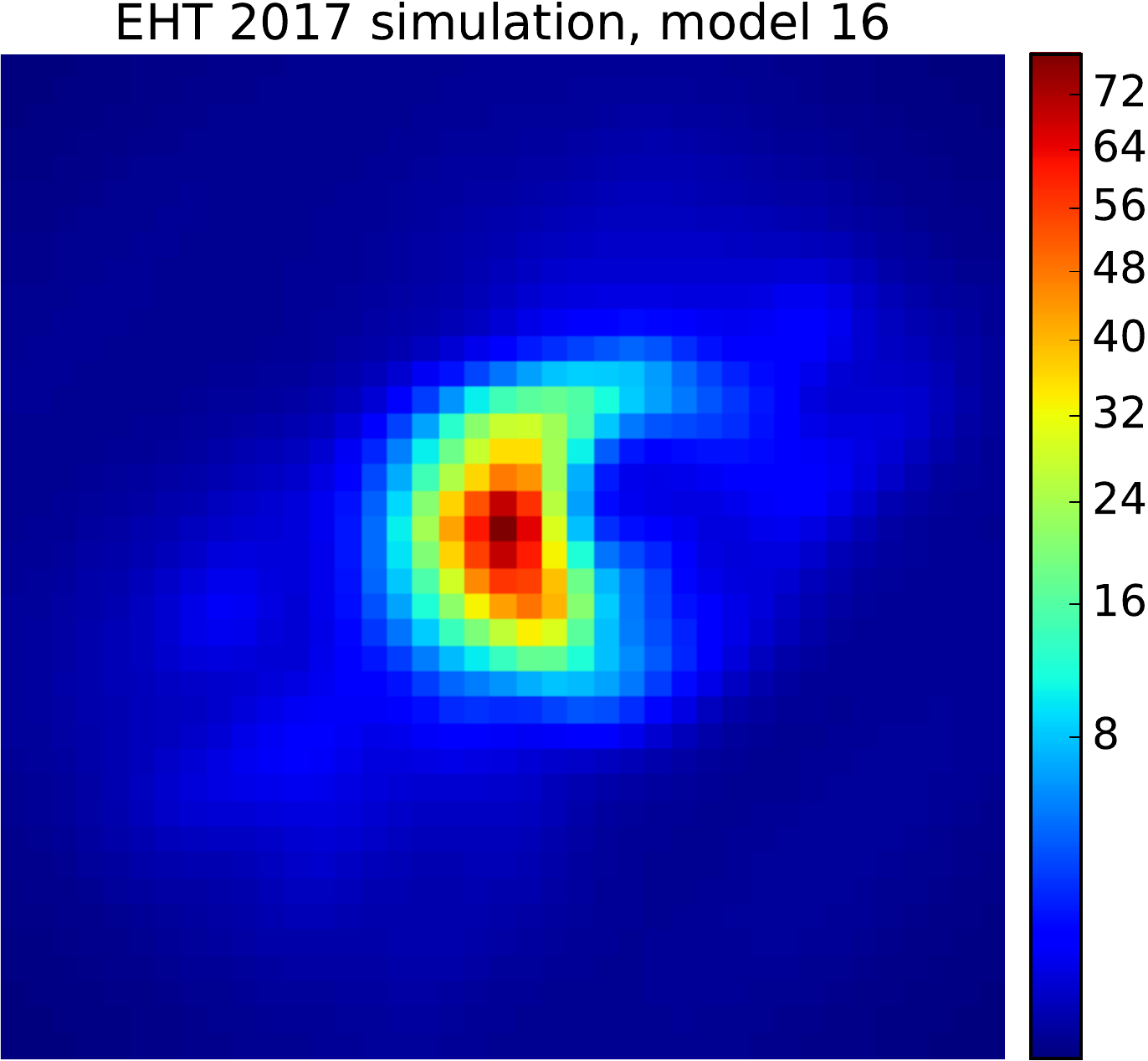}
\includegraphics[scale=0.32]{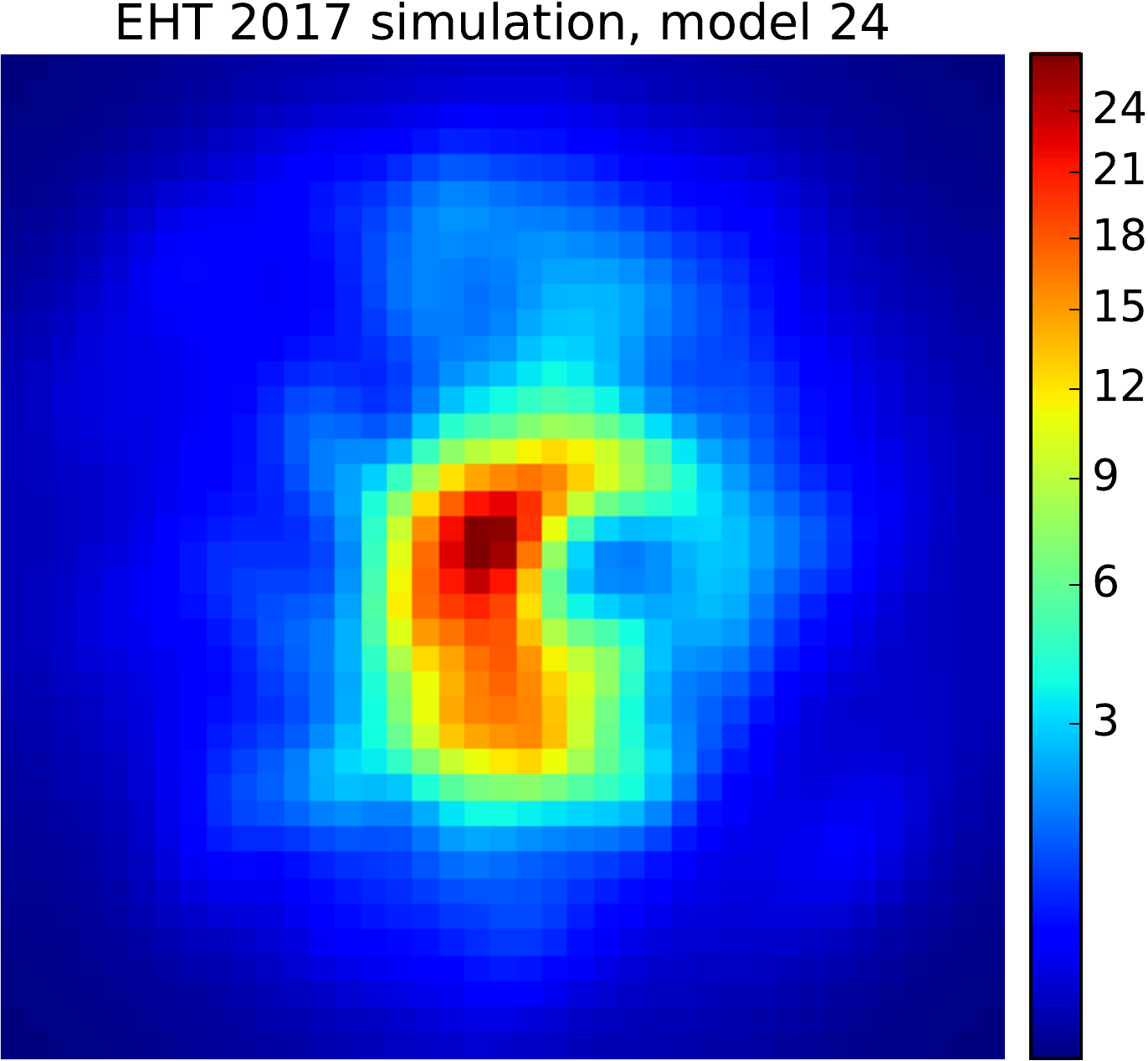}
\includegraphics[scale=0.32]{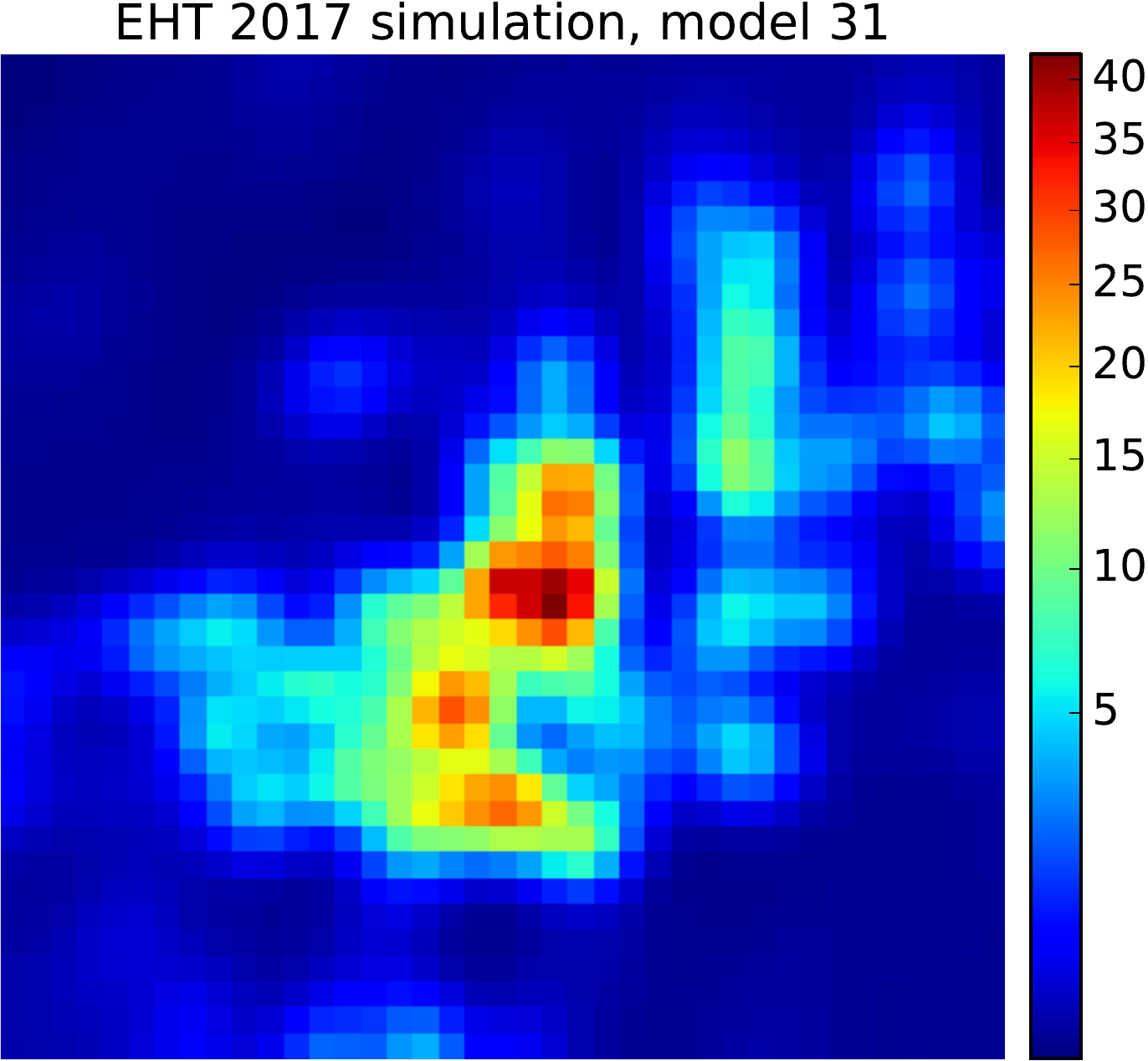}
\includegraphics[scale=0.32]{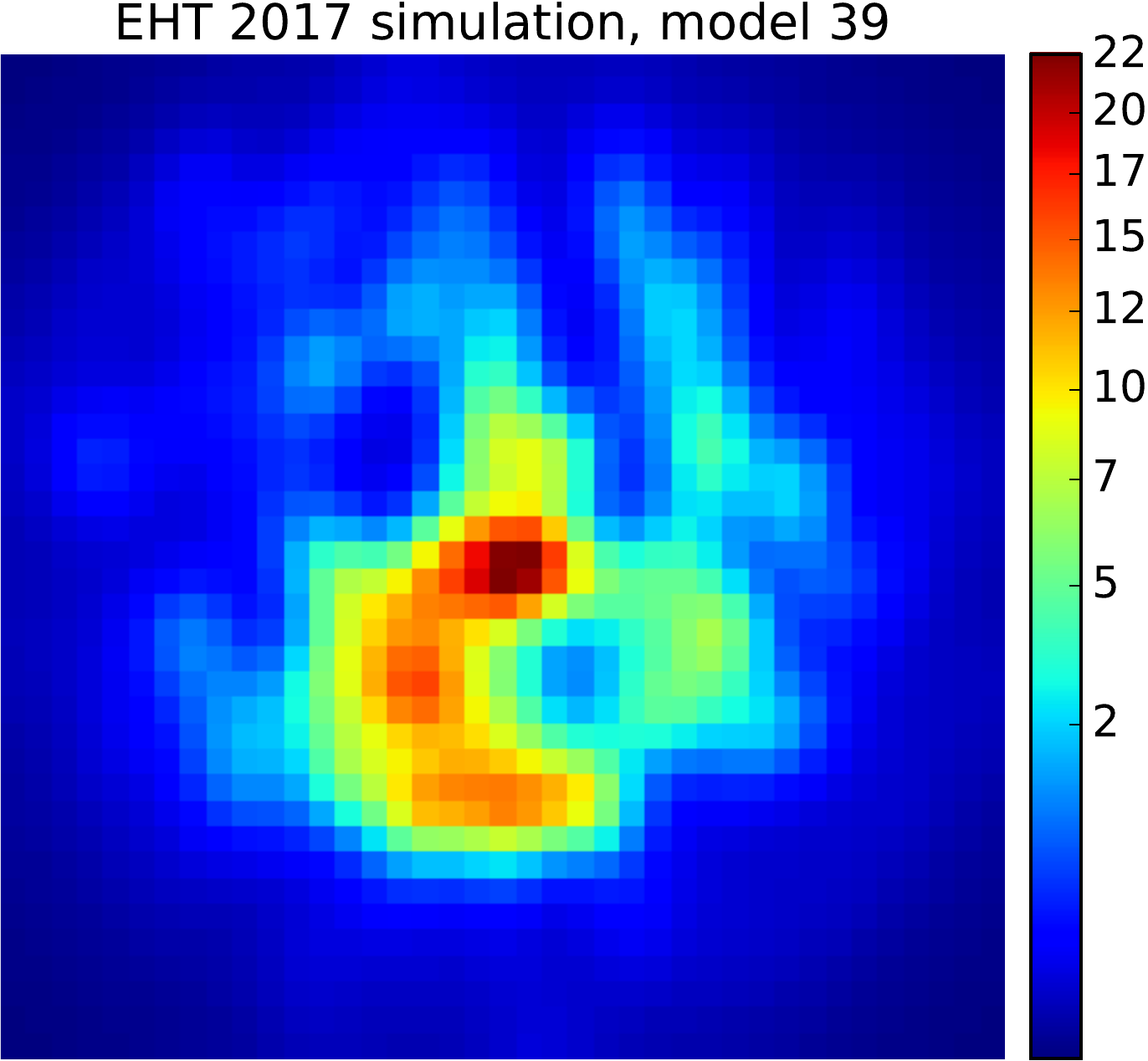}
\caption{Image reconstructions of simulated EHT 2017 observations of all models including ensemble-average scattering at 230\,GHz. Images were reconstructed using a MEM algorithm implemented in \texttt{eht-imaging} \citep{Chael2016, Chael2018}. The NRMSE-values when comparing to the input models (Fig. \ref{fig:models}) are shown in Table \ref{tab:nrmsemodels}.}
\label{fig:eht}
\end{figure*}

\begin{table}
\centering
\caption{NRMSE values for image reconstructions in Figures \ref{fig:fft230} and \ref{fig:fft690}}
\label{tab:nrmse}
\begin{tabular}{ll|l}
$\nu$ (GHz) & $t_{\mathrm{int,tot}}$ (months) & NRMSE  \\ \hline
230             & 1                         & 0.22     \\
                & 6                         & 0.19      \\
                & 24                        & 0.18       \\ \hline
690             & 1                         & 0.72      \\
                & 6                         & 0.52     \\
                & 24                        & 0.38  
\end{tabular}
\end{table}

\subsubsection{Different source models}
Figure \ref{fig:fftmodels} shows the image reconstructions for the different source models (Fig. \ref{fig:models}) and frequencies for a 6-month observation with 4.4-meter reflectors. NRMSE values are shown in Table \ref{tab:nrmsemodels}. For the low-inclination models (31 and 39), the black hole shadow can be traced more easily than for the high-inclination models. At 690\,GHz, the apparent difference between the disk (models 16 and 31) and jet (models 24 and 39) image reconstructions is small, because the image morphology is dominated by general \\\\\\ relativistic effects such as gravitational lensing and Doppler boosting, and the jet feature in the simulation is relatively faint. At 230\,GHz however, extended structure associated with the jet can be seen more clearly in the reconstructions of the jet models, especially for model 24. 230\,GHz observations may thus be more useful for discriminating between disk and jet models, but 690\,GHz observations will allow for significantly sharper reconstructions of the black hole shadow.

For comparison, Figure \ref{fig:eht} shows image reconstructions of simulated EHT observations of all scattered models made with the \texttt{eht-imaging} software \citep{Chael2016, Chael2018}. The stations included are the same as for the April 2017 EHT observations (Sec. \ref{sec:eht}). The integration time per measurement was set to 30\,s, with a measurement cadence of 300\,s, observing for one day in total. The bandwidth was set to 4\,GHz. No atmospheric or instrumental effects were included except thermal noise. Images were reconstructed from the simulated visibility amplitudes and closure phases using a maximum entropy (MEM) algorithm with the Gull-Skilling entropy function \citep{Chael2016}. Closure phases were used for image reconstruction instead of visibility phases because atmospheric corruptions would severely corrupt visibility phases, while closure phases are immune to these. The images were blurred with a Gaussian with a size of half the beam corresponding to the array resolution, in order to mitigate spurious superresolved structures. The NRMSE values (Tab. \ref{tab:nrmsemodels}) are higher than the NRMSE values of the high-S/N\,230 GHz simulations of the SVLBI array, indicating less similarity between the input models and reconstructions. Comparing the reconstructions by eye, the SVLBI reconstructions are more robust in that they contain less spurious structure than most of the EHT reconstructions, especially for the low-inclination models 31 and 39. Comparing Fig. \ref{fig:eht} to the bottom row of Fig. \ref{fig:fftmodels}, observing Sgr\,A* for multiple months with two 4.4-meter reflectors in space at 690\,GHz could produce an image of the black hole shadow with a quality that significantly surpasses the image quality that can be expected for the EHT.

\begin{table}[]
\centering
\caption{NRMSE values for image reconstructions in Figures \ref{fig:fftmodels} and \ref{fig:eht}}
\label{tab:nrmsemodels}
\begin{tabular}{ll|ll}
$\nu$ (GHz) & Model  & EHI & EHT \\ \hline
230         & 16     & 0.29 & 0.25                 \\
            & 24     & 0.18 & 0.31               \\
            & 31     & 0.26 & 0.63               \\
            & 39     & 0.19 & 0.41               \\ \hline
690         & 16     & 0.49 & -              \\
            & 24     & 0.27 & -              \\
            & 31     & 0.66 & -              \\
            & 39     & 0.52 & -             
\end{tabular}
\end{table}

\begin{figure*}[!h]
\centering
\includegraphics[scale=0.35]{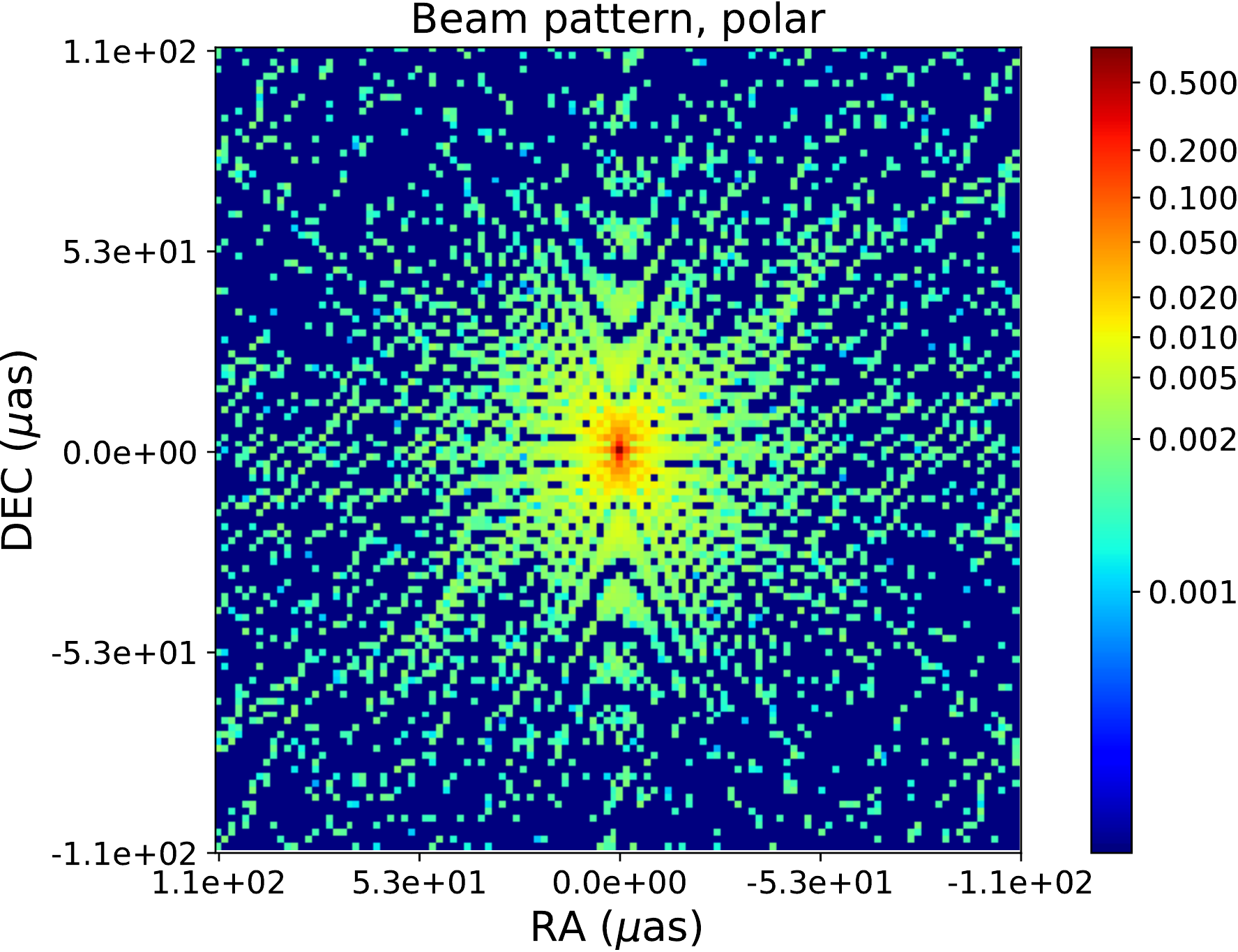}
\includegraphics[scale=0.35]{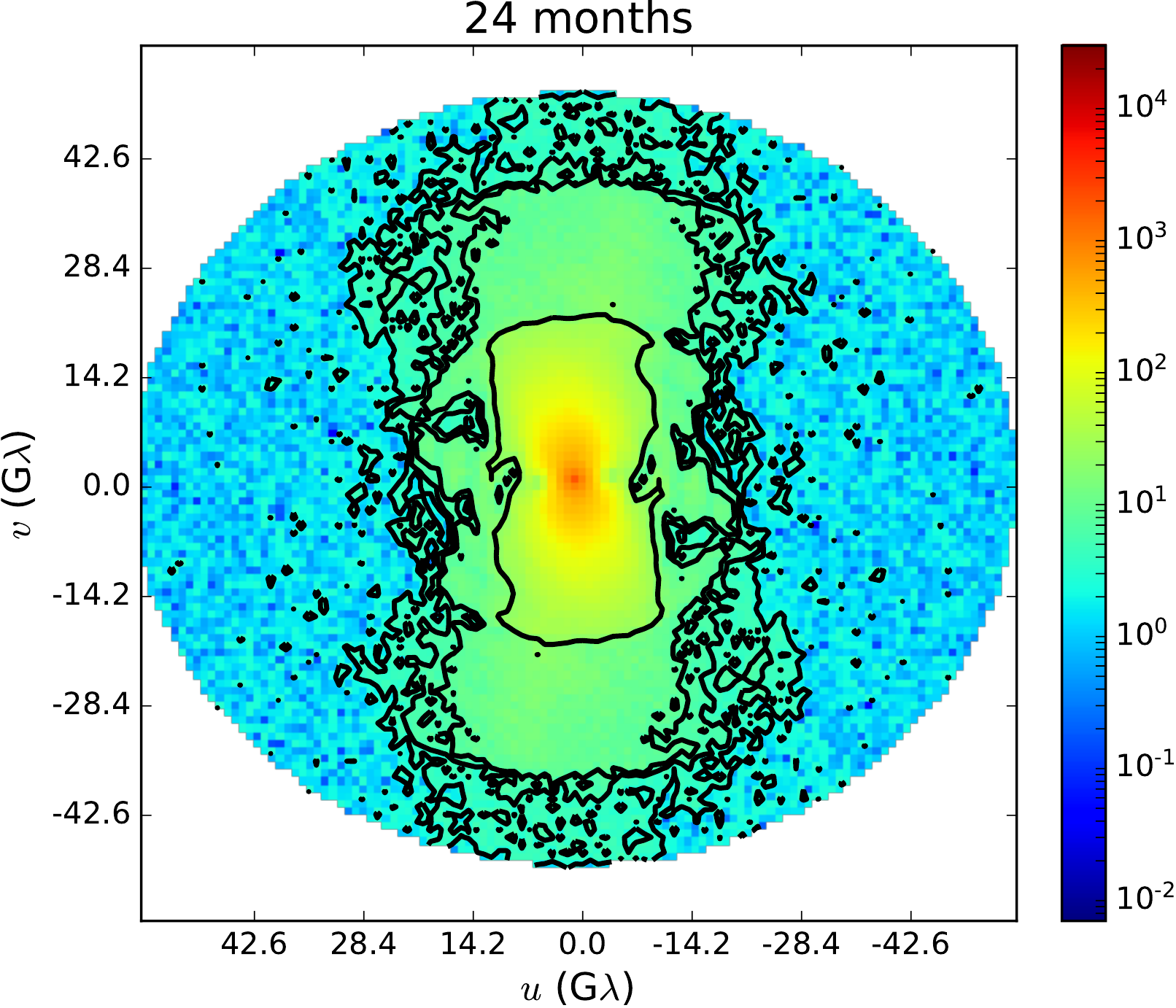}
\includegraphics[scale=0.35]{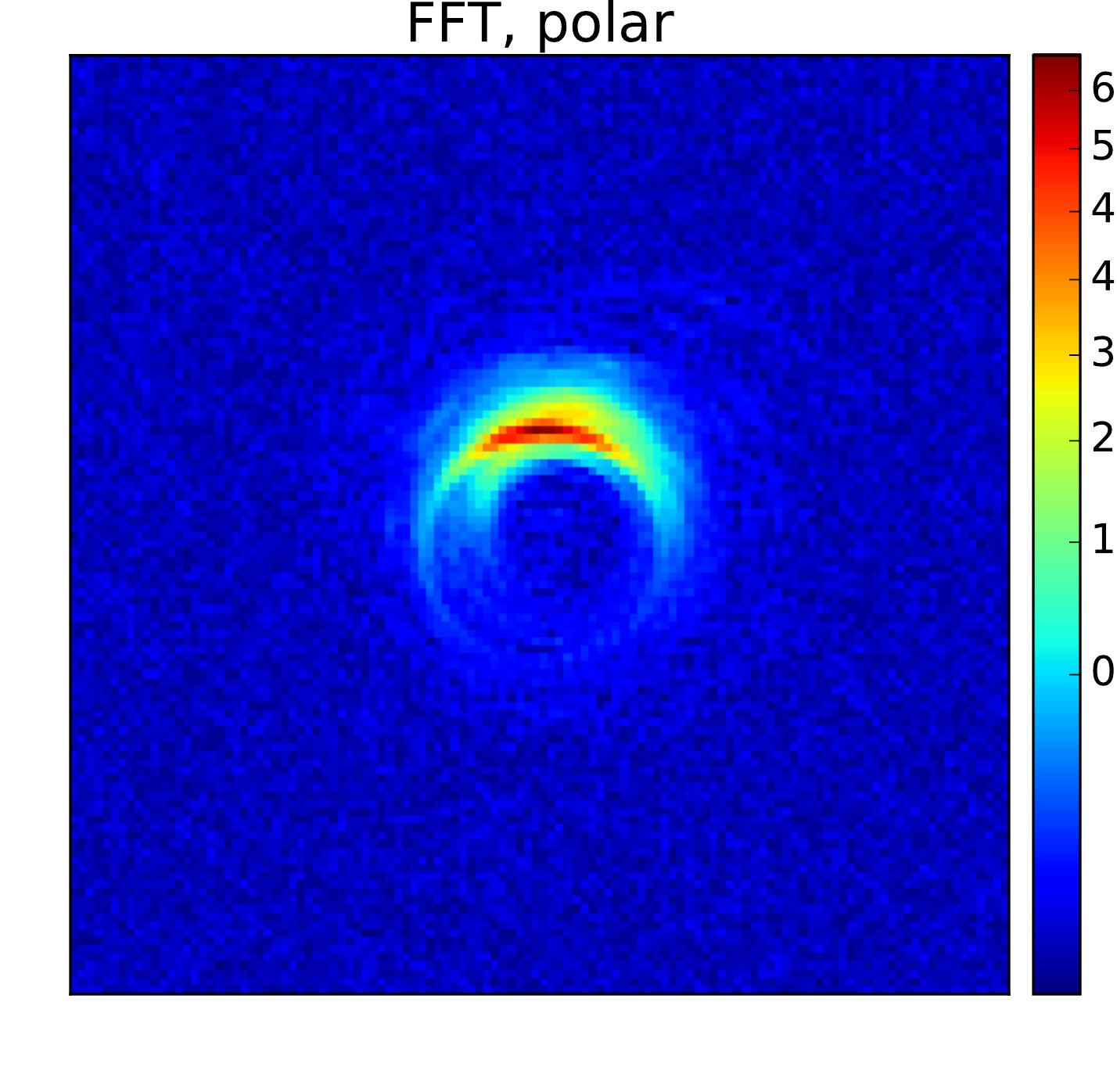}\\
\includegraphics[scale=0.35]{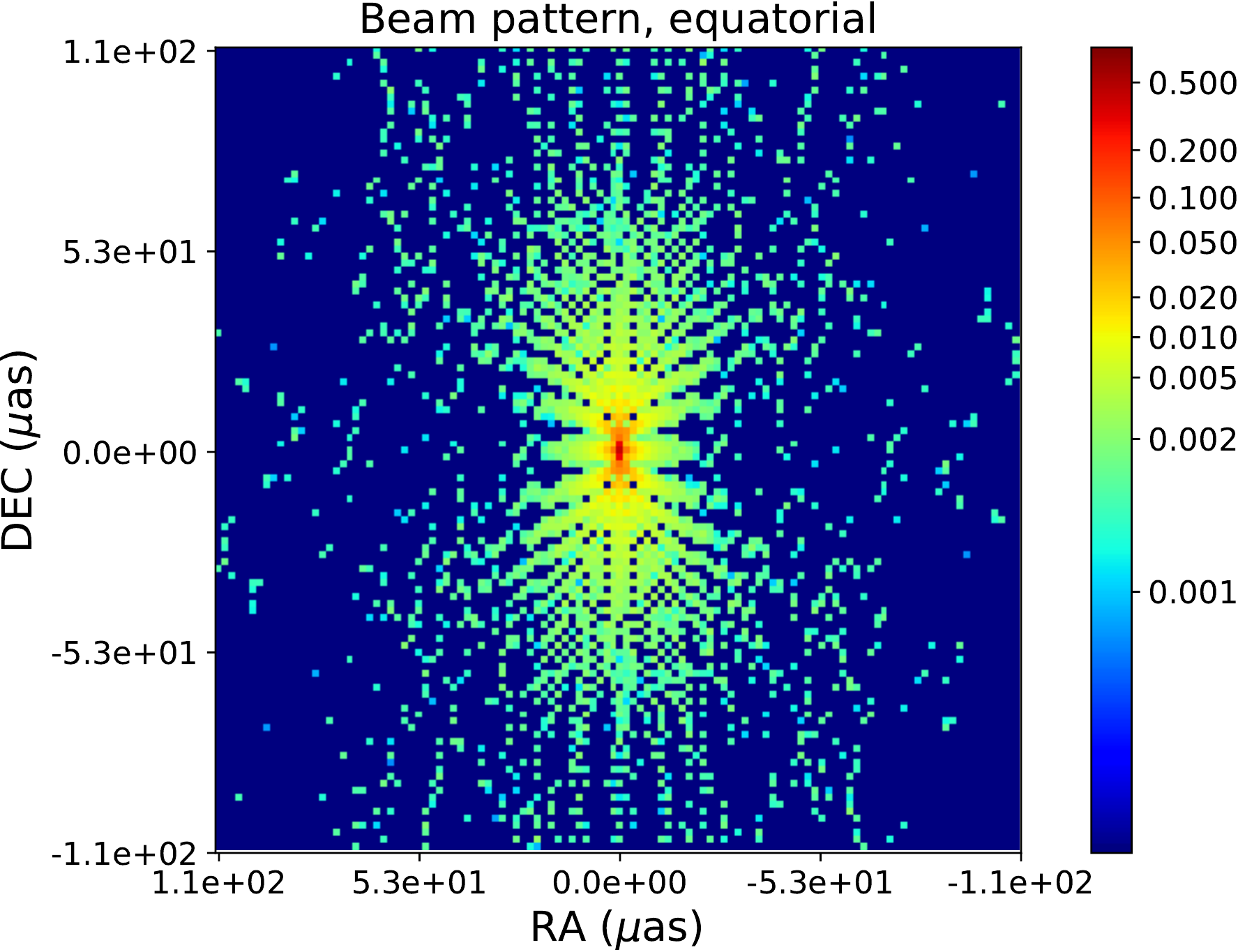}
\includegraphics[scale=0.35]{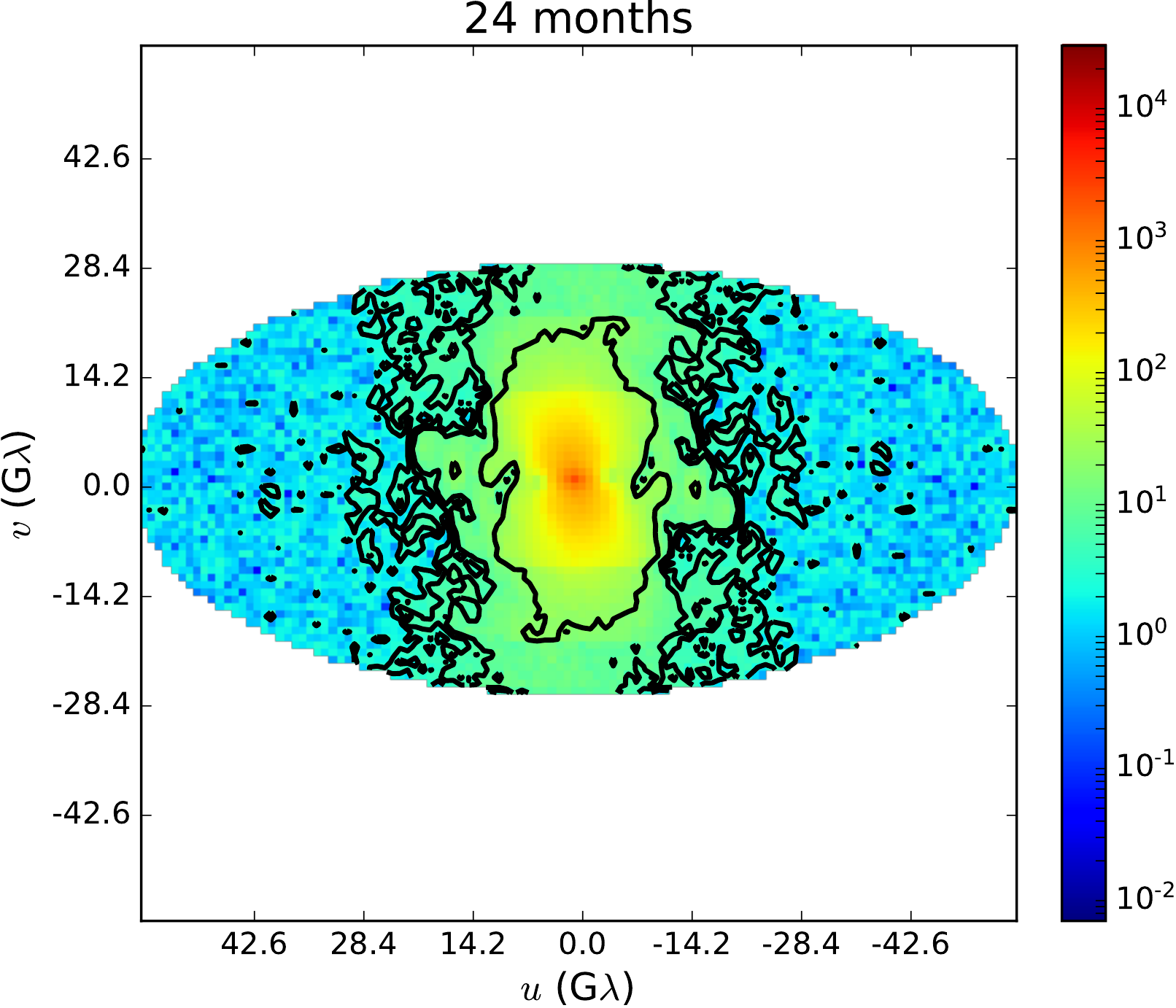}
\includegraphics[scale=0.35]{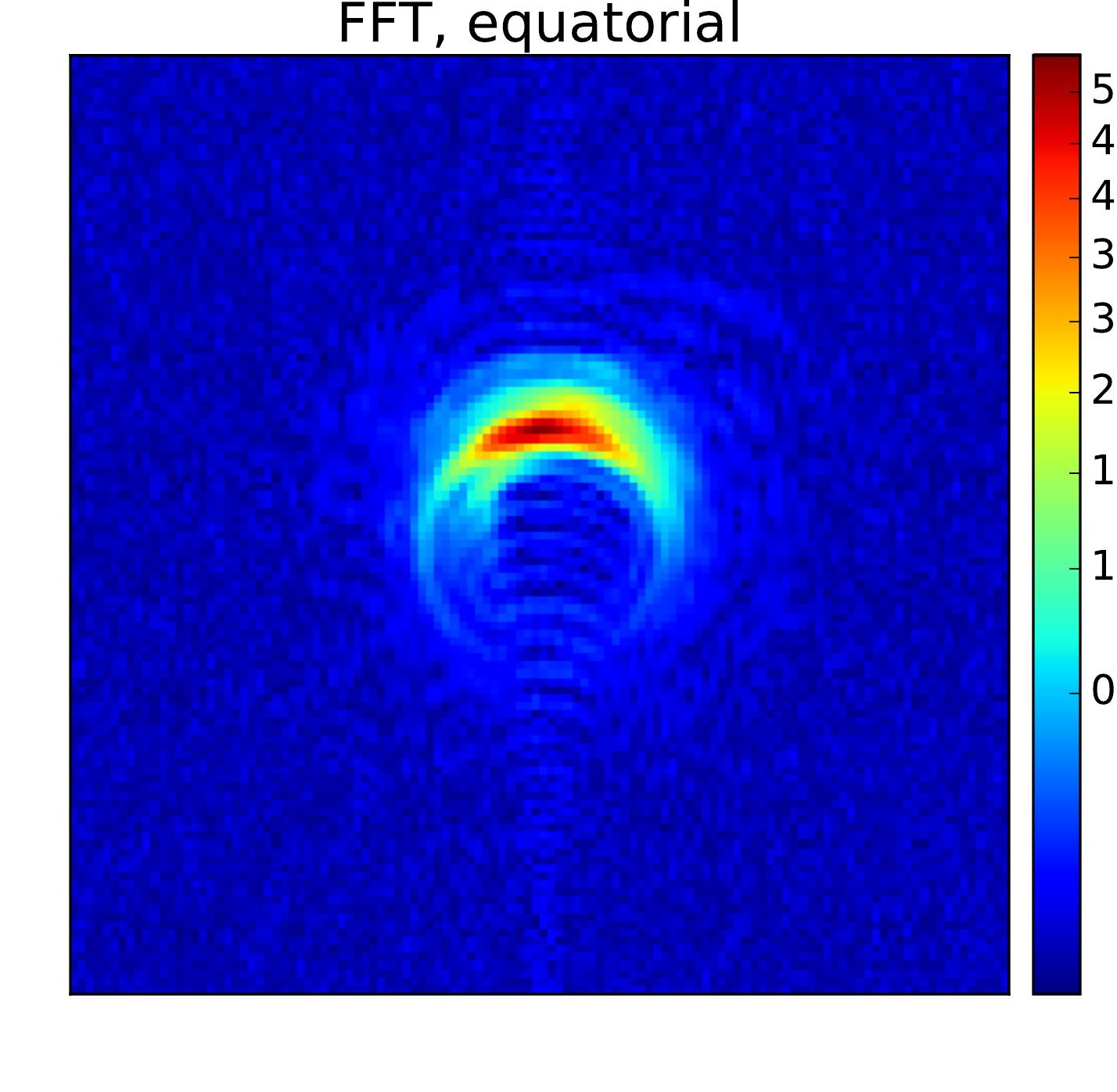}
\caption{Beam pattern (left), S/N (middle), and FFT (right) of the gridded visibilities of model 39 at a frequency of 690\,GHz, rotated by 90 degrees on the sky and observed for 24 months with a 4.4-meter reflector. The upper panels show the simulations in case of polar satellite orbits, and the lower panels correspond to equatorial satellite orbits, taking the declination of Sgr\,A* ($-29^{\circ}$) into account.}
\label{fig:dec}
\end{figure*}

\subsubsection{Source declination}
\label{sec:dec}
In previous simulations, the orbital plane of the satellites was set perpendicular to the line of sight to the source, so that the $uv$-coverage had the shape of a circular spiral. This orientation was chosen in order to keep the simulations free from any preferential directions with respect to the source geometry. In practice the orbits will be polar or equatorial, while the source is at a certain angle with respect to the orbital plane. The line of nodes will remain perpendicular to the line of sight, so that in case of a polar orbit the projected east-west baselines can be maximized (although a compromising orientation may be chosen if multiple sources are observed from the same orbit). The declination of Sgr\,A* is $-29^{\circ}$, so that the angle between the line of sight and the orbital plane is $29^{\circ}$ in case of an equatorial orbit, and $61^{\circ}$ if the orbit is polar. The effect of source declination on the $uv$-coverage is two-fold. Due to the different projection of the orbital plane as seen from the source, the baselines will get shortened in the north-south direction by a factor $\sin{\alpha}$, where $\alpha$ is the angle between the orbital plane and the line of sight. Also, depending on $\alpha$ and the orbital radius, the satellites may traverse the Earth's shadow during their orbit, so that source visibility is temporarily lost and gaps occur in the $uv$-plane. For an orbital radius $R$, this will occur if $\alpha$ is smaller than
\begin{equation}
\alpha_{\text{crit}}=\arcsin\left(\frac{R_{\text{E}}}{R}\right),
\end{equation}
where $R_{\text{E}}$ is the radius of the Earth. Using $R=13,892$\,km (as considered for the simulations above) and $R_{\text{E}}=6378$\,km, $\alpha_{\text{crit}}=27.3^{\circ}$. Thus, in the case of Sgr\,A*, the satellites will not be in the Earth's shadow for the orbital radius considered here, for both the polar and equatorial configuration. The only effect of the source declination on the $uv$-coverage will be the foreshortening of the baselines as described above.

\begin{figure*}[!h]
\centering
\includegraphics[scale=0.35]{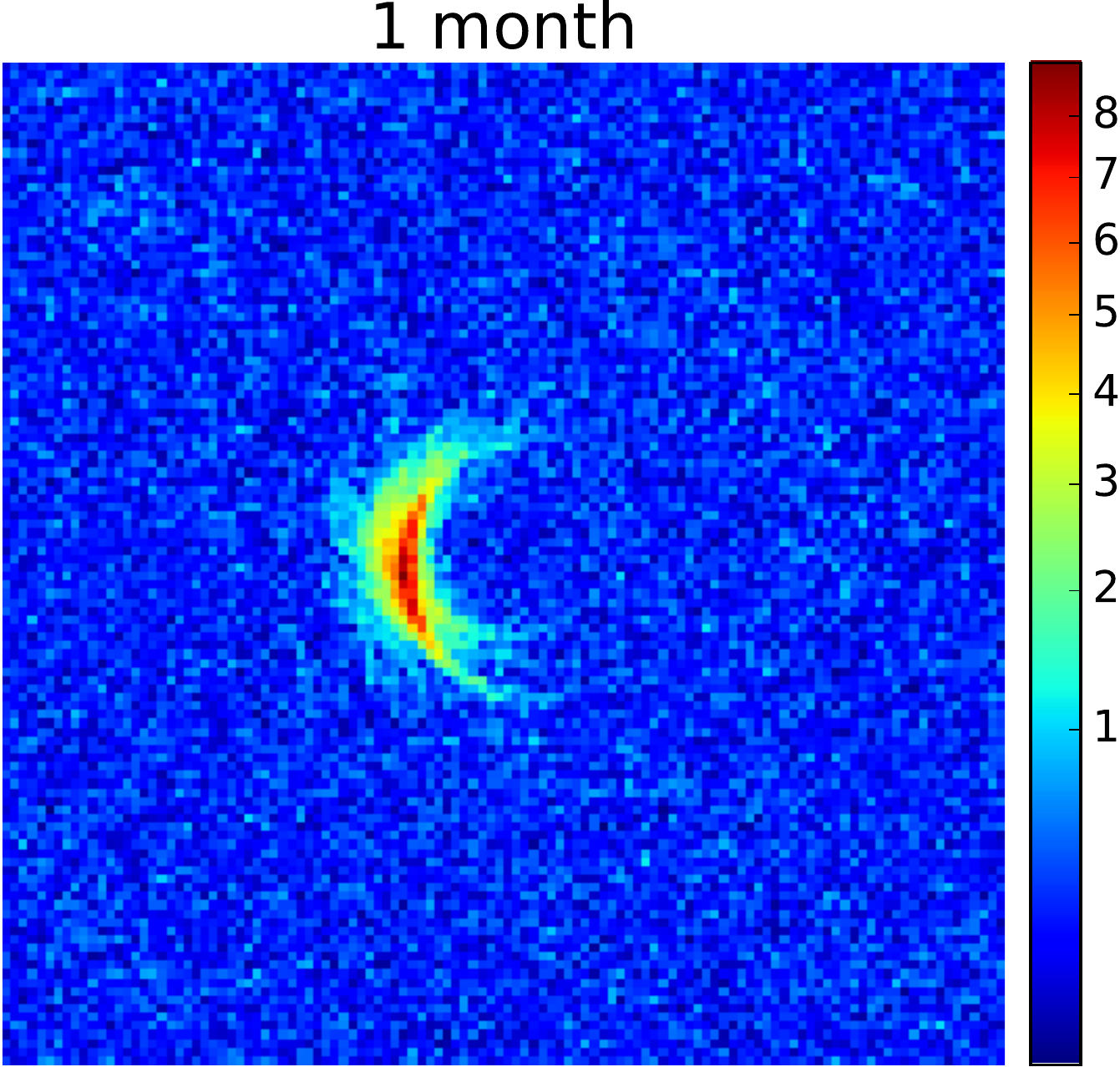}
\includegraphics[scale=0.35]{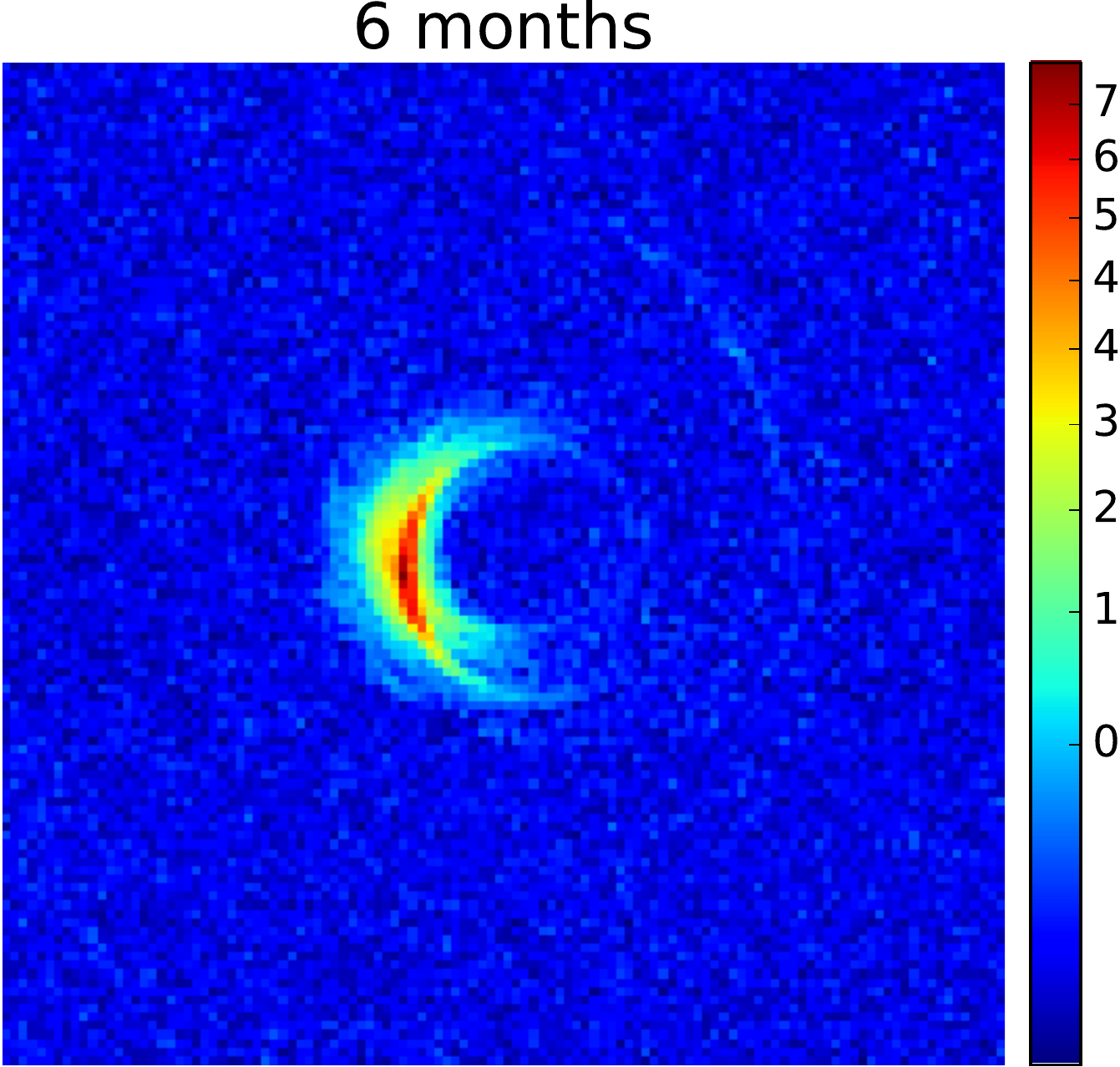}
\includegraphics[scale=0.35]{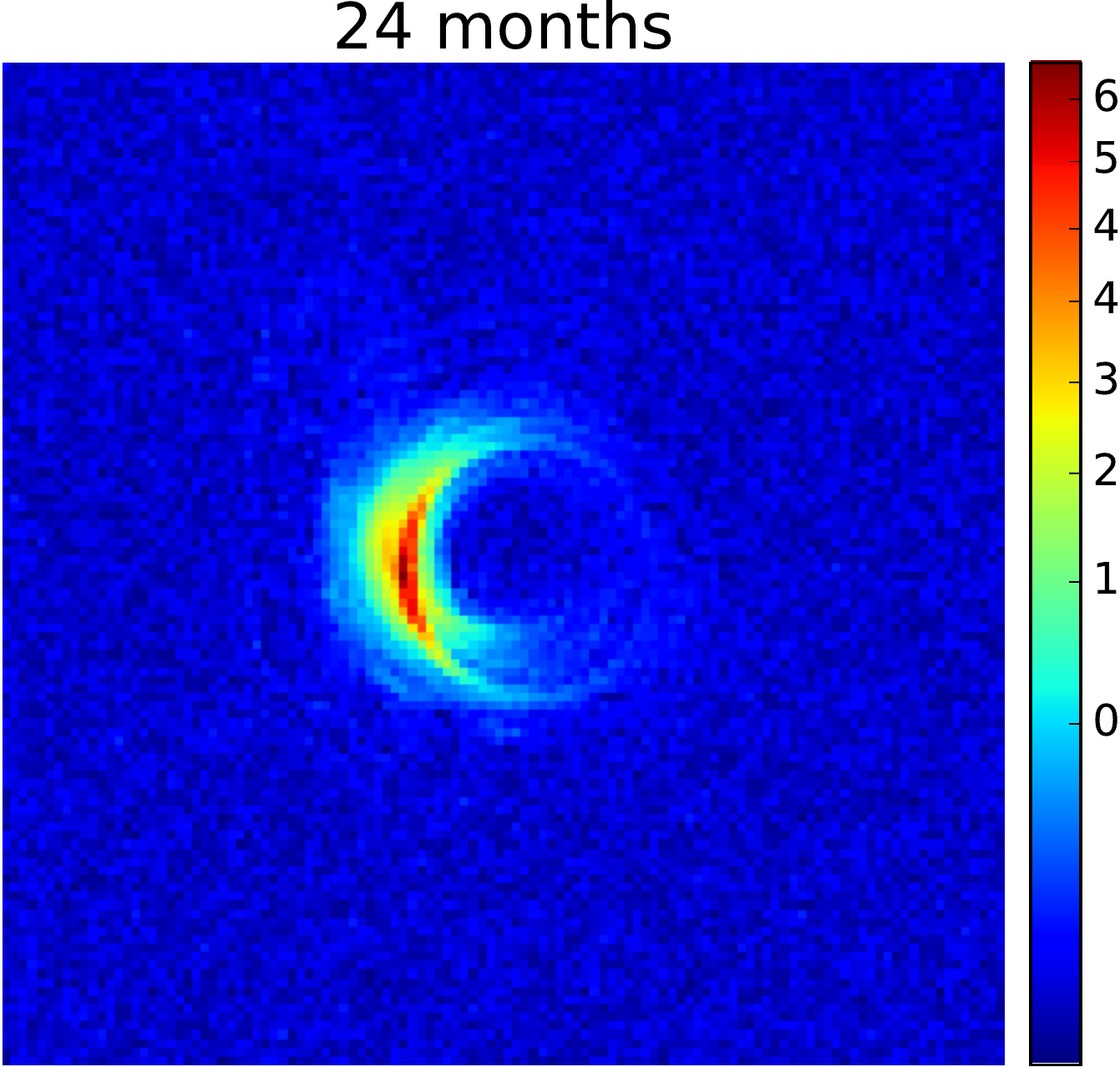}
\caption{FFT of the gridded visibilities of model 39 observed as a movie with a 4.4-meter reflector for 1, 6, and 24 months (left to right) at 690\,GHz. The movie was scattered with a scattering screen traversing in front of the source \citep{Johnson2016so, Johnson2018}. NRMSE values from left to right are 0.75, 0.56 and 0.39.}
\label{fig:tv}
\end{figure*}

We simulated observations of model 39 with a 4.4-meter reflector and 24 months of integration considering the declination of Sgr\,A* in either polar or equatorial orbits. The model image was rotated on the sky by 90 degrees to maximize the effect of baseline shortening in the direction where the S/N is highest. Figure \ref{fig:dec} shows the resulting beam pattern, S/N map, and dirty images. The beam pattern (left panel) is the FFT of the $uv$-coverage, assigning a value of 1 to the grid cells containing data points, and zeroes to the empty grid cells. The beam pattern is indeed more elongated for an equatorial orbit, and the baselines get shortened by a factor $\sin{29^{\circ}}\sim 0.5$ in the $v$-direction. This factor is $\cos{29^{\circ}}$\,$\sim$\,0.87 for a polar orbit. Projected baseline shortening leads to a slight increase in S/N (middle panel) in the $u$-direction as the grid cells contain more points. The reconstructed image for an equatorial orbit (right panel) shows some artefacts due to the beam pattern, which may be taken out by using e.g. a CLEAN algorithm \citep{Hogbom1974}. For both the polar and equatorial orbits, the black hole shadow is still well visible in the dirty images. Because of the low S/N on long baselines and extremely high resolution, a foreshortening factor of 2 does not severly limit our ability to image the source. With a single orbital setup, it is therefore possible to observe sources in a wide angular range on the sky.

\subsubsection{Source and scattering variability}
\label{sec:tv}

GRMHD simulations of Sgr\,A* and VLBI data products simulated from these exhibit variability on short time scales \citep[$\sim$ minutes, e.g.][]{Moscibrodzka2009, Dexter2010, Shiokawa2013, Dexter2013, Kim2016, Medeiros2016, Roelofs2017} at mm-wavelengths due to orbital dynamics of the turbulent structure. Integrating observations for multiple months thus strongly violates the static source assumption in standard aperture synthesis imaging.

In case of the EHT, attempting to image one day of observations of Sgr\,A* as a variable source with standard imaging methods will indeed lead to unsatisfactory results because of this violation: the measured visibilities correspond to different images at different $uv$-points \citep{Lu2016}. However, \citet{Lu2016} also show that one can make use of the linearity of the Fourier transform to reconstruct the average image of the time-variable source. Averaging multiple images is equivalent to averaging the corresponding visibilities in $uv$-space. The important features imprinted on the observed source by general relativistic effects, such as the size of the lensed photon ring and crescent shape caused by Doppler boosting, are continuously present in the image and will therefore remain prominent in the average image of the source. The turbulent substructure can then be averaged out if one observes enough epochs, provided that the variability indeed occurs on small spatial scales. In combination with additional methods such as normalizing the visibilities to the total flux of the source and applying a smoothing algorithm in the $uv$-domain, averaging 8 days of observations before imaging leads to a reconstruction that is almost equally similar to the input model as observing the time-averaged source for one day \citep{Lu2016}. Furthermore, if sufficient $uv$-coverage is obtained on source variability time scales, dynamical imaging methods \citep{Johnson2017, Bouman2017} could be used to reconstruct movies of the source and solve for a time-averaged image simultaneously.

In the case of our space VLBI system, the situation is similar to the EHT in that the total integration time is much longer than the variability time scale of the source. Dynamical imaging methods would likely not work here because there are only one or three baselines at each time. With two small dishes, observing multiple epochs and averaging visibilities is already necessary in order to obtain an S/N that is sufficient for imaging on the longest baselines (Fig. \ref{fig:snr_spiral}). Hence, the method from \citet{Lu2016} described above could be used to mitigate source variability.

GRMHD simulations of source variability over time scales of months are not available. However, we demonstrate here that this method works in principle by simulating SVLBI observations of the 81 GRMHD movie frames of model 39 from \citet{Moscibrodzka2014} that were used to obtain the averaged image in Figure \ref{fig:models}. The frames were spaced by 10\,$GM/c^3$, corresponding to 221 seconds for Sgr\,A*, resulting in a total movie duration of 5 hours. To include the effect of refractive substructure rather than just the ensemble average scattering kernel, the movie was scattered with a a scattering screen traversing in front of the source using the \texttt{stochastic optics} module in \texttt{eht-imaging} \citep{Johnson2016so, Johnson2018}. The frames were observed with two 4.4-meter reflectors, and the movie was repeated every time the last frame was reached. The resulting visibilities were gridded and averaged as described in Section \ref{sec:conint}. 

Figure \ref{fig:tv} shows reconstructed images after integrating for 1, 6, and 24 months. The average source structure showing the size and shape of the black hole shadow can in principle be recovered using this method. After one iteration of the spiral, the source structure is well visible already. Of course, the 5-hour GRMHD movie may not be representative of the source variability over multiple months. Due to e.g. flaring activity, the source may undergo more radical and large-scale changes over this time period, and recovering the average quiescent structure may be more challenging. In future studies, the limitations of imaging a possibly more strongly varying source should be investigated more deeply.  

\begin{figure*}[!h]
\centering
\includegraphics[width=\textwidth]{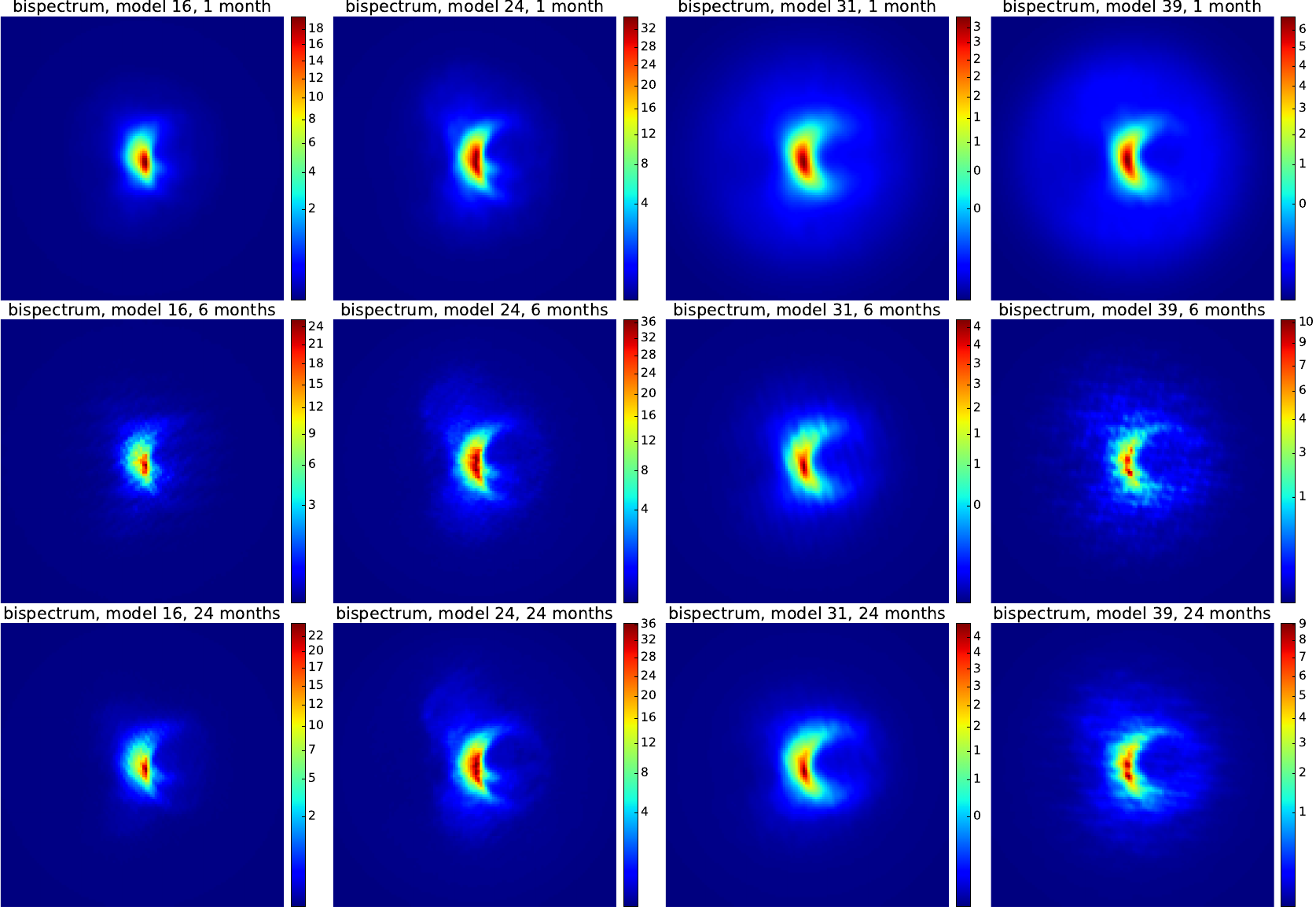}

\caption{Image reconstructions of simulated observations of all 690\,GHz models (left to right) with a three-satellite MEO system using the bispectrum alone. The bispectrum was accumulated over 1, 6, and 24 months (top to bottom). Images were reconstructed using a MEM algorithm implemented in \texttt{eht-imaging} \citep{Chael2016, Chael2018}. NRMSE-values are shown in Table \ref{tab:nrmse3sat}.}
\label{fig:3sat}
\end{figure*}

\subsection{Three-satellite system}
\label{sec:alt}
As an alternative to building a system that behaves like a connected interferometer, a third satellite could be added to the system so that closure phases can be formed. Closure phase is the phase of the bispectrum, which is the product of complex visibilities on a triangle of baselines \citep{Jennison1958, Rogers1974}. Hence, closure phase is the sum of the individual phases on the triangle baselines. They are immune to station-based phase errors due to e.g. positioning offsets in the reconstructed orbital model, as these cancel out when the phases are summed.

\subsubsection{Static source}
Figure \ref{fig:3sat} shows image reconstructions for a three-satellite system observing the time-averaged models, where the third satellite was put at a radius of 13899\,km, which is at one third of the distance between the inner- and outermost satellite. Apart from the reflector diameter, which was set to 4.0\,m so that an Ariane 6 spacecraft could fit three, the noise parameters were kept equal to the two satellite system. The images were reconstructed with \texttt{eht-imaging} using the bispectrum accumulated over 1, 6, and 24\,months. For the averaging of 6 or 24 iterations of the $uv$-spiral, we assumed a system for which the individual phases are corrupted (i.e. no connected interferometer-like behavior as outlined in Section \ref{sec:conint}). Hence, we did not average complex visibilities, but we averaged the bispectrum, which still has coherent (closure) phases. Thermal noise on closure phases is Gaussian down to an S/N of $\sim$\,3, where it starts to deviate \citep[e.g.][]{Rogers1995, Roelofs2017}. It is still close to Gaussian (at the level of $\sim$\,a few percent) down to an S/N of 1, which makes multi-epoch averaging of the bispectrum a viable method to accumulate S/N. This is also reflected in the images, which show an increasing amount of detailed structure as the number of observing epochs increases. 

The image quality after one month of integration with three 4.0-meter satellites is better than for the two 4.4 meter satellites that only use visibilities with S/N\,$>7$ (Fig. \ref{fig:highsnr}). NRMSE values for the three-satellite reconstructions (Table \ref{tab:nrmse3sat}) are generally in between those for the two 4.4 and 25-meter dishes using S/N $>7$ visibilities (Table \ref{tab:nrmsehighsnr}) after integrating for one month. Comparing these NRMSE values to the ones for images made with gridded visibilities is not reliable because the images were made in a different way (maximum entropy versus a simple FFT). The latter have a systematically higher NRMSE as the noise on the gridded visibilities is transferred to the image plane, while the MEM algorithm fits a model image to the data, resulting in an artificially high dynamic range. As the total integration time increases, the three-satellite images visually do not become quite as sharp as the images made with complex visibility gridding (Fig. \ref{fig:fftmodels}). Possible contributing causes are a higher noise level due to smaller dishes, systematics caused by averaging of data points with S/N\,$<1$ where the error distrubution starts to significantly deviate from a Gaussian, and the fact that less information was used for image reconstruction (bispectrum vs complex visibilities).

\begin{table}[]
\centering
\caption{NRMSE values for image reconstructions in Figure \ref{fig:3sat}}
\label{tab:nrmse3sat}
\begin{tabular}{l|lll}
Model & 1 month & 6 months & 24 months \\ \hline
16   &  0.30 & 0.29 & 0.25                \\
24   &  0.17 & 0.16 & 0.14               \\
31   &  0.38 & 0.31 & 0.28               \\
39   &  0.36 & 0.34 & 0.31     
\end{tabular}
\end{table}

\begin{figure*}[!h]
\centering
\includegraphics[scale=0.35]{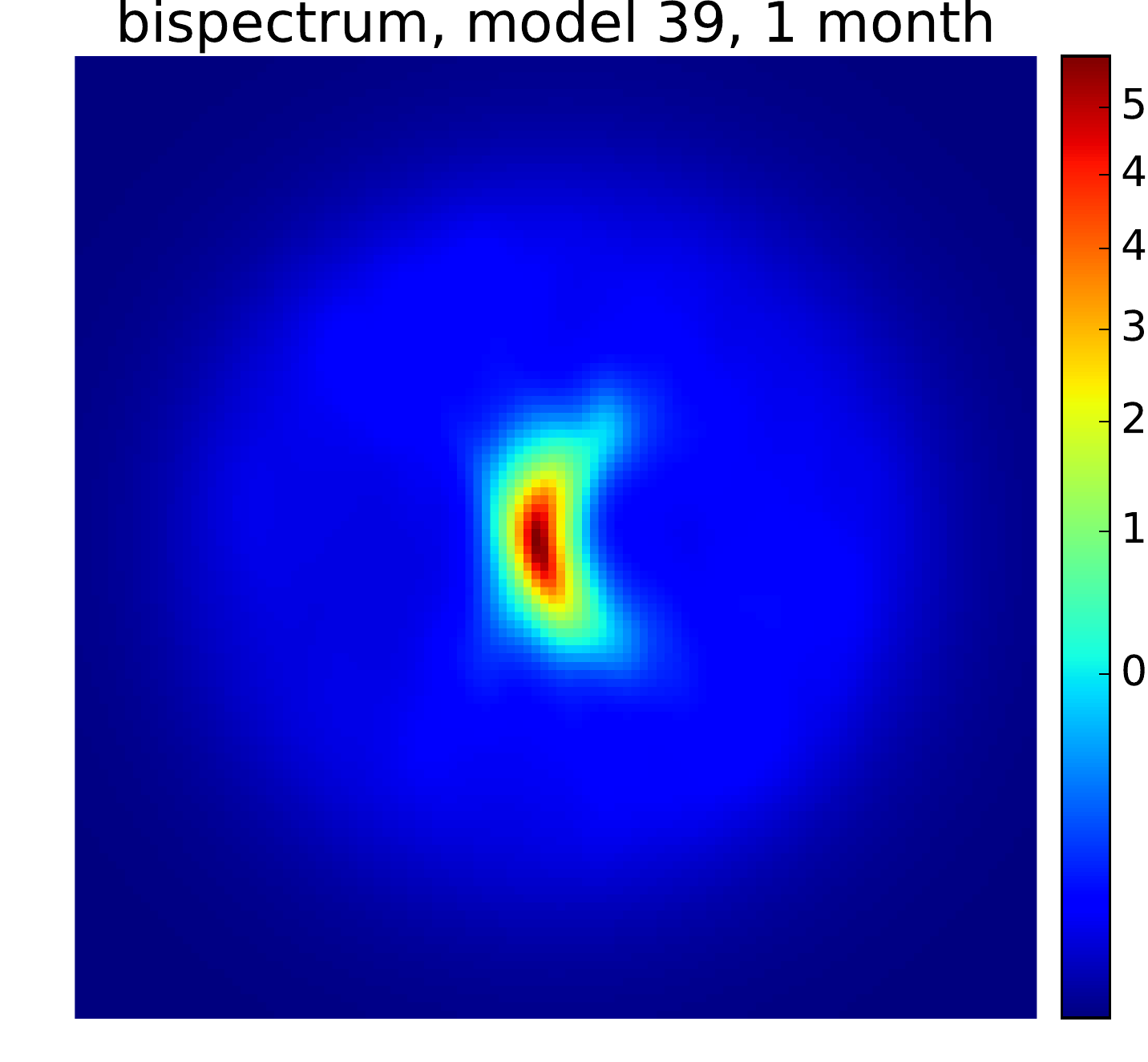}
\includegraphics[scale=0.35]{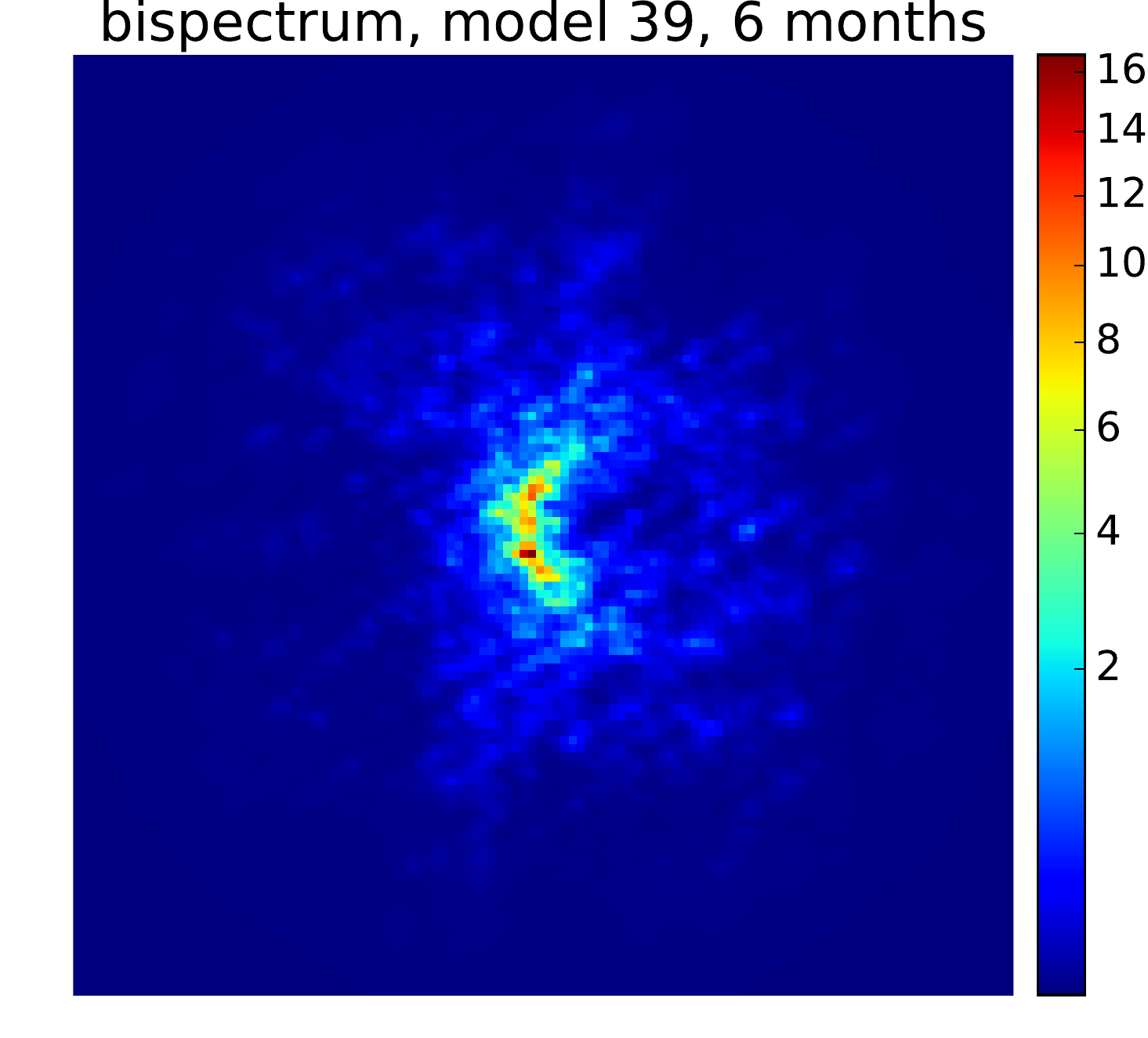}
\includegraphics[scale=0.35]{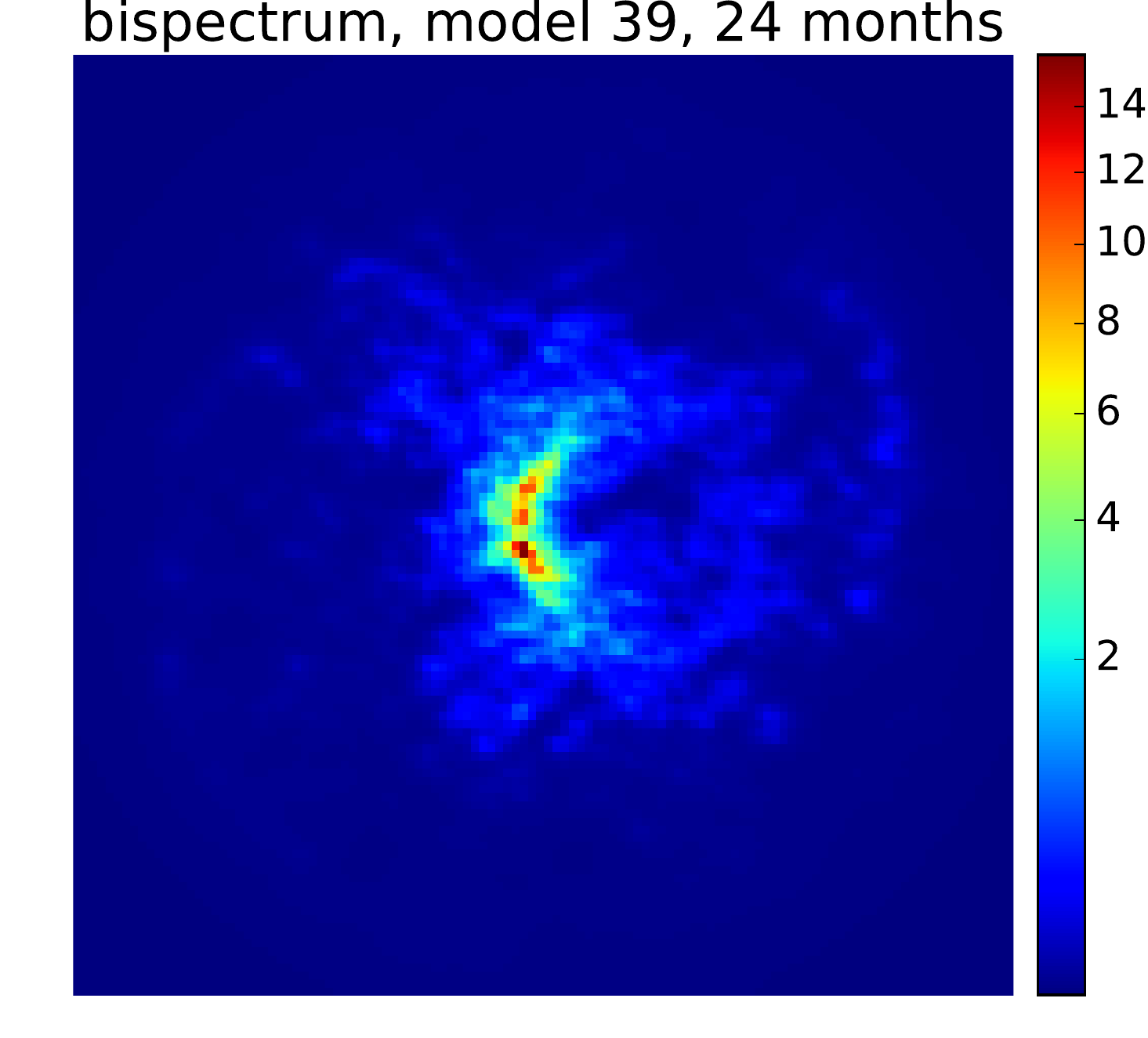}
\caption{Same as Fig. \ref{fig:3sat} but for model 39 at 690 GHz observed as a movie. NRMSE-values are 0.39, 0.52, and 0.45 from left to right.}
\label{fig:3satmov}
\end{figure*}

\subsubsection{Variable source}
Figure \ref{fig:3satmov} shows the same as Figure \ref{fig:3sat}, but for the GRMHD model 39 observed as a movie instead of a time-averaged image. Here, the image quality is significantly worse than for the two-satellite observation of the time-variable source (Fig. \ref{fig:tv}), with more spurious substructure. The reason for this difference is the fact that the average of a set bispectra does not correspond to the Fourier transform of the average of the set of images corresponding to those bispectra, which was also noted by \citet{Lu2016}. The relation holds for complex visibilities, but not for the triple product of these. So, for a time-variable source, the bispectrum alone will not be sufficient to reconstruct a static image with a quality similar to the two-satellite system employing a two-stage correlation scheme. Either a combination of these techniques should be used, or more advanced (dynamical) imaging techniques should be developed for this purpose.

\section{Summary and outlook}
\label{sec:sum}
In this paper, we have presented imaging simulations of the EHI SVLBI system consisting of two MEO satellites in circular orbits at slightly different radii, as discussed by \citet{Martin2017} and \citet{Kudriashov2017}. The EHI could be used to image the black hole shadow of Sgr\,A* up to frequencies of about 690\,GHz. Such high observing frequencies can be reached in space because of the absence of atmospheric corruptions. The setup allows for long baselines (up to $\sim$\,60\,G$\lambda$ at 690 GHz) resulting in a maximum image resolution of 4\,$\mu$as, which is a significant improvement compared to the $\sim$\,23\,$\mu$as resolution that can be obtained with EHT baselines at 230\,GHz. The two-element interferometer setup results in a spiral-shaped sampling of the $uv$-plane with a density that cannot be obtained with Earth-based VLBI, so that high-fidelity images can be reconstructed. Apart from the higher resolution, advantages of observing at higher frequencies are the small interstellar scattering and source variability effects at 690\,GHz compared to 230\,GHz, and the closer origin of the emission to the event horizon. 

Using GRMHD simulations of Sgr\,A* and model system parameters, we have performed simulated observations in order to assess the image quality that can be expected. The signal-to-noise ratio of the measured visibilities is expected to be <7 on baselines longer than 10-20\,G$\lambda$, preventing robust fringe detection on these baselines using conventional VLBI methods. However, the detection threshold may be decreased by using a system with excellent clock and orbit reconstruction ($\lesssim$\,0.1\,mm) accuracy. Higher-S/N measurements may then be obtained by averaging visibilities measured in different iterations of the $uv$-spiral. If such a system cannot be built within a reasonable budget, one would need to launch two 25-meter antennas rather than 4.4-meter antennas, in order to obtain sufficent S/N for conventional fringe fitting on long baselines.

At 230\,GHz, the expected image resolution is comparable to the expected resolution of the images produced by the EHT because of stronger scattering effects on long baselines, although the reconstructed SVLBI images are more robust due to the dense and uniform $uv$-coverage. At 690\,GHz, interstellar scattering only has a small effect on the observed image, and the proposed setup could allow for reconstructed images of Sgr\,A* with unprecedented angular resolution and fidelity within one or a few months of integration.

We have shown that source variability can be averaged out to reconstruct an image of the quiescent source structure showing the photon ring and Doppler-boosted emission. We note that the ability to reconstruct an average image from a time-variable source using this method depends on the nature of the variability. If the variability is caused by small-scale turbulent structures while the large-scale features remain prominent, such as in the GRMHD simulations we have considered, variability can indeed be averaged out. If, on the other hand, there are large-scale structural changes in the source, this will become more difficult. Since Sgr\,A* is a variable source, future studies leading to a full mission proposal should further investigate the ability to reconstruct an image under the assumption of different variability scenarios within the parameter space allowed by existing (EHT) measurements. 

If the phase stability and orbit reconstruction accuracy of the two-satellite system are not sufficient to obtain detections on long baselines, three 4-meter antennas could be launched so that closure phases could be formed and used for imaging. Since closure phases are immune to station-based phase errors, such a system could relax the orbit reconstruction and stability requirements. However, a system solely relying on measurements of the bispectrum poses challenges for imaging a time-variable source. 

There are still significant technical challenges to overcome for the concept to be turned into an actual mission. The main issues to be worked out are the maximum orbit reconstruction accuracy that can be obtained, the complexity of the on-board correlation and processing that would be needed to send reduced data to the ground, and the frequency reference stability for 690\,GHz observations. These challenges should be addressed in future engineering studies. More investigations should also be made into the possibilities of reducing the system noise as this is an important determining factor for the image quality. 

The EHI concept could be of great astrophysical interest as it allows for precise tests of general relativity and accretion models. A quantitative comparison of the precision of these tests between the EHT and the SVLBI experiments discussed here, with inclusion of all instrumental corruptions, should be done as the project develops further. Furthermore, observations of GRMHD data that are ray-traced in full Stokes \citep[e.g][]{Gold2017, Moscibrodzka2018} could be simulated in order to infer what could be learned about the magnetic field structure near the event horizon.

Apart from Sgr\,A*, other sources will be interesting to observe with the SVLBI concept presented here as well. Emission from M\,87, the black hole with the second largest apparent size on the sky and also a prime EHT target, is not affected by interstellar scattering. Imaging it at 230\,GHz with the long baselines of the MEO SVLBI experiment could thus have a significant advantage over imaging it from the ground at the same frequency. Another advantage of imaging M\,87 is that it is variable at $\sim$\,10$^3$ times longer time scales than Sgr\,A*, possibly allowing for static snapshot reconstructions and multi-epoch dynamical reconstructions depending on the satellite separation which sets the radial $uv$-filling speed. Since GRMHD simulations of M\,87 exhibit similar features as GRMHD simulations of Sgr\,A* at millimeter wavelengths \citep[e.g.][]{Moscibrodzka2016}, the static imaging results presented here for Sgr\ A* may be largely applicable to M\,87 as well, provided that its mass is close to the $6.6\times 10^9M_{\odot}$ measured by \citet{Gebhardt2011}.

The Sobrero Galaxy M\,104 hosts a supermassive black hole of $\sim$ $10^9M_{\odot}$ \citep{Kormendy1996} at a distance of $9.55 \pm 0.13 \pm 0.31$ Mpc \citep{McQuinn2016}, yielding an apparent event horizon size of $\sim$\,11\,$\mu$as, which can be resolved by EHI baselines. The black hole at the center of the elliptical galaxy M84 has a mass of $8.5^{+0.9}_{-0.8}\times10^8M_{\odot}$ \citep{Walsh2010}. At a distance of 17\,Mpc, the size of the event horizon is $\sim 5$\,$\mu$as on the sky, which is comparable to the EHI resolution at 690\,GHz. M\,81* has a black hole with mass $7.0^{+2}_{-1}\times10^7M_{\odot}$ \citep{Devereux2003} at a distance of $3.63 \pm 0.34$\,Mpc \citep{Freedman1994}, yielding an apparent event horizon diameter of $\sim$\,2\,$\mu$as. Another close active galactic nucleus (AGN) is Centaurus A, at a distance of $3.8 \pm 0.1$\,Mpc \citep{Harris2010} and with a black hole mass of $(5.5 \pm 3)\times10^7M_{\odot}$ \citep{Neumayer2010}. For this black hole, the event horizon ($\sim$\,1\,$\mu$as) may be too small to resolve with the setup discussed here, but at this distance the 4\,$\mu$as angular resolution at 690\,GHz corresponds to a linear scale of only 2 light hours. This would enable one to image the structure of the relativistic jet on a length scale that is two orders of magnitude shorter than what has been achieved earlier \citep{Muller2014}. Similarly, jets of several other AGN could be studied in detail in order to improve our understanding of jet launching and collimation.

Different variations of the presented concept could be explored. Depending on the technical possibilities, one could try to push for even higher frequencies, which would increase the resolution further. A shorter separation of the orbits might enable studying various objects, such as protoplanetary disks \citep[e.g.][]{Hogerheijde2011}, at lower resolution for many orbits before the satellites separate. Further studies should assess whether this could lead to valuable science. Another possibility is investigating a space-space-ground hybrid system that can perform both lower-frequency space-ground observations for dynamical imaging \citep{Palumbo2019} and higher-frequency space-space observations for high-resolution static imaging.  

\begin{acknowledgements}
This work is supported by the ERC Synergy Grant “BlackHoleCam: Imaging the Event Horizon of Black Holes” (Grant 610058). We thank Andrew Chael and Katie Bouman for making the eht-imaging software publicly available, and for providing the code for regridding and aligning the images and calculating the NRMSE. The design study was the result of a colloquium on space interferometry by H. Falcke at ESTEC on Sep 2, 2015 and an internal ESA study organized by M. Martin-Neira. We are grateful to the anonymous referee for useful and constructive comments. M. Mo\'scibrodzka acknowledges H. Shiokawa for providing the HARM3D GRMHD simulation data used in our ray-tracing simulations. We thank Alan Roy, Michael Bremer, Jason Dexter, Itziar Barat, Thijs de Graauw, Vincent Fish, Andrey Baryshev, Andr\'e Young, and Daniel Palumbo for useful comments and discussions on this work.
\end{acknowledgements}

\bibliographystyle{aa} 
\bibliography{bibliography}

\end{document}